\documentclass[fleqn,usenatbib]{mnras}

\usepackage{newtxtext,newtxmath}
\usepackage{subcaption}
\usepackage[T1]{fontenc}

\DeclareRobustCommand{\VAN}[3]{#2}
\let\VANthebibliography\thebibliography
\def\thebibliography{\DeclareRobustCommand{\VAN}[3]{##3}\VANthebibliography}

\usepackage{graphicx}	% Including figure files
\usepackage{amsmath}	% Advanced maths commands

\title[The UV Luminosity Function Turnover with JWST]{Tilting at the Turnover: Modeling the Faint-End of the UV Luminosity Function Behind  Abell s1063 with JWST}

\author[C. M. Goolsby et al.]{
Caio Moreira Goolsby,$^{1}$\thanks{E-mail: caio.goolsby@postgrad.manchester.ac.uk}
Christopher J. Conselice,$^{1}$
Duncan Austin,$^{1}$
Tom Harvey,$^{1}$
Jose Diego,$^{2}$
\newauthor
Nathan Adams$^{1}$
Julien Marabotto,$^{1}$
Jordan D'Silva,$^{1}$
and Qiong Li$^{1}$
\\
$^{1}$Jodrell Bank Centre for Astrophysics, University of Manchester, Oxford Road, Manchester UK\\
$^{2}$Instituto de Física de Cantabria (CSIC-UC). Avda. Los Castros s/n. 39005 Santander, Spain\\
}

\date{Accepted XXX. Received YYY; in original form ZZZ}

\pubyear{\the\year{}}

\begin{document}
\label{firstpage}
\pagerange{\pageref{firstpage}--\pageref{lastpage}}
\maketitle

\begin{abstract}
 In this work we leverage the strong gravitational field of Abell S1063 to identify faint, highly magnified galaxies using ultra-deep James Webb Space Telescope (JWST)/NIRCam imaging from the GLIMPSE survey and ancillary Hubble Space Telescope (HST)/ACS imaging from the Hubble Frontier Fields program. We construct a photometric catalogue of lensed high-redshift candidates and use these sources to constrain the faint end of the rest-frame UV luminosity function (UVLF) over $z\simeq6$--11. Rather than treating the UVLF turnover ($M_{\rm t}$) as a hard cutoff, we model it as a gradual quadratic suppression and explicitly account for the potential continued contribution of galaxies beyond the turnover. In a shallow-turnover scenario, up to one-third of the UV luminosity density can arise from sources fainter than $M_{\rm t}$. While we find no direct evidence for a turnover down to $M_{\rm UV}=-13.5$ at $z=6$, our analysis can only confidently exclude weak, medium, and strong turnover models down to $M_{\rm t}=-15.9$, $-15.1$, and $-14.8$, respectively. Across these models, we infer conservative lower limits of the UV luminosity and star formation density as well as the ionization rate as: $\rho_{\rm UV}\geq22\times10^{25}\,{\rm erg\,s^{-1}\,Hz^{-1}\,Mpc^{-3}}$, ${\rm SFRD}\geq25\times10^{-3}\,M_\odot\,{\rm yr^{-1}\,Mpc^{-3}}$, and $\log_{10}(\dot{n}_{\rm ion}/{\rm s^{-1}\,Mpc^{-3}})\geq51.02$. We find that galaxies fainter than the conventional $M_{\rm UV}=-17$ limit contribute more than half of the UV luminosity density and at least $\sim64\%$ of the ionizing photons produced by star-forming galaxies at $z=6$. Because our turnover model permits a suppressed, but non-zero, galaxy population beyond $M_{\rm t}$, sources fainter than the turnover remain important contributors to both $\rho_{\rm UV}$ and $\dot{n}_{\rm ion}$, emphasizing the need to consider the turnover and its shape when assessing the population of galaxies during reionization.
\end{abstract}


%%%%%%%%%%%%%%%%%%%%%%%%%%%%%%%%%%%%%%%%%%%%%%%%%%

%%%%%%%%%%%%%%%%% BODY OF PAPER %%%%%%%%%%%%%%%%%%

\section{Introduction}
\label{sec:intro}

% --- Introduction text block (no equations) ---

Roughly $(3$--$4)\times10^5$~years after the Big Bang, the expanding Universe cooled enough for electrons and protons to combine into neutral hydrogen \citep{Planck2018}. For the next $\sim\!10^8$~years, neutral hydrogen and helium gas dominated the cosmic baryons during the so-called Dark Ages, until the formation of the first stars and galaxies, whose ultraviolet photons began to carve out patchy ionized regions in the surrounding hydrogen clouds   \citep{BarkanaLoeb2001,PritchardLoeb2012}. Between $\!\sim\!40 > z >  \!\sim\!6$, these patches slowly expanded and grew in number during the 'Epoch of Reionization,' until, by $z  \!\sim\!6$, the Universe had been almost entirely reionized as enough highly-energetic photons were being produced to keep the hydrogen throughout the inter-galactic medium (IGM) in a state of almost constant ionization. \citep{Planck2018,Keating2020}.
 
The identity of the sources that drove cosmic reionization has remained a central and evolving question, with three main classes of candidates repeatedly emerging in the literature. One possibility is that accreting black holes, observed as active galactic nuclei (AGN) or quasars, supplied a substantial fraction of the ionizing background. This theory has gained attention in the JWST era following the discovery of faint broad-line AGN candidates at high redshifts, prompting renewed arguments that AGN may have played a more important role than previously assumed \citep{Madau2024}. However, both observational constraints and radiation-hydrodynamic simulations generally continue to disfavour AGN as the dominant drivers of reionization, instead finding that quasars contribute only a small fraction of the required photon budget at $z\sim6$ \citep{Trebitsch2021,Robertson2022,Jiang2025}.

A second possibility is that relatively bright and moderately massive star-forming galaxies provided a large, and perhaps dominant, share of the escaped ionizing emissivity, $\dot{n}_{\rm ion}$. In this picture, the key is not extreme abundance but the combination of substantial UV luminosities with elevated ionizing-photon production efficiencies and escape fractions, particularly in systems selected as Ly$\alpha$ emitters or inferred Lyman-continuum leakers \citep[e.g.,][]{Matthee2022, Mascia_2023}. Recent JWST-based work has strengthened this possible channel of reionization further by showing that as galaxies at magnitudes roughly from $M_{\rm UV}\sim-16$ to $-20$ may, for reasonable escape fractions, supply enough ionizing photons to reionize the Universe without exceeding observational constraints \citep{Simmonds2024}. Nevertheless, this interpretation remains sensitive to how representative such bright or line-emitting systems are of the broader galaxy population, and to whether their inferred escape fractions can be robustly extended to the redshifts and number densities relevant for the reionization era \citep[e.g.,][]{Duncan2015,Robertson2022,Matthee2022}.

A third, and now highly influential, theoretical framework is that the dominant contribution came from the numerous population of faint, dwarf-like galaxies. In this framework, individually modest systems collectively dominate $\dot{n}_{\rm ion}$ through their steeply rising number densities toward low luminosities, potentially aided by high ionizing efficiencies and bursty star-formation histories \citep{Robertson2022,Atek2024Nature}. This picture is supported by recent observational and simulation-based work suggesting that low-mass galaxies dominate the escaped ionizing budget at earlier times, even if more massive systems become increasingly important toward lower redshift \citep{bera2023,Atek2024Nature}. The principal uncertainty for this channel is that it depends critically on the poorly constrained ultra-faint end of the galaxy population, including the shape of the UV luminosity function (UVLF), the location of any faint-end turnover, and the luminosity dependence of both $\xi_{\rm ion}$ and $f_{\rm esc}$ \citep{Robertson2022, Duncan2025_inprep}.

In this context, the ultra-violet luminosity function provides one of the central empirical routes for identifying the sources of reionization, since it links the abundance of galaxies to their ultraviolet luminosities and therefore to the available photon budget. Integrating the UVLF yields the non-ionizing UV luminosity density, $\rho_{\rm UV}$, which can then be converted into an estimate of the escaped ionizing emissivity, $\dot{n}_{\rm ion}$, through assumptions about the ionizing-photon production efficiency, $\xi_{\rm ion}$, and the escape fraction, $f_{\rm esc}$. Thus, reionization depends heavily on the shape of the UVLF and on any luminosity dependence in $\xi_{\rm ion}$ and $f_{\rm esc}$. A steep faint-end slope would favour a substantial contribution from numerous low-luminosity galaxies, whereas a shallower slope, a turnover, or enhanced escape efficiency in brighter systems could redistribute the ionizing photon budget toward more massive galaxies. 

However, the presence of a turnover does not necessarily imply the disappearance of faint galaxies; if the turnover is shallow, a significant population of low-luminosity systems may persist below the turnover magnitude and continue to contribute meaningfully to $\rho_{\rm UV}$. The faint-end form of the UVLF is therefore particularly important, since its normalization, slope, turnover magnitude, and turnover shape directly regulate how much of the total $\rho_{\rm UV}$, and hence the inferred $\dot{n}_{\rm ion}$, can plausibly arise from galaxies below current blank-field detection limits \citep{Robertson2022,Matthee2022,Simmonds2024,Atek2024Nature}.

The shape of the UVLF also encodes the baryonic physics governing galaxy formation in low-mass haloes. Theoretical models generally predict that star formation should become increasingly inefficient at low halo masses as a result of inefficient gas cooling, stellar feedback, radiative feedback, and environmental suppression, producing a downturn or turnover in the abundance of the faintest galaxies. This turnover need not correspond to an abrupt truncation of the galaxy population. Instead, its astrophysical impact depends on both the turnover magnitude and the turnover slope: a gradual suppression can still permit a substantial population of faint galaxies below the turnover, whereas a sharper decline would rapidly reduce their contribution to $\rho_{\rm UV}$. Many simulations place this, still unseen, turnover at approximately $M_{\rm UV}\!\sim\!-8$ to $-12$, although both its location and shape remain strongly model-dependent \citep[e.g.][]{Jaacks2013,2016ocvirk,Yue2016,Gnedin2016}. Moreover, because some of the principal mechanisms driving this suppression are themselves coupled to the evolving radiation background, the turnover may evolve throughout reionization. Photoheating feedback from the ionized intergalactic medium can inhibit gas accretion and retention in shallow potential wells \citep{Gnedin2000,Okamoto2008}, while Lyman--Werner radiation can dissociate molecular hydrogen and raise the minimum halo mass required for cooling and star formation \citep{Wise2008,Visbal2014}. Consequently, the UVLF turnover is not only a measure of the limiting abundance of faint galaxies, but also a potential probe of the time-dependent feedback processes that regulate galaxy formation during reionization.

Deep HST imaging first established the UVLF at $z\!\gtrsim\!6$ to moderately faint limits and provided the initial empirical backbone for reionization models \citep[e.g.][]{Duncan2014,Bouwens2015,Finkelstein2015, Oesch2018,Bhatawdekar2019}. JWST has since extended UVLF measurements to earlier times with programs such as CEERS, JADES, PRIMER, COSMOS-Web, and UNCOVER, but even the deepest blank fields reach only at the very best $\simeq30$–31~AB in the near-IR and typically probe to intrinsic magnitudes of $M_{\rm UV}\!\gtrsim\!-17$ at $z\!\sim\!6$–10 once selection and completeness are fully accounted for \citep{Finklestein2023, Endsley2023_1, Adams2024EPOCHS}. This leaves the ultra-faint regime largely inaccessible in blank fields, and this will remain true likely even during the lifetime of JWST.

Strong gravitational lensing by massive $(z\!\lesssim\!1)$ clusters offers a powerful route to overcome these depth limitations. Magnification boosts the observed fluxes of background galaxies by factors of a few to tens, effectively pushing surveys to much fainter intrinsic luminosities while preserving surface brightness. This approach, pioneered with projects such as the Hubble Frontier Fields, enabled UVLF measurements to be made down to $M_{\rm UV}\!\lesssim\!-15$ and, in favourable regions, potentially even fainter, thereby directly probing the population most relevant for the photon budget \citep{Livermore2017,Bouwens2017,Atek2018}.

\begin{figure*}
    \centering
    \includegraphics[width=\textwidth]{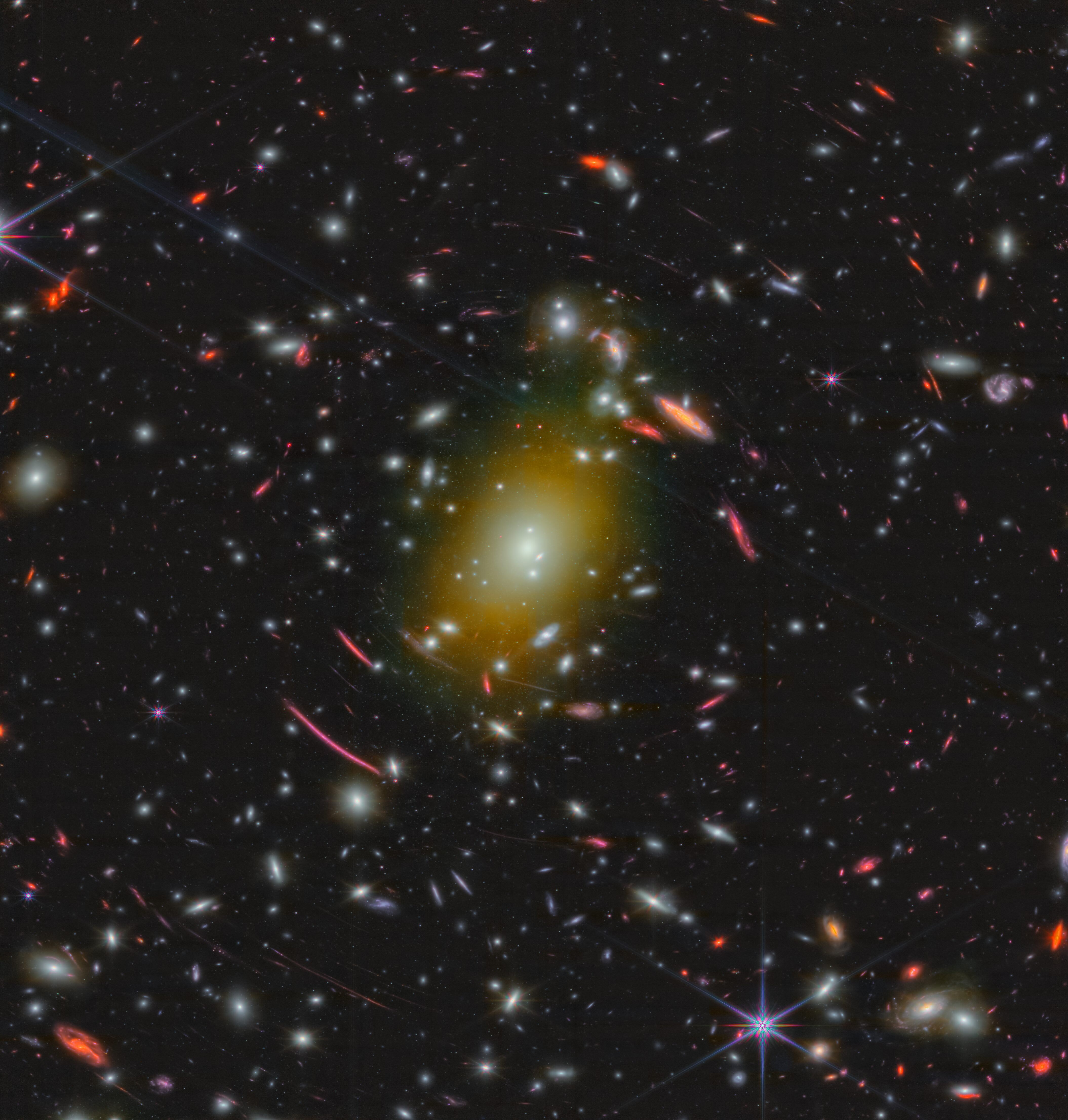}
    
    \caption{RGB Composite of the processed, calibrated, and background subtracted F200W, F356W, and F444W cluster module GLIMPSE NIRCam observations of the Abell S1063 Cluster. The mosaic images are combined non-linearly using a hyperbolic stretch to highlight the details of the cluster galaxies, ICL and background lensed arcs.}
    \label{fig:RGB}
\end{figure*}
However, the incredible benefits that gravitational lensing provide must be significantly tempered by the complexities involved with the foreground cluster lensing and interference. Most significantly, lensing concentrates sensitivity into small, highly magnified source-plane areas, so survey volumes are reduced and strongly non-uniform. This amplifies Poisson noise and sample variance and complicates completeness estimates. Second, uncertainties in the lensing model itself propagates into calculations of both demagnified luminosities of individual galaxies and the effective survey volume. Fractional errors can be tens of percent in the high-magnification regime and increase rapidly near critical curves \citep{Meneghetti2017,Priewe2017,Raney2020}. Third, crowded cluster cores suffer from bright intracluster and brightest-cluster-galaxy light, requiring careful modelling and subtraction to avoid photometric biases and spurious detections \citep[e.g.,][]{Bhatawdekar2019}. Finally, multiply imaged systems, differential magnification across extended sources, and contamination by cluster members introduce additional systematics that can bias both individual observations and the overall UVLF. Despite this, cluster-scale lensing is the only method currently available to probe the population of faint galaxies, and as such, astronomers have continued to try to use lensing to increase our scientific capabilities with JWST. 

Recent JWST observations, particularly when combined with the magnification afforded by gravitational lensing, have substantially revised empirical estimates of the high-redshift galaxy population and its potential contribution to reionization. Measurements of steep faint-end slopes in the rest-frame UVLF, together with evidence for elevated ionizing-photon production efficiencies, $\xi_{\rm ion}$, in low-mass systems, suggest that faint galaxies may contribute a larger share of the ionizing photon budget than previously inferred \citep{Endsley2023_1,Donnan2024,Adams2024EPOCHS}. These results place renewed emphasis on the broader problem of balancing the ionizing photon budget: galaxy-based estimates of the escaped ionizing emissivity, $\dot{n}_{\rm ion}$, must be reconciled with IGM-based constraints on the timing, duration, and topology of reionization, including limits from the CMB optical depth, Ly$\alpha$ forest opacity, and the evolving neutral hydrogen fraction \citep{Munoz2024,Simmonds2024,Atek2024Nature,melia2024}. At the same time, they sharpen the question of which sources dominate the production and escape of ionizing photons, whether numerous low-luminosity galaxies, rarer but more massive systems, AGN, or some luminosity- and redshift-dependent combination of these populations. Addressing these questions requires improved constraints on the faint-end shape of the UVLF, tighter empirical measurements of $\xi_{\rm ion}$ and $f_{\rm esc}$, and a consistent treatment of IGM clumping, recombinations, and reionization topology. Since modest changes in any of these quantities can substantially alter the inferred emissivity, the emerging observational picture is best framed as a problem of ionizing photon accounting: determining whether the available sources can supply the required number of photons over the observed reionization timescale, and identifying the physical populations primarily responsible for doing so.

In this paper we revisit the form and shape of the faint end of the $z\!\sim\!6$--11 UVLF using our own reduction of the JWST Cycle~2 program \textsc{GLIMPSE} (PID~3293; PIs H.~Atek \& J.~Chisholm, see \citet{atek2025}), which targets Abell~S1063, one of the best-studied and most powerful cluster lenses on the sky, with ultra-deep NIRCam imaging. Our goals are twofold: (i) to obtain new constraints on the faint-end slope $\alpha$ by exploiting extreme magnifications, and (ii) to test for statistical preference of a faint-end turnover, translating any bounds on $M_{\rm t}$ into limits on $\rho_{\rm UV}$. We combine custom intracluster-light subtraction, completeness and volume estimates that account for spatially varying magnification, and an information-criterion comparison between standard Schechter fits and smoothly suppressed (``turnover'') models. 

The paper is organized as follows. $\S$ \ref{sec:data} summarizes the imaging data, reduction, and cluster-light subtraction. $\S$ \ref{sec:phot} describes our methodology for selecting galaxies, calculating galaxy properties, and accounting for gravitational effects. In $\S$ \ref{subsec:candidates}, we briefly discusses the ultra-faint galaxies we use to build our UVLFs. $\S$ \ref{subsec:uvlf} details our construction of the UVLF, including survey volumes, completeness, and uncertainty propagation, and presents measurements alongside literature comparisons. We introduce a flexible turnover parameterization and uses an information criterion to assess model preference. Finally, $\S$ \ref{sec:disc} discusses the implications of these results, both on reionisation and on galaxy evolution.

Throughout this paper, we adopt a flat $\Lambda$CDM cosmology with $\Omega_{\rm m}=0.3$, $\Omega_\Lambda=0.7$, $\sigma_8=0.8$, and $H_0=70~{\rm km\,s^{-1}\,Mpc^{-1}}$, and all magnitudes are on the AB system \citep{OkeGunn1983}.

\section{Data Reduction}
\label{sec:data}

\begin{figure}
    \centering
    \includegraphics[width=\columnwidth]{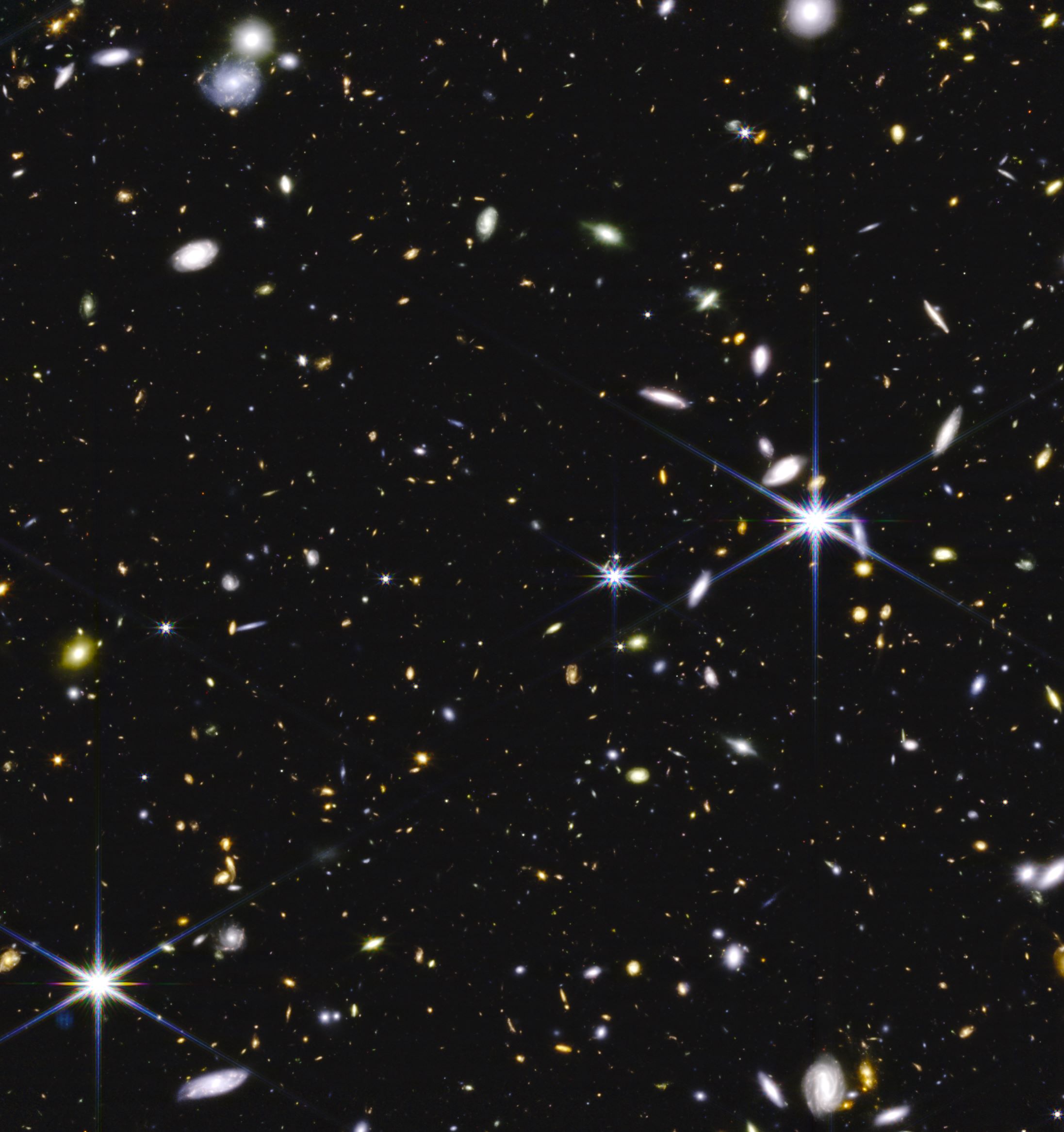}
    
    \caption{RGB Composite of the processed, calibrated, and background subtracted F200W, F356W, and F444W non-cluster module GLIMPSE NIRCam data products. The mosaic images are combined non-linearly using a hyperbolic stretch.}
    \label{fig:rgb2}
\end{figure}

    \subsection{JWST GLIMPSE Survey}
    \label{subsec:glmpse}

The primary source of data for this paper is the publicly available Gravitational Lensing In Massive Clusters with JWST (GLIMPSE) program (PID~3293; PIs H.~Atek \& J.~Chisholm). GLIMPSE is a Cycle~2 JWST program designed to leverage strong gravitational lensing by the Abell S1063 cluster at redshift $z = 0.34-0.35$ to study faint and magnified high-redshift objects with 154.6 hours of both wide and medium-band NIRCam imaging. It is comprised of a left and right hand module, of which the right hand targets the cluster proper and the left hand surveys a parallel, slightly-magnified field. A full discussion of this survey can be seen in \citet{atek2025}. It should be noted that most of the JWST Data Reduction was conducted by the methods outlined in \citet{Adams2024EPOCHS} and \citet{conselice2024}.

The data itself consists of one JWST pointing in the F090W, F115W, F150W, F200W, F277W, F356W, and F444W wide band filters, along with additional medium-band coverage in F410M and F480M. The uncalibrated images were downloaded from the MAST portal, and subsequently reduced using the JWST pipeline version 1.18.0. The reduction was slightly modified to exclude the 1/$f$ correction commonly used  to remove low-frequency, detector-correlated striping along the readout direction as it would lead to regions of over-subtraction around the Brightest Cluster Galaxy (BCG). Additionally, due to the extreme depth and sensitivity of the GLIMPSE survey, the background subtraction was done on a more aggressive 32x32 pixel grid rather than the more common 64x64 background calibration. This was done with a custom two-dimensional background modeling procedure (implemented via \texttt{photutils}; \citealt{Bradley2022}). The results of this are visible in  Figure \ref{fig:RGB} and Figure \ref{fig:rgb2}.

    \subsection{HST Ancillary Data}
To complement the NIRCam observations and to extend the wavelength coverage blueward of 0.9~$\mu$m, we incorporate archival HST ACS/WFC imaging. Abell S1063, the target of GLIMPSE, was covered by the Hubble Legacy Fields \citep{Illingworth2016HLF, Whitaker2019HLF}. We retrieved the latest v2.5 mosaics from MAST for the relevant ACS/WFC filters: F435W, F606W, and F814W, projected to a $0.\!''03$ pixel scale.

We reprojected the ACS/WFC mosaics onto the NIRCam World Coordinate System (WCS) using the \texttt{reproject} package \citep{robitaille2020}. Although the HST data are significantly shallower than the JWST imaging, they are valuable in constraining Lyman breaks for bright sources at $z<7.5$. However, the HST data does not cover the entirety of the GLIMPSE survey, meaning that only $z<7.5$ galaxies in a portion of the larger JWST survey can be robustly identified. 
    
\begin{figure}
    \centering
    \includegraphics[width=\columnwidth]{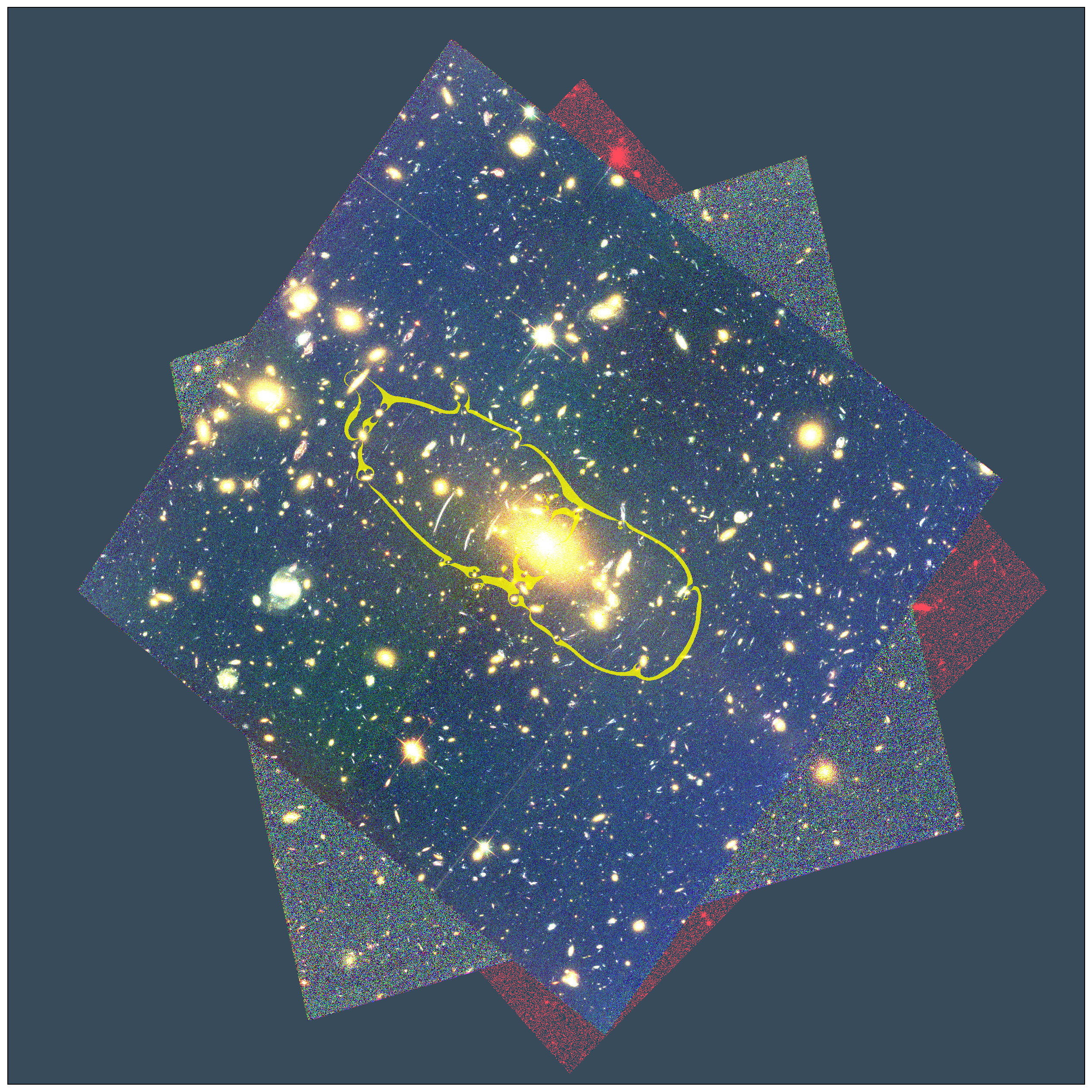}
    \label{fig:hst}
    \caption{RGB composite of processed, calibrated, and background-subtracted Hubble Fronteir Fields F814W, F606W, and F435W imaging of AS1063. Yellow region represents the critical curve (magnification greater than 100) at redshift 8 in the image plane.}
\end{figure}

\subsection{Cluster Light Removal}

To remove the light from the foreground cluster, we leverage a custom cluster-light subtraction pipeline, in addition to the traditional background subtraction in the JWST pipeline. This pipeline is presented and discussed in-depth in Appendix \ref{app:icl_subtraction}, and our methods employs wavelet decomposition of the science image to attempt to extract diffuse, large-scale structure. However, due to the nature of Abell S1063, the vast majority of the structure identified and removed was diffuse light from the brightest central galaxy of the cluster, rather than the intra-cluster light that dominates discussions of cluster imaging. Moreover, much of this diffuse light was removed in earlier stages of the image reduction pipeline discussed in $\S$ \ref{subsec:glmpse}. Moreover, because Abell S1063 is dominated by its BCG, the vast majority of the cluster-light does not overlap with the regions of high-magnification. However, we manually mask regions that are highly-subtracted, especially those around individual cluster members, to ensure a robust selection of galaxies.

\section{Photometry, Galaxy Selection, and Property Estimation}
\label{sec:phot}

\begin{figure*}
    \centering

    \includegraphics[width=0.7\textwidth]{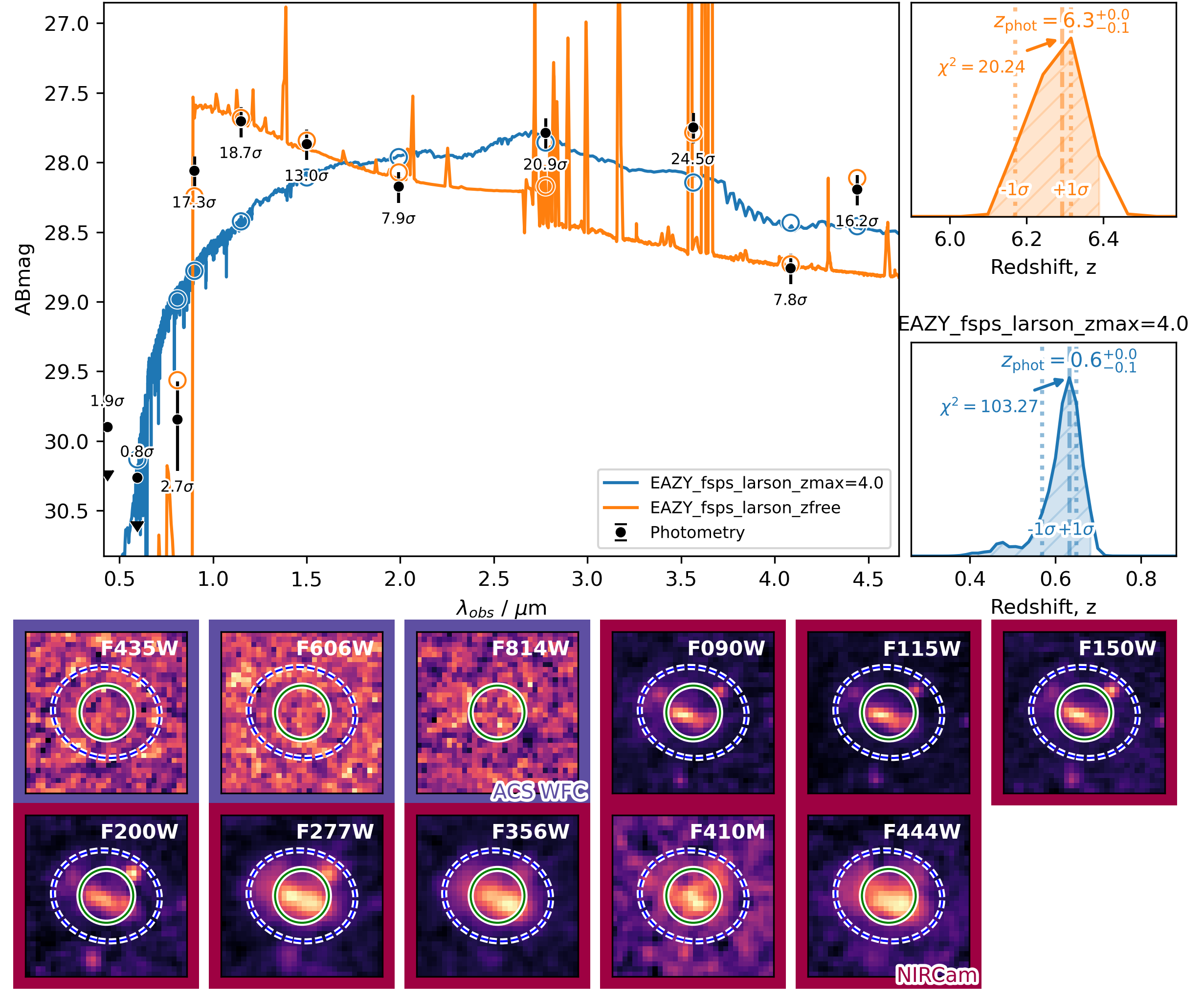}

    \vspace{0.5cm}

    \includegraphics[width=0.7\textwidth]{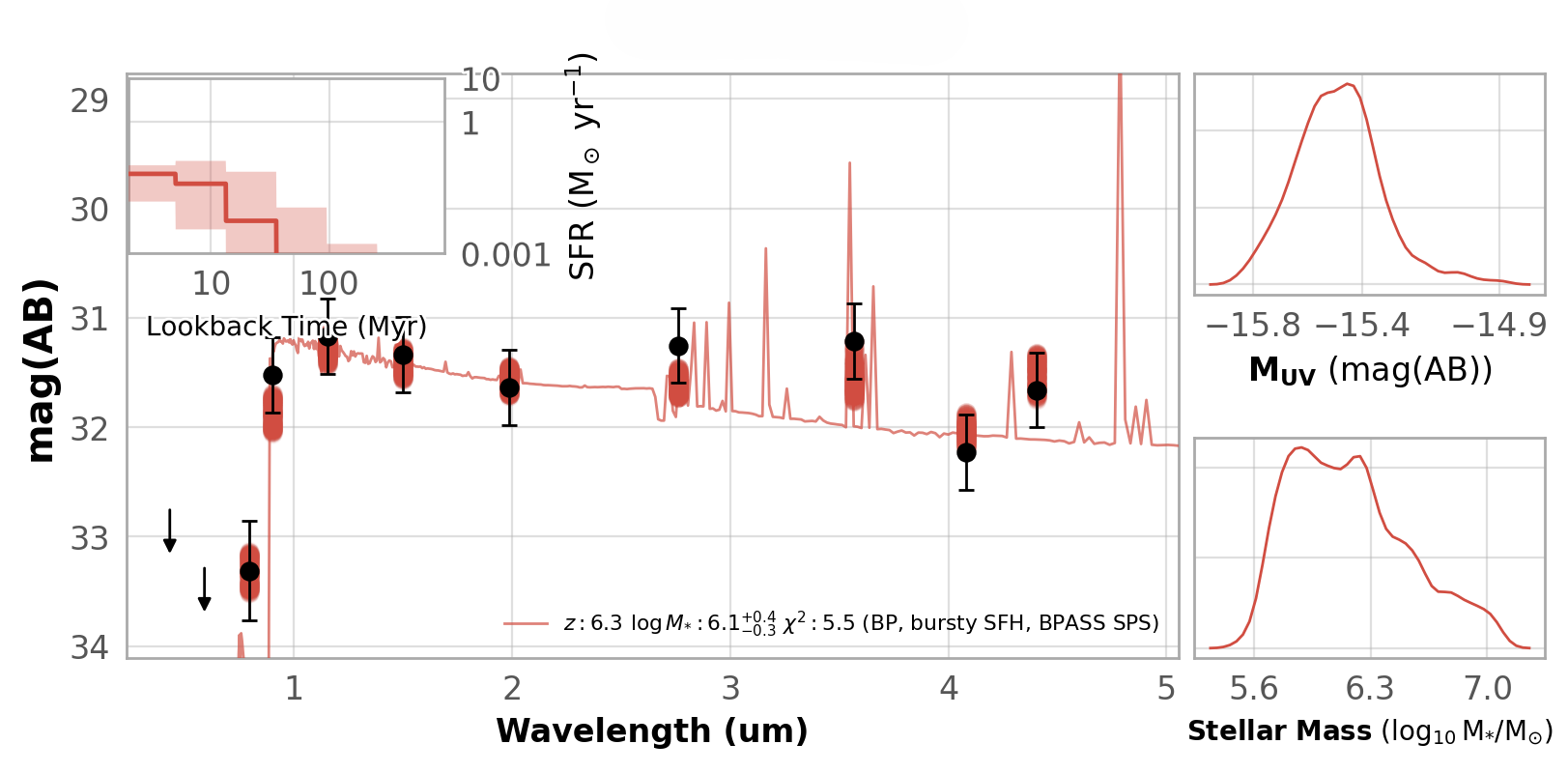}

    \caption{Best-fit SED for galaxy ID:8777, one of a multiply-imaged lensed galaxy in AS1063. The top panel shows the best high-redshift and low-redshift EAZY SED fit, along side the probability distribution for each solution on the top right. Below, the apertures for our photometery (green circle) are overlaid on the cutout for the galaxy in each filter. On the bottom is the BAGPIPES SED fit for galaxy 8777, alongside its BAGPIPES star formation history (top left), de-lensed UV-Magnitude PDF (top right), and stellar-mass PDF (bottom right). This image of the galaxy is highly magnified, with our lens model producing an expected $\mu$ of 23.9.  }
    \label{fig:sed_8777}
\end{figure*}

Source detection and photometric catalogues were produced with the \texttt{GALFIND}\footnote{\url{https://github.com/duncanaustin98/galfind}} package, which has been developed within the EPOCHS collaboration. Object detection and segmentation are carried out with \texttt{SExtractor} \citep{bertin1996} in dual-image mode, with detections measured in $0.\!''32$ diameter apertures. We adopted a minimum detection area of 9 pixels, a threshold of $1.8\sigma$, and a Gaussian detection filter with 2.5~pixel FWHM. 

Forced aperture photometry is measured at the segmentation-centroid positions in every band using circular 0\farcs32-diameter apertures on the PSF-matched images. Local depths, and thus noise, are calculated from the 200 closest of many 0\farcs32 apertures which have been placed in source-free regions; the per-band $1\sigma$ noise is taken as the NMAD of these blank-aperture fluxes, which naturally captures drizzle-induced correlated noise. From these measurements we also report per-source and field-median 5$\sigma$ point-source depths in 0\farcs32 apertures, and impose a minimum 10\% photometric floor to account for potential zero-point systematics. Aperture corrections, computed from the publically available WebbPSFs, convert 0\farcs32 fluxes to an estimate of total flux; the applied correction and its uncertainty are tracked per band. In practice, the encircled-energy fraction within a 0.16\arcsec\ radius (0\farcs32 diameter) is $\sim$70–80\% for the NIRCam LW bands, and we adopt band-specific correction factors derived from the F444W PSF model (cross-checked against empirical stars where available). All of this information is saved in a photometric catalog. In the case of Abell S1063, this automatic catalog contains 22,423 sources from the cluster-light-subtracted images. 

While the GLIMPSE survey took observations with 7 wide band and 2 medium band filters, when running photometry a discrepancy was noted with one of the medium band filters: F480M. When looking at the 5-sigma depths of the images taken, it was observed that one of the NIRCam modules, the one with the cluster, was artificially deeper (observationally, prior to both cluster-light subtraction and Gravitational Lensing effects) than the flat module without the cluster. Possible explanations of this were considered and studied, such as the impact of background subtraction or not having included all of the F480M observations during the image reduction pipeline. However, none resolved the issue satisfactorily, leading to the data being removed from the sample used for this research. Since it was a medium band filter, the removal of this filter should not, and does not, have a significant impact in our galaxy selection procedure, although this decision may have adversely impacted the galaxy properties inferred from SED fitting as F480M could have been used to slightly constrain metal-line emissions from the continuum slope.

\subsection{EAZY Photometric Estimation}
\label{subsec:eazy}
We estimate photometric redshifts with \textsc{EaZy-py} \citep{Brammer2008} by fitting the full redshift range, $z\in[0,25]$, and extracting the best-fitting solution for each galaxy within three redshift intervals: $z\in[0,4]$, $[0,6]$, and $[0,25]$. This allows us to compare the preferred low- and high-redshift solutions while retaining a consistent fitting framework. Our template basis combines the default FSPS library \texttt{tweak\_fsps\_QSF\_12\_v3} with six additional high-$z$ templates from \citet{Larson2023}. FSPS \citep{ConroyGunn2010} provides a physically motivated grid that varies stellar metallicity, age, and star-formation history (SFH), enabling fits to rapidly assembling systems common at early times. Metallicity coverage emphasizes sub-solar values, appropriate for primordial to moderately enriched stellar populations. Dust attenuation is bracketed by curves that range from Calzetti-like to SMC-like behavior, allowing the models to reproduce both modestly reddened systems and blue, low-dust galaxies. Critically, FSPS couples stellar continua to nebular line and continuum emission in a self-consistent manner tied to the ionizing photon budget. This treatment is essential for JWST-era photometry at $z\gtrsim6$, where large equivalent-width optical lines (e.g., [\ion{O}{iii}] and H$\beta$) can substantially boost broadband fluxes if not modeled.

The \citet{Larson2023} additions extend the library toward extremely blue, low-metallicity, low-dust populations with young ages and steep UV slopes ($\beta$). These templates better track the bluest SEDs now uncovered by JWST, improving fits in the rest-UV and reducing biases in inferred redshifts and continuum slopes. Practically, we find the combined FSPS+$\,$Larson basis captures a wide diversity of early-galaxy SEDs without over-specialization.

\subsection{EPOCHS Selection}
\label{subsec:selection}
To assemble a robust high–redshift sample we adopt a modified version of the EPOCHS selection used in \citet{Adams2024EPOCHS, conselice2024} and adapt it to our reductions. Photometric redshifts are computed with \textsc{EaZy-py} as described above; candidates are then required to satisfy the following criteria:

\begin{enumerate}
    \item \textbf{Blueward non-detections:} $\leq 3\sigma$ in all bands blueward of the expected Lyman break.
    \item \textbf{Bluest non-detection:} $\leq 2\sigma$ in the bluest available band.
    \item \textbf{Redward detections:} $\geq 4\sigma$ in the first two bands immediately redward of the break and $\geq 2\sigma$ in all other redward bands (excluding F410M). 
    \item \textbf{Redshift–PDF concentration:} the integrated posterior around the primary solution satisfies
    \[
      \int_{0.90\,z_{\rm phot}}^{1.10\,z_{\rm phot}} P(z)\,dz \ge 0.60,
    \]
    ensuring the majority of probability lies within $\pm10\%$ of $z_{\rm phot}$.
    \item \textbf{Fit quality:} For the best fit high-redshift solution from \textsc{EAZY}, the reduced $\chi^2_{\rm red} < 3$ 
    \item \textbf{High--$z$ vs.\ low--$z$ preference:} a separation of $\Delta\chi^2 \ge 2$ between the best high–$z$ fit and the best low–$z$ solution obtained from an \textsc{EAZY} run capped at $z_{\max}=6$. This suppresses Balmer/4000\,\AA–break interlopers.
    \item \textbf{Point–source:} if the 50\% encircled–flux radius (\texttt{FLUX\_RADIUS}) is smaller than the F444W PSF FWHM (i.e.\ consistent with a point source).
    \item \textbf{Hot–pixel/artefact guard:} the 50\% encircled–flux radius must be $\geq 1.5$ pixels in the LW widebands (F277W, F356W, F444W) to reject spurious compact features (e.g.\ hot pixels) that can masquerade as F200W dropouts.
\end{enumerate}

Candidates must satisfy all criteria to enter the parent high–$z$ sample, which we restrict to $z>7.5$ and $z>5.5$ where there is ancillary HST data. Furthermore, we limit our selection to between  $5.5<z<12.5$ to remove objects that are only visible in the long-wavelength filters, as many of these objects seem to be detector artifacts. For further discussion and validation of these thresholds and diagnostics, including extensive tests on public JWST fields, see \citet{Adams2024EPOCHS, conselice2024}. 
After accounting for multiply-lensed systems, which will be discussed in sections \ref{subsec:grav} and \ref{subsec:candidates}, we were left with a total of 249 galaxies between $5.5<z<12.5$ selected for further analysis.

\subsection{Gravitational Lensing}
\label{subsec:grav}

AS1063 is one of the best studied gravitational lenses. We used the set of 43 spectroscopically confirmed systems compiled in \cite{Diego2026a}, resulting in over 100 lensing constraints. The lens model is derived with the hybrid lens reconstruction algorithm {\small WSLAP+} \citep{Diego2005,Diego2007}, which makes minimal assumptions about the distribution of dark matter in the galaxy cluster. The lens model is discussed in detail in \cite{Diego2026a}.

To calculate the magnification of each galaxy selected, we produce a magnification map at the best-fitted redshift of the galaxy, and find the median magnification value within the \texttt{Sextractor} region of the individual galaxy. If the median magnification, $\mu$, is below 1.05, we do not model any magnification for that galaxy as our flat photometric flux error of 10 \% would dominate.  We assume a 15 \% error on $\mu$ if $ 1.05 < \mu < 5 $, and we assume a 40 \% error when $\mu > 5$. This is consistent with the larger error expected in galaxies with greater magnification \citep[see for instance][]{Meneghetti2017}.

Moreover, the gravitational effects of the foreground cluster can also produce multiple images of individual galaxies. In order to remove these from our sample, primarily to ensure we do not over count the number of galaxies, especially in faint bins, we ray trace each selected galaxy, searching for potential counter images. Among those that we have selected in our redshift range, we identified 2 pairs of galaxies, or galaxy systems, which seem to be multiply imaged. These were then removed from our sample of galaxies. One of the images of one of the pairs can be seen in Figure \ref{fig:sed_8777}.

\begin{figure}
    \centering
    \includegraphics[width=\columnwidth]{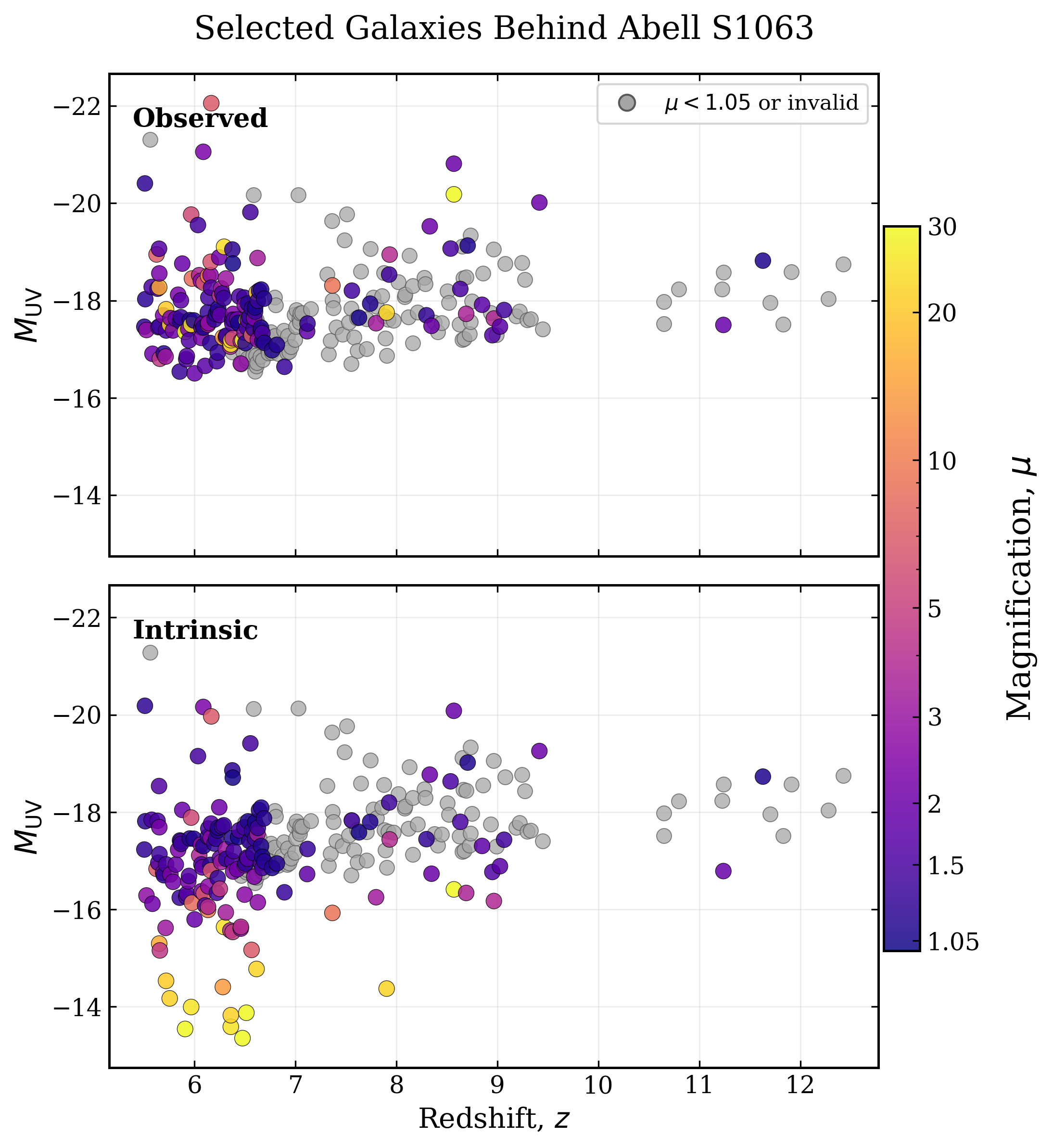}
    \caption{Redshift--$M_{\rm UV}$ distribution of the selected galaxy sample behind Abell S1063. For each galaxy, the observed UV absolute magnitude is shown together with the corresponding intrinsic, lensing-corrected value, with faint connecting lines linking the two measurements. Points are coloured by the gravitational magnification factor, $\mu$, shown on a logarithmic scale. The lensing correction shifts galaxies to intrinsically fainter $M_{\rm UV}$ values, illustrating how strong magnification enables the recovery of galaxies that would otherwise lie below the blank-field detection limit.}

    \label{fig:zphot_vs_zspec}
\end{figure}
\subsection{Galaxy SED Properties}
\label{subsec:properties}

\begin{table*}
\centering
\caption{De-lensed properties of highly-magnified galaxies in our high-redshift sample. Galaxy properties are taken from \textsc{BAGPIPES} assuming \citet{Calzetti2000} dust attenuation and a continuity bursty star formation history. Magnification uncertainties are assumed to be 40\% and are not derived from the lens model posterior.}
\label{tab:candidates_final}
\small
\begin{tabular}{lcccccccc}
\hline
ID & RA & Dec. & $z$ & $\mu$ & $M_{\rm UV,obs}$ & $M_{\rm UV,int}$ & $\log_{10}(M_\star/M_\odot)$ & $\beta_{\rm UV}$ \\
\hline
4339  & 342.161859 & -44.535195 & $5.623^{+0.133}_{-0.088}$ & $7.07^{+2.83}_{-2.83}$ & $-18.95^{+0.19}_{-0.17}$ & $-16.82^{+0.41}_{-0.58}$ & $7.45^{+0.45}_{-0.76}$ & $-2.28^{+0.37}_{-0.19}$ \\
7695  & 342.164349 & -44.530321 & $6.091^{+0.110}_{-0.121}$ & $6.51^{+2.60}_{-2.60}$ & $-18.37^{+0.15}_{-0.14}$ & $-16.33^{+0.40}_{-0.57}$ & $7.11^{+0.43}_{-0.77}$ & $-2.31^{+0.26}_{-0.16}$ \\
7870  & 342.182972 & -44.537012 & $5.966^{+0.174}_{-0.235}$ & $25.35^{+10.14}_{-10.14}$ & $-17.50^{+0.27}_{-0.23}$ & $-14.00^{+0.45}_{-0.60}$ & $5.40^{+1.04}_{-1.01}$ & $-1.78^{+1.55}_{-0.56}$ \\
8718  & 342.190863 & -44.537480 & $6.165^{+0.055}_{-0.056}$ & $6.78^{+2.71}_{-2.71}$ & $-22.05^{+0.15}_{-0.13}$ & $-19.98^{+0.39}_{-0.57}$ & $8.28^{+0.54}_{-0.51}$ & $-2.32^{+0.23}_{-0.20}$ \\
8777  & 342.183894 & -44.535330 & $6.292^{+0.087}_{-0.105}$ & $23.94^{+9.58}_{-9.58}$ & $-19.10^{+0.14}_{-0.12}$ & $-15.66^{+0.39}_{-0.57}$ & $5.82^{+1.10}_{-1.30}$ & $-1.85^{+1.49}_{-0.49}$ \\
9307  & 342.188447 & -44.536193 & $5.650^{+0.089}_{-0.023}$ & $15.42^{+6.17}_{-6.17}$ & $-18.27^{+0.18}_{-0.17}$ & $-15.30^{+0.41}_{-0.58}$ & $6.01^{+1.04}_{-1.41}$ & $-1.91^{+1.28}_{-0.46}$ \\
10033 & 342.189062 & -44.535237 & $6.359^{+0.173}_{-0.231}$ & $21.69^{+8.68}_{-8.68}$ & $-17.19^{+0.21}_{-0.18}$ & $-13.84^{+0.42}_{-0.58}$ & $5.24^{+1.06}_{-0.91}$ & $-1.81^{+1.74}_{-0.55}$ \\
10044 & 342.188848 & -44.535143 & $6.359^{+0.224}_{-0.258}$ & $25.32^{+10.13}_{-10.13}$ & $-17.10^{+0.24}_{-0.20}$ & $-13.59^{+0.44}_{-0.59}$ & $5.38^{+0.99}_{-0.99}$ & $-1.67^{+1.92}_{-0.66}$ \\
10177 & 342.189384 & -44.535172 & $7.903^{+0.209}_{-0.483}$ & $22.45^{+8.98}_{-8.98}$ & $-17.75^{+0.22}_{-0.20}$ & $-14.38^{+0.43}_{-0.59}$ & $5.50^{+1.05}_{-1.07}$ & $-1.77^{+1.81}_{-0.64}$ \\
11346 & 342.193342 & -44.534256 & $6.280^{+0.213}_{-0.417}$ & $13.53^{+5.41}_{-5.41}$ & $-17.24^{+0.24}_{-0.21}$ & $-14.41^{+0.44}_{-0.59}$ & $5.41^{+1.06}_{-1.01}$ & $-1.67^{+1.60}_{-0.71}$ \\
11582 & 342.173310 & -44.525949 & $5.966^{+0.086}_{-0.090}$ & $5.66^{+2.26}_{-2.26}$ & $-19.77^{+0.16}_{-0.15}$ & $-17.88^{+0.40}_{-0.57}$ & $7.77^{+0.42}_{-0.78}$ & $-2.35^{+0.18}_{-0.17}$ \\
11612 & 342.195348 & -44.534142 & $5.974^{+0.147}_{-0.179}$ & $8.42^{+3.37}_{-3.37}$ & $-18.45^{+0.15}_{-0.14}$ & $-16.14^{+0.40}_{-0.57}$ & $6.84^{+0.51}_{-0.69}$ & $-2.31^{+0.40}_{-0.21}$ \\
12840 & 342.193861 & -44.531610 & $5.906^{+0.210}_{-0.189}$ & $34.00^{+13.60}_{-13.60}$ & $-17.36^{+0.32}_{-0.28}$ & $-13.53^{+0.48}_{-0.62}$ & $5.40^{+0.95}_{-1.06}$ & $-1.51^{+1.84}_{-0.78}$ \\
12892 & 342.182239 & -44.526096 & $8.569^{+0.117}_{-0.144}$ & $32.23^{+12.89}_{-12.89}$ & $-20.18^{+0.13}_{-0.12}$ & $-16.41^{+0.39}_{-0.57}$ & $6.12^{+1.13}_{-1.52}$ & $-1.81^{+2.09}_{-0.60}$ \\
13881 & 342.194032 & -44.529449 & $6.160^{+0.124}_{-0.100}$ & $6.28^{+2.51}_{-2.51}$ & $-18.80^{+0.15}_{-0.13}$ & $-16.80^{+0.39}_{-0.57}$ & $6.91^{+0.49}_{-0.56}$ & $-2.35^{+0.15}_{-0.16}$ \\
14394 & 342.199724 & -44.530735 & $6.133^{+0.093}_{-0.088}$ & $9.97^{+3.99}_{-3.99}$ & $-18.51^{+0.14}_{-0.13}$ & $-16.01^{+0.39}_{-0.57}$ & $6.76^{+0.55}_{-0.70}$ & $-2.25^{+0.88}_{-0.22}$ \\
\hline
\end{tabular}
\end{table*}

We use the Bayesian SED–fitting code \textsc{BAGPIPES} \citep{Carnall2018} to infer stellar masses and star–formation rates (SFRs) for GLIMPSE galaxies from the demagnified multi–band photometry following the setup of \citet{Harvey2025}. Redshifts are fixed to the catalogued best–fitting \textsc{EAZY} values \citep{Ilbert2006}, so that the inference marginalizes only over the galaxy–evolution parameters. We adopt the BPASS stellar population synthesis models \citep{EldridgeStanway2017,StanwayEldridge2018} rather than BC03, as these models better capture binary–evolution pathways relevant at low metallicity and young ages. A Kroupa–type IMF is assumed \citep{Kroupa2002Sci}.

Our fiducial model uses the ``continuity–bursty'' SFH, implemented as six lookback–time bins: the first spans $5$\,Myr, and the remaining five are logarithmically spaced back to $z=25$. The bursty prior allows substantial bin–to–bin variation in SFR, enabling stochastic recent activity while preserving long–timescale continuity \citep{Leja2019}. This choice has been validated in recent high–$z$ studies \citep{conselice2024, Adams2024EPOCHS}. We impose broad, weakly informative priors: $\log_{10} M_{\star}/{\rm M_\odot}\in[5,12]$, formation redshift prior extending to $z=25$, and a log–uniform prior on metallicity motivated by expectations of sub–solar abundances at early times.

We assume a single–component Calzetti attenuation law \citep{Calzetti2000} with a log–uniform prior on $A_V$ in $[0.001,5.0]$, consistent with IRX–$\beta$ trends inferred for high–$z$ star–forming systems \citep{bowler2024alma}. Nebular emission is included with ionization parameter $\log U\in[-4,-1]$ and is coupled consistently to the ionizing photon production of the BPASS templates. Stellar metallicity follows a log prior spanning $10^{-4}$–$2\,Z_\odot$.

We run \textsc{BAGPIPES} in photometry–only mode, using \texttt{MultiNest} \citep{Feroz2009} for evidence–aware posterior sampling and drawing $N\!=\!500$ posterior samples per object, with redshift fixed from EAZY. 

Following \citet{Bhatawdekar2019}, we also estimate the rest–UV continuum slope, $\beta$, and the monochromatic UV absolute magnitude, $M_{\mathrm{UV}}$, using an SED–template–lightweight approach. At a fixed redshift $z_{\rm fix}$ (set to the best–fitting \textsc{EAZY} solution), we fit a simple power law to the rest–frame UV photometry in the wavelength window $1216<\lambda_{\rm rest}/{\rm \AA}<3000$ using an affine–invariant MCMC sampler (\texttt{emcee}) to obtain posterior uncertainties on $\beta$ and the normalization. From the posterior continuum, we compute $M_{\mathrm{UV}}$ by averaging the model flux within a top–hat of width $100$\,\AA\ centered at $1500$\,\AA\ in the rest frame. Uncertainties on $M_{\rm UV}$ reflect the full posterior of $(\beta,\mathrm{norm})$ and the photometric errors, and retain the contribution from the lensing–magnification uncertainty propagated in \S\ref{subsec:grav}.

\begin{figure}
    \centering
    \includegraphics[width=\columnwidth]{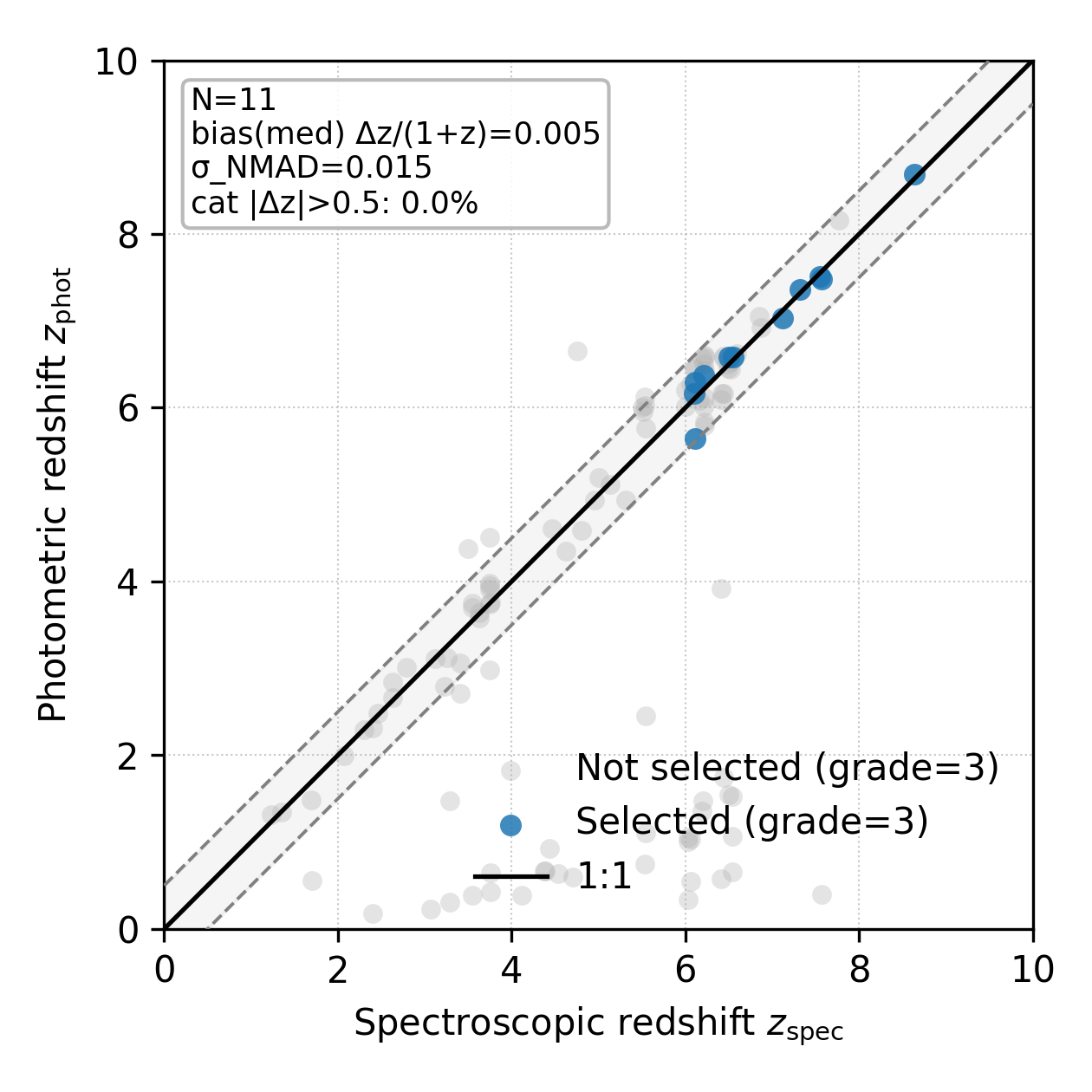}
    \caption{Photometric vs.\ spectroscopic redshifts for all grade\,=\,3 objects in our matching of spectroscopic redshifts with our catalog.  Sources which were not selected by our methods to be reliable galaxies are shown in gray; selected sources are highlighted. The solid line indicates $z_{\rm phot}=z_{\rm spec}$ and the dashed lines show $\pm0.5$ in $|\Delta z|$.  For the selected, grade\,=\,3 subset with robust spectroscopic matches, we obtain $\mathrm{median}[\Delta z/(1+z_{\rm spec})]=0.005$, $\sigma_{\rm NMAD}=0.0147$, $\mathrm{std}=0.024$, and $\mathrm{RMSE}=0.023$, with zero catastrophic outliers under either criterion.}
    \label{fig:zphot_vs_zspec}
\end{figure}

\subsection{Spectroscopic Comparison}

In order to test the robustness of our selection criteria, we compare our selected sample both to sources reported previously and to those confirmed spectroscopically. A 2025 JWST/NIRSpec DDT program (PID~9223; PI: S.~Fujimoto and R.~P.~Naidu) obtained ultra-deep spectroscopy of 262 unique objects with the G395M grating and F290LP filter. A more complete discussion of this survey can be found in \citet{asada2026}. Using the publicly available DAWN JWST Archive (v4) NIRSpec reductions, we matched these to 180 distinct sources in our photometric catalogue, of which 102 have robust spectroscopic redshifts. Many of these are not selected due to the objects not laying on the GLIMPSE footprint. Among these, 45 lie at sufficiently high redshift to fall within our catalogue footprint, and 10 are admitted by our final selection criteria. Objects that failed our selection were primarily rejected due to insufficient $\Delta\chi^{2}$ preference for the high-$z$ solution and/or a $\leq 2\sigma$ non-detection in the bluest band.

For the subset used to quantify photo-$z$, we measure a median bias in $\Delta z/(1+z_{\rm spec})$ of $0.005$, a scatter of $\sigma_{\rm NMAD}=0.0147$, a sample standard deviation of $0.024$, and an $\mathrm{RMSE}$ of $0.023$. We find no catastrophic outliers under either $|\Delta z|>0.5$ or $|\Delta z|/(1+z_{\rm spec})>0.15$ (Figure~\ref{fig:zphot_vs_zspec}). None of our selected galaxies are identified as low-redshift interlopers. While this does not constrain the number of galaxies that we potentially are missing, it provides some evidence as to the accuracy of the galaxy sample we are providing.

\section{Results}

\subsection{Highly-Magnified Galaxy Candidates}
\label{subsec:candidates}

Among the 249 galaxies we selected in the \(5.5 < z < 12.5\) GLIMPSE field, 18 of them appear to be highly magnified ($\mu>5$) and 7 extremely magnified ($\mu>20$). While the JWST GLIMPSE data is one of the deepest images taken to-date, the intrinsic depth of the GLIMPSE field would not be enough to probe the extreme-faint end of the UVLF. Thus, we rely heavily on these individual galaxies to provide us with an understanding of the faint-galaxy population at this high redshift. The properties of these galaxies can be seen in Table~\ref{tab:candidates_final}. However, due to high lensing uncertainties and the known difficulties in retrieving galaxy properties accurately from \textsc{BAGPIPES}, these results should be taken very carefully, both to ensure the robustness of the UVLF and to understand their impact on the Epoch of Reionization.

\begin{figure*}
    \centering
    \includegraphics[width=\textwidth]{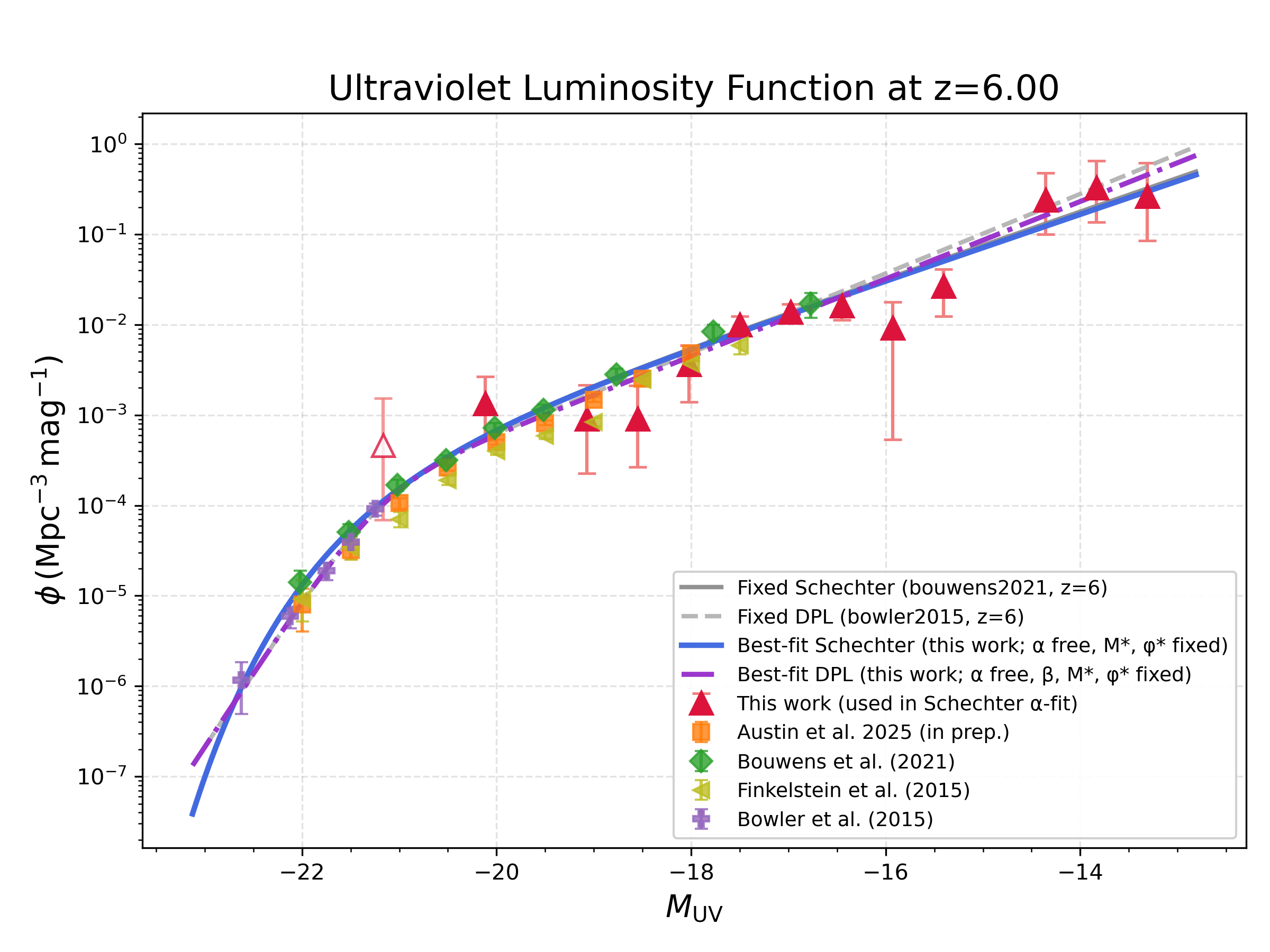}
    \caption{Ultraviolet luminosity function for galaxies at redshifts z = 6, where we probe the deepest into the faintest systems down to $M_{UV} \sim -13$. Our data points are red triangles. The solid and dashed grey lines represent Schechter and Double Power Law (DPL) fits to the UVLF at redshift $z \sim 6$ respectively. The solid blue line and the dashed purple line represent the same Schechter and DPL, but with the faint end slope fitted to our data. The outline of a red triangle is one of our data points, but brighter than the knee of the Schechter and DPL function, and therefore excluded when the faint-end is being fitted.   }
    \label{fig:uvlf_1}
\end{figure*}
% This physical picture, that numerous, metal-poor dwarfs with intermittent starbursts powers the photon budget, would require that the contribution of these faint galaxies to the ionizing photon budget be at least considered, despite ongoing discussions of potential suppression mechanisms such as supernovae feedback. Thus, incorporating the GLIMPSE lensed sources (see Tab. \ref{tab:candidates_final}) into the UVLF samples this consequential population for reionization.

\begin{figure*}
    \centering
    \includegraphics[width=0.48\textwidth]{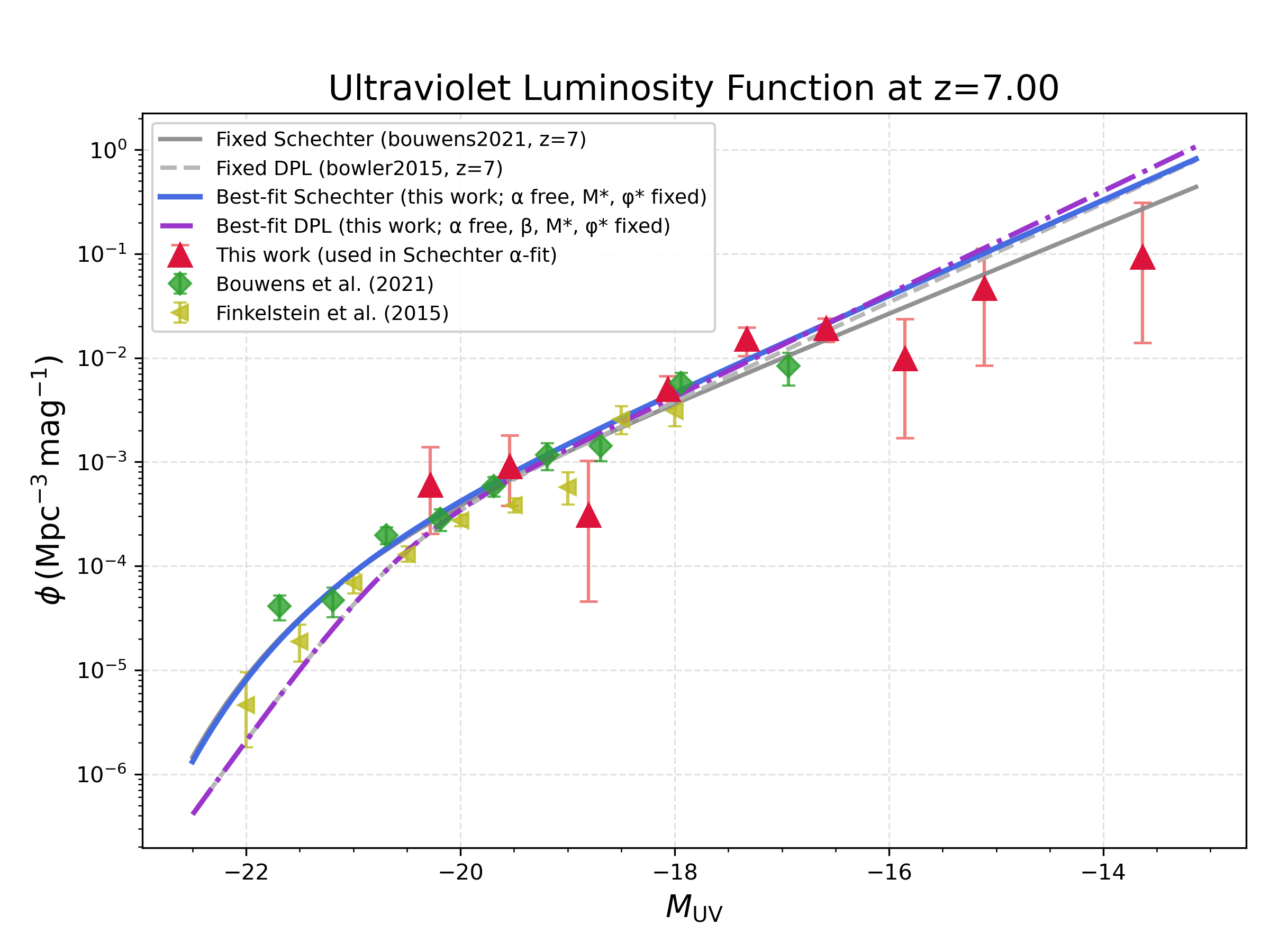}
    \includegraphics[width=0.48\textwidth]{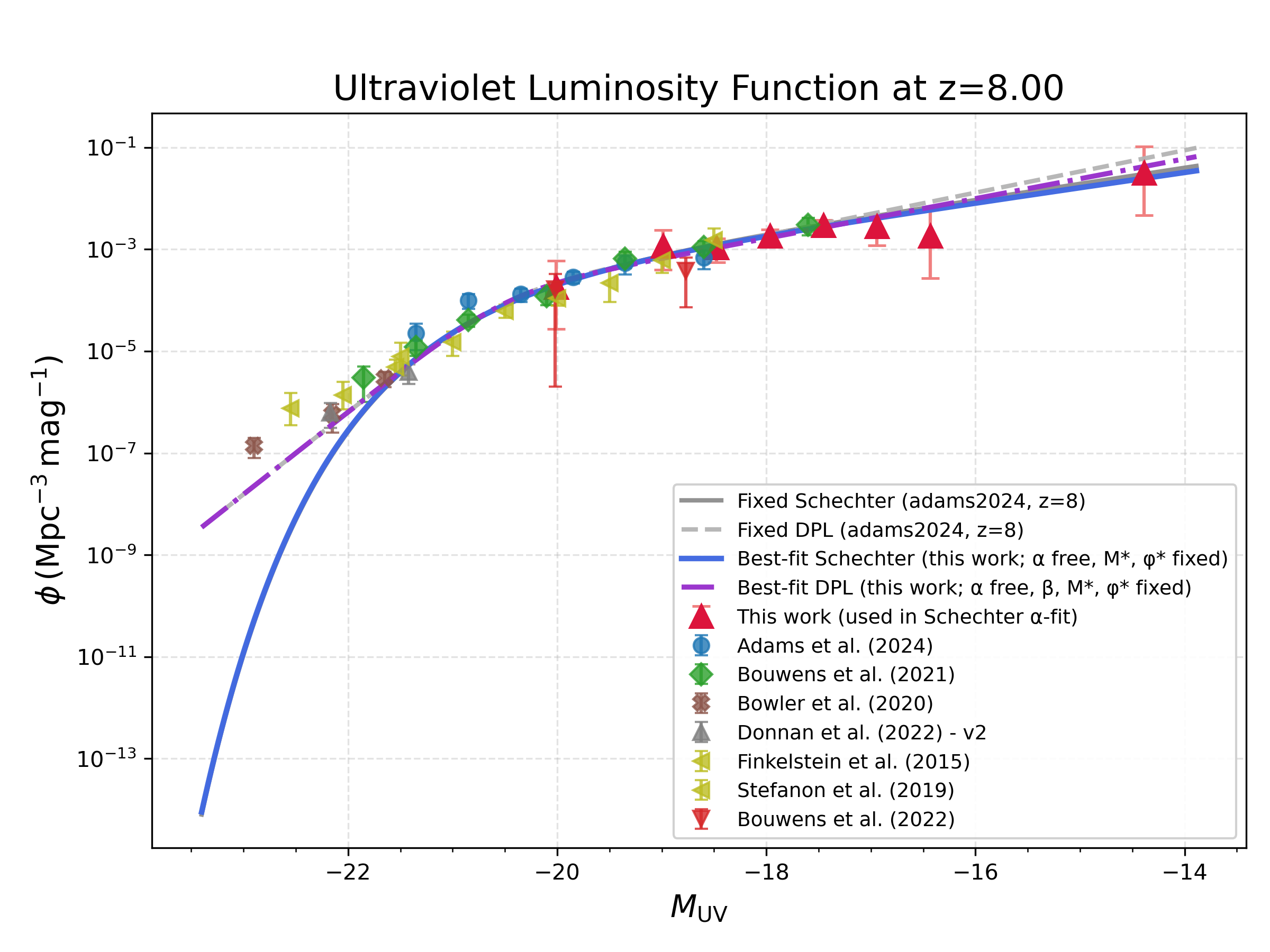}

    \vspace{0.5em}

    \includegraphics[width=0.48\textwidth]{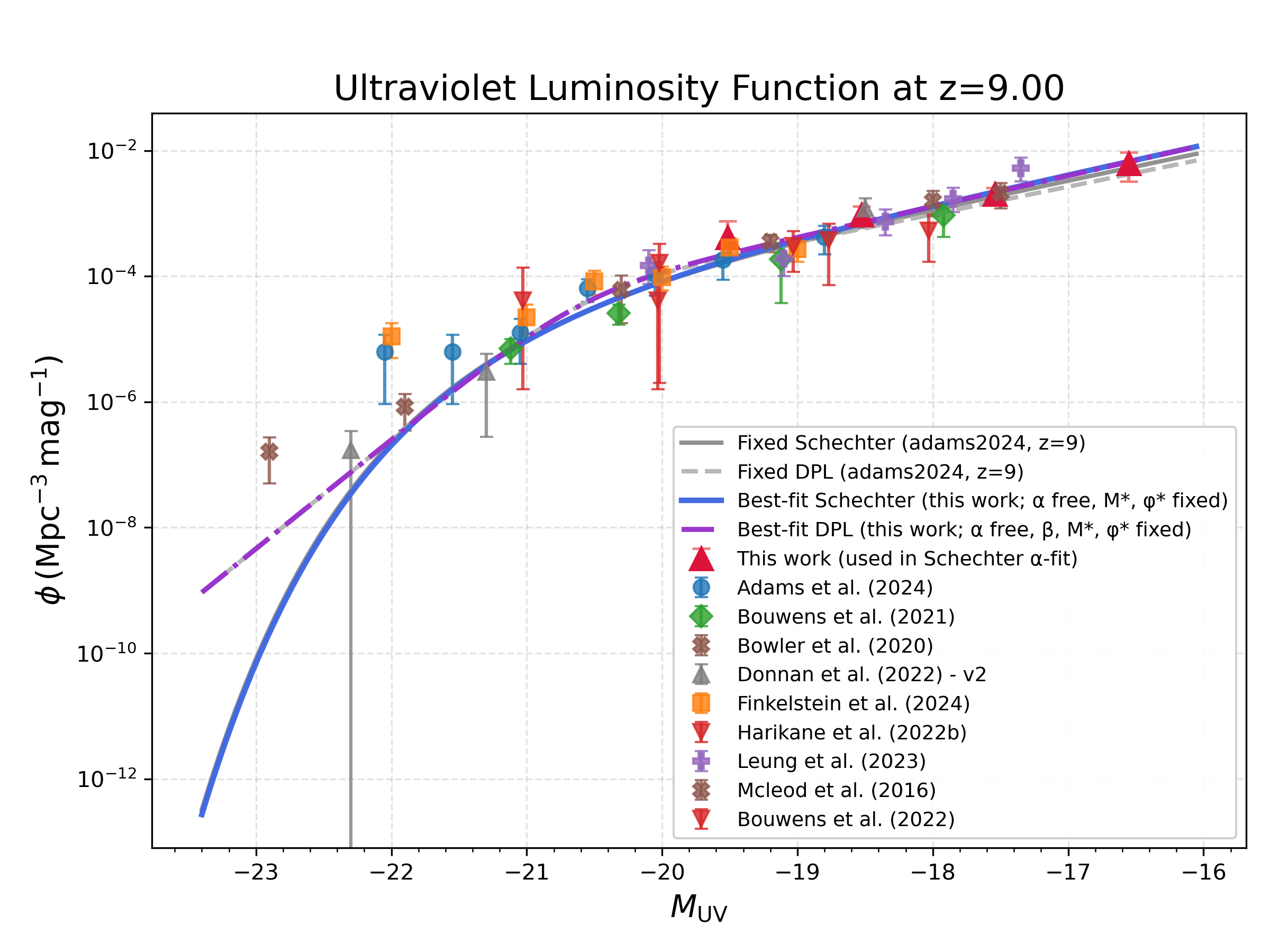}
    \includegraphics[width=0.48\textwidth]{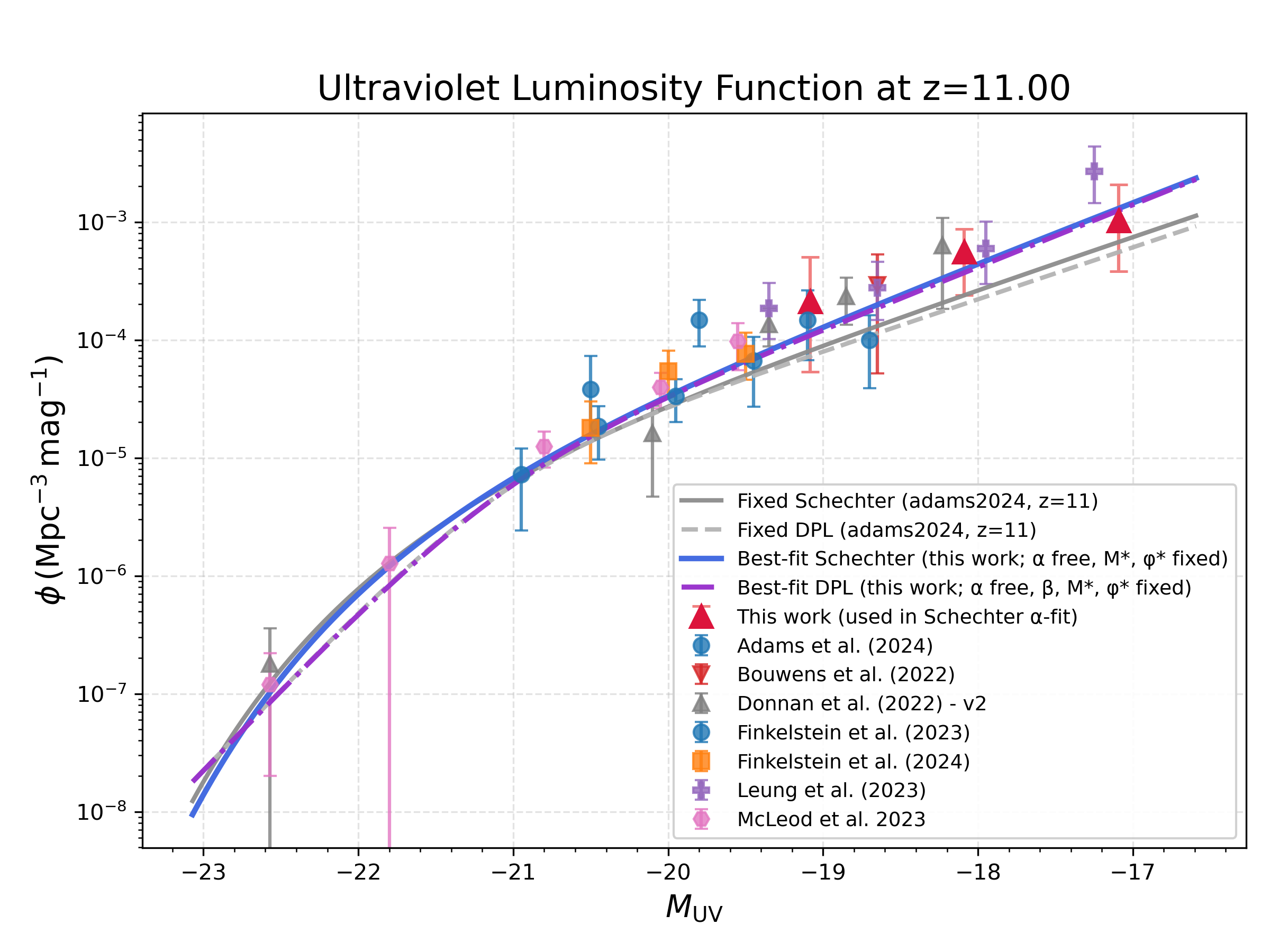}

    \caption{
     Ultraviolet Luminosity function for various redshift bins
    (top left) Redshift z=7;
    (top right) Redshift z=8;
    (bottom left) Redshift z=9;
    (bottom right) Redshift z=11. Our data points are red triangles. The solid and dashed grey lines represent Schechter and Double Power Law (DPL) fits to the UVLF at redshift z=6 respectively. The solid blue line and the dashed purple line represent the same Schechter and DPL, but with the faint end slope fitted to our data.  Comparisons to previous work on these measurements are also shown.
    }
\label{fig:uvlf_secondary}
\end{figure*}

\subsection{Ultraviolet Luminosity Function}
\label{subsec:uvlf}

In this section we investigate the distribution of the UV luminosities in our sample. This rest–frame UVLF quantifies the comoving number density of galaxies per unit magnitude at $M_{\rm UV}$ and redshift $z$, $\Phi(M_{\rm UV},z)$ (e.g. \citealt{Schmidt1968,RowanRobinson1968}). In binned form, for magnitude bin $k$ of width $\Delta M_{\rm UV}$ centered at $M_k$, the standard $1/V_{\max}$ estimator is
\begin{equation}
\Phi(M_k) \;=\; \frac{1}{\Delta M_{\rm UV}} \sum_{i \in \mathcal{B}_k} \frac{1}{C_i\,V_{\max,i}},
\label{eq:uvlf_vmax}
\end{equation}
where the sum runs over galaxies $i$ in bin $k$, $C_i$ is the total completeness applicable to galaxy $i$ (defined in section \ref{subsubsec:completeness}), and $V_{\max,i}$ (defined in section \ref{subsubsec:volume}) is the effective comoving volume in which $i$ would have entered our selection. The completeness appears in the denominator because each detected source statistically represents $1/C_i$ true sources in the survey volume.

We adopt the following redshift bins throughout: $z\in[5.5,6.5]$, $[6.5,7.5]$, $[7.5,8.5]$, $[8.5,9.5]$, and $[9.5,12.5]$.

\subsubsection{Survey Volume Calculations}
\label{subsubsec:volume}

Because our survey leverages strong lensing, the accessible volume depends on the spatially varying magnification $\mu(\boldsymbol{\theta},z)$. Building on the effective area in magnification bins described previously, we compute $V_{\max,i}$ for each galaxy by explicitly integrating over redshift and magnification:

\begin{equation}
V_{\max,i}
\,=\, \sum_{j} \sum_{m} \sum_{n}
\left[ A_{\rm src}^{(j)}(\mu_m, z_n)\;
\left.\frac{{\rm d}V_c}{{\rm d}z\,{\rm d}\Omega}\right|_{z_n}
\;\Delta z \right]
\,\mathcal{S}_i(\mu_m,z_n),
\label{eq:vmax_sum}
\end{equation}

\noindent where $j$ indexes fields (cluster and any parallel), $m$ indexes magnification bins, and $n$ indexes redshift steps. The term $A_{\rm src}^{(j)}(\mu_m, z_n)$ is the source–plane solid angle (per field) effectively surveyed at magnification bin $\mu_m$ and redshift $z_n$ after all masks, with $A_{\rm src}=A_{\rm img}/\mu$; ${\rm d}V_c/({\rm d}z\,{\rm d}\Omega)$ is the comoving volume element for our fiducial cosmology; and $\mathcal{S}_i(\mu,z)\in\{0,1\}$ indicates whether galaxy $i$ would be selected if placed at $(\mu,z)$ (defined below). In practice we evaluate Eq.~\ref{eq:vmax_sum} on a grid and sum the qualified cells.

For each $z$ between 5.5 and 12.5 in steps of $\Delta z=0.01$, we construct maps of $\mu(\boldsymbol{\theta},z)$ and histogram them into $N_\mu=1000$ logarithmically spaced $\mu$ bins (consistent with our area–in–$\mu$ formalism). We reproject the cluster–member and stellar/bright–artifact masks onto these maps, and to suppress hot–pixel / small–scale artifacts we tile each map into $10\times10$-pixel cells, replacing each cell with the mean $\mu$ over the unmasked pixels. The effective image–plane area $A_{\rm img}^{(j)}(\mu_m,z_n)$ is then the masked area in cells whose mean magnification falls in bin $m$; converting to the source plane gives $A_{\rm src}^{(j)}=A_{\rm img}^{(j)}/\mu_m$.

For each cell $(\mu_m,z_n)$ we scale the best–fit rest–frame SED of galaxy $i$ (from \textsc{EAZY}; \citealt{Brammer2008}) to redshift $z_n$ and amplify fluxes by $\mu_m$ to predict observed magnitudes and S/N in our filters. We then apply the identical colour and S/N cuts used for the real sample, with one modification to avoid unphysical rejections of high-$z$ objects: when testing $\mathcal{S}_i$ we do not enforce the stringent $<2\sigma$ non–detection requirement blueward of the break, so that noise fluctuations or confusion do not produce false failures at high $z$. If the predicted photometry passes the selection, $\mathcal{S}_i=1$; otherwise $\mathcal{S}_i=0$. Finally, for each galaxy we restrict the sum in Eq.~\ref{eq:vmax_sum} to its analysis redshift bin (e.g. $[7.5,8.5]$ for $z\!\sim\!8$).

\begin{figure*}
    \centering
    \includegraphics[width=0.9\textwidth]{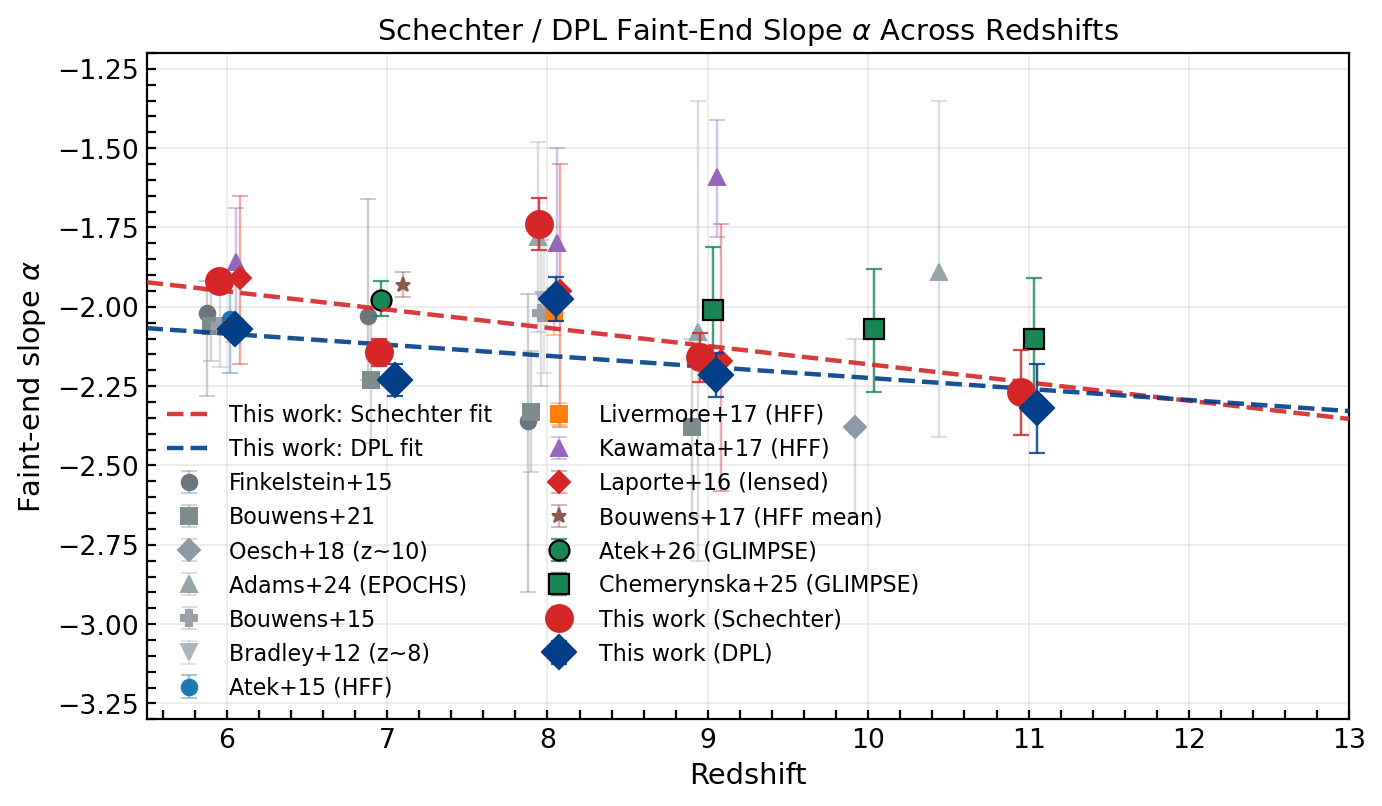}
    \caption{Evolution of the faint end slope across the Epoch of Reionization. The best-fitted faint end slopes for each of our redshift bins are displayed in red for the Schechter fit and in blue for the DPL fit. Comparable literature faint-end slope values are also visibile, with the points being grey for non-lensing surveys and colour for surveys that leveraged gravitational lensing. Results from previous work using the GLIMPSE data are green.}
    \label{fig:alpha}
\end{figure*}

\begin{table*}
\centering
\caption[UVLF model parameters]{Model parameters and re-fitted faint-end slopes at each redshift. The upper panel shows the Schechter parameterization(Eq. \ref{eq:schechter_mag}, while the lower panel shows the double power-law (DPL) parameterization (Eq. \ref{eq:dpl_mag}) Entries are quoted as posterior medians with $1\sigma$ uncertainties. All values except the re-fitted faint-end slope are taken directly from the literature \citep{Bowler2015, Bouwens2021, Adams2024EPOCHS}.}
\label{tab:MCMC}
\small

%======================
% Schechter
%======================
\begin{tabular}{cccccc}
\hline
\multicolumn{6}{c}{Schechter} \\
\hline
$z$ & $M_\ast$ & $\log_{10}\phi_\ast$ & $\alpha$ & $\beta$ & Re-fitted $\alpha$ \\
 & (AB) & (Mpc$^{-3}$\,mag$^{-1}$) &  &  &  \\
\hline
 6.0  & $-20.93 \pm 0.09$              & $-3.29_{-0.09}^{+0.09}$ & $-1.93 \pm 0.08$              & --- & $-1.919 \pm 0.034$ \\
 7.0  & $-21.15 \pm 0.13$              & $-3.29_{-0.07}^{+0.07}$ & $-2.06 \pm 0.011$             & --- & $-2.144 \pm 0.044$ \\
 8.0  & $-20.02_{-0.35}^{+0.27}$       & $-3.24_{-0.37}^{+0.25}$ & $-1.78_{-0.30}^{+0.30}$       & --- & $-1.741 \pm 0.083$ \\
 9.0  & $-20.35_{-0.31}^{+0.27}$       & $-3.90_{-0.34}^{+0.26}$ & $-2.1$                        & --- & $-2.159 \pm 0.077$ \\
11.0  & $-21.35_{-0.45}^{+0.64}$       & $-5.00_{-0.34}^{+0.38}$ & $-2.1$                        & --- & $-2.471 \pm 0.083$ \\
\hline
\end{tabular}

\vspace{0.4cm}

%======================
% DPL
%======================
\begin{tabular}{cccccc}
\hline
\multicolumn{6}{c}{Double power law (DPL)} \\
\hline
$z$ & $M_\ast$ & $\log_{10}\phi_\ast$ & $\alpha$ & $\beta$ & Re-fitted $\alpha$ \\
 & (AB) & (Mpc$^{-3}$\,mag$^{-1}$) &  &  &  \\
\hline
 6.0  & $-21.20 \pm 0.22$              & $-3.72_{-0.24}^{+0.21}$ & $-2.10_{-0.14}^{+0.16}$ & $-5.1_{-0.6}^{+0.5}$   & $-2.071 \pm 0.032$ \\
 7.0  & $-20.61_{-0.26}^{+0.31}$       & $-3.66_{-0.23}^{+0.25}$ & $-2.19_{-0.10}^{+0.12}$ & $-4.6_{-0.5}^{+0.4}$   & $-2.230 \pm 0.050$ \\
 8.0  & $-20.50_{-0.26}^{+0.39}$       & $-3.76_{-0.27}^{+0.31}$ & $-2.04_{-0.20}^{+0.24}$ & $-5.05_{-0.62}^{+0.64}$ & $-1.975 \pm 0.070$ \\
 9.0  & $-20.60_{-0.24}^{+0.43}$       & $-4.16_{-0.25}^{+0.28}$ & $-2.1$                  & $-5.35_{-1.08}^{+1.00}$ & $-2.214 \pm 0.069$ \\
11.0  & $-21.10_{-0.64}^{+0.78}$       & $-5.02_{-0.39}^{+0.47}$ & $-2.1$                  & $-4.45_{-1.02}^{+0.97}$ & $-2.535 \pm 0.087$ \\
\hline
\end{tabular}

\end{table*}

\subsubsection{Completeness}
\label{subsubsec:completeness}

Our total completeness factor for galaxy $i$ is the product of selection and detection terms, where the total completion captures the probability that a real galaxy meeting an observed ($M_{\rm UV},z$) would satisfy our colour/S/N selection after photometric redshift fitting, and $C_{\rm det}$ captures the probability that it is found and measured by our imaging pipeline in the presence of the ICL and instrumental systematics. We find that $C_{\rm sel}$ dominates the total incompleteness.

We estimate $C_{\rm sel}$ using five realizations of the JAdes extraGalactic Ultradeep Artificial Realizations (JAGUAR; \citealt{Williams2018}) mock catalog of star–forming galaxies. We place the mock images uniformly at random over the GLIMPSE footprint, run them through our photometry and \textsc{EAZY} workflow with the identical selection criteria, and compute, for each $(M_{\rm UV},z)$ bin, the fraction that are recovered by the selection. This yields $C_{\rm sel}(M_{\rm UV},z)$.   These completeness methods are outlined in detail in the EPOCHS papers, especially \citet{Adams2024EPOCHS} and \citet{conselice2024}.

To isolate detection losses from bright foreground sources and local backgrounds, we further inject suites of galaxy cutouts into the source plane (thus preserving lensing surface–brightness effects), forward–model them through our ICL–subtraction pipeline, and run the standard photometric catalog creation pipeline to measure recovery. The recovery fraction as a function of $(M_{\rm UV},z)$ defines $C_{\rm det}$. Because lensing boosts total flux while also increasing the observed surface-area, this completion method caputres both S/N and surface–brightness selection effects.

While in flat-field photometery completion dominates the faint-end bins, our sample is highly complete, with the least-complete bin in the sample being 64 \% complete.

\subsubsection{Error Propagation}
\label{subsubsec:uvlf_primary}

Perhaps the most crucial part of the UVLF, especially in the strong-lensing regime, are its uncertainties. The UVLF uncertainties combine three terms: (i) Poisson counting statistics, (ii) uncertainty in the accessible comoving volume, and (iii) lensing–induced magnitude errors. Throughout, the per–galaxy weight is

\begin{equation}
w_i \equiv \frac{C_i}{V_{\mathrm{max},i}},
\end{equation}

\noindent where $C_i$ is the selection/completeness factor and $V_{\mathrm{max},i}$ is the object’s accessible volume. All errors are accumulated in linear $\phi$ and then displayed as asymmetric low/high bars on log–$y$ plots.

For bins with $N\ge5$ objects, the statistical uncertainty is well described by the variance of the weighted $1/V_{\max}$ estimator. However,
For bins with $N<5$, the count distribution is strongly asymmetric and a symmetric approximation can severely misestimate the error (e.g.\ yielding unphysical $1\pm 1$ detections). In this regime we adopt the $\chi^2$--based Poisson confidence intervals of \citet{Ulm1990}.

Uncertainties in the effective comoving volume were propagated by applying a fractional error to each $V_{\mathrm{max},i}$. We adopt an error estimate of 20 \% for the effective survey volume for every galaxy which is then combined across objects in quadrature within each bin.

Magnification uncertainties were propagated by combining the initial observed $M_{\rm UV}$ error described in section \ref{subsec:properties} with the assumed lensing errors described in section \ref{subsec:grav}. For moderately lensed galaxies, this means a magnification error of 15 \%, and for highly magnified galaxies, a magnification error of 40 \%. 

The total per--bin errors are obtained by combining these terms in quadrature. For likelihood calculations and bin optimization, which require a single symmetric variance in log--space, we convert these asymmetric errors into an effective $\log_{10}\phi$ uncertainty with a floor of $0.08$\,dex to avoid overweighting bins with very small errors.

It is also important to note that cosmic variance was not included in the quoted uncertainties. The GLIMPSE cluster fields probe highly magnified lines of sight through massive foreground clusters, with effective survey areas set by lensing geometry rather than contiguous blank fields, making standard cosmic variance prescriptions poorly applicable. Furthermore, in such narrow, strongly lensed sight-lines the galaxy population will be dominated by high-redshift faint dwarf galaxies where uncertainties in faint galaxy bias and populations mean that current cosmic variance prescriptions often fail, especially in small volumes \citep{Moster2011,Dawoodbhoy2023}. This is discussed more in $\S$ \ref{subsubsec:cosmicv}.

\subsubsection{Redshift 6-10 UVLFs}
\label{subsubsec:uvlf_fit}

To place our UVLF measurements in context, we compare them directly to published determinations at comparable redshift and evaluate consistency within a common parametric framework. For each redshift slice, we compile literature points whose reported redshifts satisfy $\lvert z_{\rm lit}-z_{\rm slice}\rvert\leq 0.6$. Because different works quote either $\phi$ or $\log_{10}\phi$, we homogenize to linear number densities, $\phi~[\mathrm{Mpc}^{-3}\,\mathrm{mag}^{-1}]$, converting logarithmic measurements and propagating their quoted asymmetric uncertainties to linear space.

We model the UVLF in magnitude space with both a Schechter function \citep{Schechter1976} and a double power-law (DPL) \citep{Bowler2015} parameterization:
\begin{equation}
\phi_{\rm Sch}
= 0.4\ln 10 \;\phi^\ast \;10^{-0.4(M-M^\ast)(\alpha+1)}\,
\exp\!\big[-10^{-0.4(M-M^\ast)}\big],
\label{eq:schechter_mag}
\end{equation}
and
\begin{equation}
\phi_{\rm DPL}
=
\frac{\phi^\ast}{
10^{0.4(\alpha+1)(M-M^\ast)} +
10^{0.4(\beta+1)(M-M^\ast)}
}.
\label{eq:dpl_mag}
\end{equation}
These are parameterized by the faint-end slope $\alpha$, characteristic magnitude $M^\ast$, normalization $\phi^\ast$, and, in the DPL case, the bright-end slope $\beta$.

For each of our redshift bins we adopt representative literature parameterizations as the baseline comparison functions. For the Schechter fits at $z=6$ and $z=7$, we use the values reported by \citet{Bouwens2021}; for the DPL fits at $z=6$ and $z=7$, we use the values reported by \citet{Bowler2015}. For $z=8$, $z=9$, and $z=11$, we adopt both the Schechter and DPL parameterizations from \citet{Adams2024EPOCHS}. 
Our GLIMPSE measurements are then confronted with this literature baseline in a like-for-like manner. We work entirely with intrinsically de-lensed UV magnitudes $M_{\rm UV}$ and corresponding number densities in the same units as the literature compilation. Holding $(M^\ast,\phi^\ast)$ fixed to the literature best values, we re-fit only the faint-end slope $\alpha$ to our measurements on the faint side of the knee, enforcing
\begin{equation}
M_i \;\ge\; M^\ast_{\rm lit} + \Delta, \qquad \Delta = 0.5~\mathrm{mag},
\label{eq:faint_cut}
\end{equation}
so that the comparison isolates differences in the faint-end behavior while anchoring the knee and normalization to the external consensus. Such comparison is necessary as the limited survey volume of the GLIMPSE survey prohibits it from adequately probing the bright end of the UVLF and constraining any of the Schechter parameters other than the faint-end slope. It is also important to note that the literature to which we compare our sample \citep[e.g.][]{  Finkelstein2015, Bowler2015,Bouwens2021,Donnan_2023,Adams2024EPOCHS} does not include the notable studies searching the extreme faint end with gravitational lensing in order to keep our results independent of theirs \citep[e.g.][]{Livermore2017,Bouwens2017,Atek2018}. 

The results of this fitting can be seen in Table \ref{tab:MCMC} and Figures \ref{fig:uvlf_1} and \ref{fig:uvlf_secondary}. These results are then compared to the literature in Figure \ref{fig:alpha}, where we find that our results are broadly in line with that of the literature, indicating a declining $\alpha$, the faint end slope, across the redshifts we sampled. Moreover, our results provide extremely tight constraints on $\alpha$ (see Tab. \ref{tab:MCMC}) in comparison to the current UVLF literature.

\begin{figure}
    \centering
    \begin{subfigure}{0.5\textwidth}
        \centering
        \includegraphics[width=\linewidth]{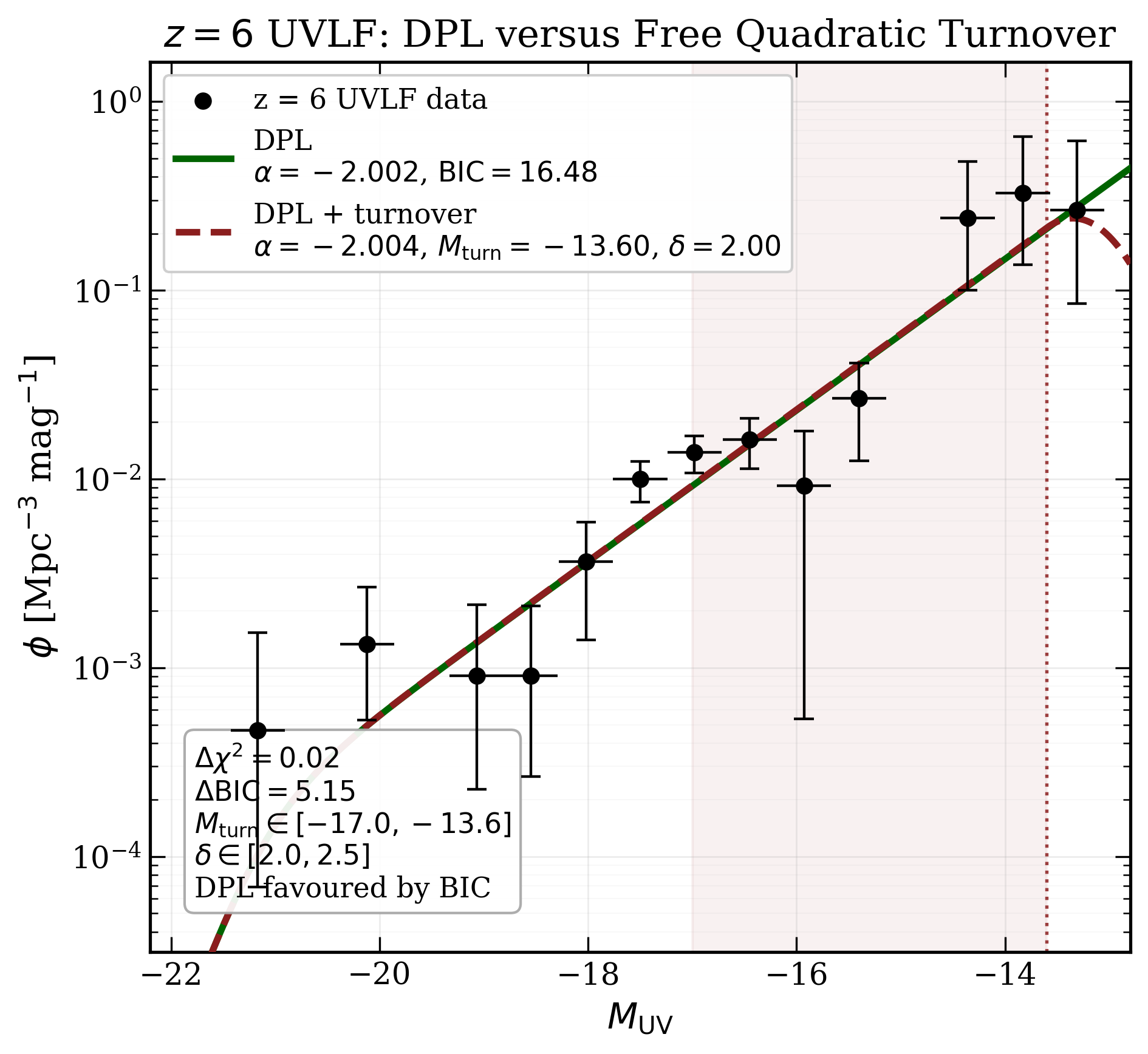}
        \label{fig:6lit}
    \end{subfigure}
    \hfill
    \begin{subfigure}{0.5\textwidth}
        \centering
        \includegraphics[width=\linewidth]{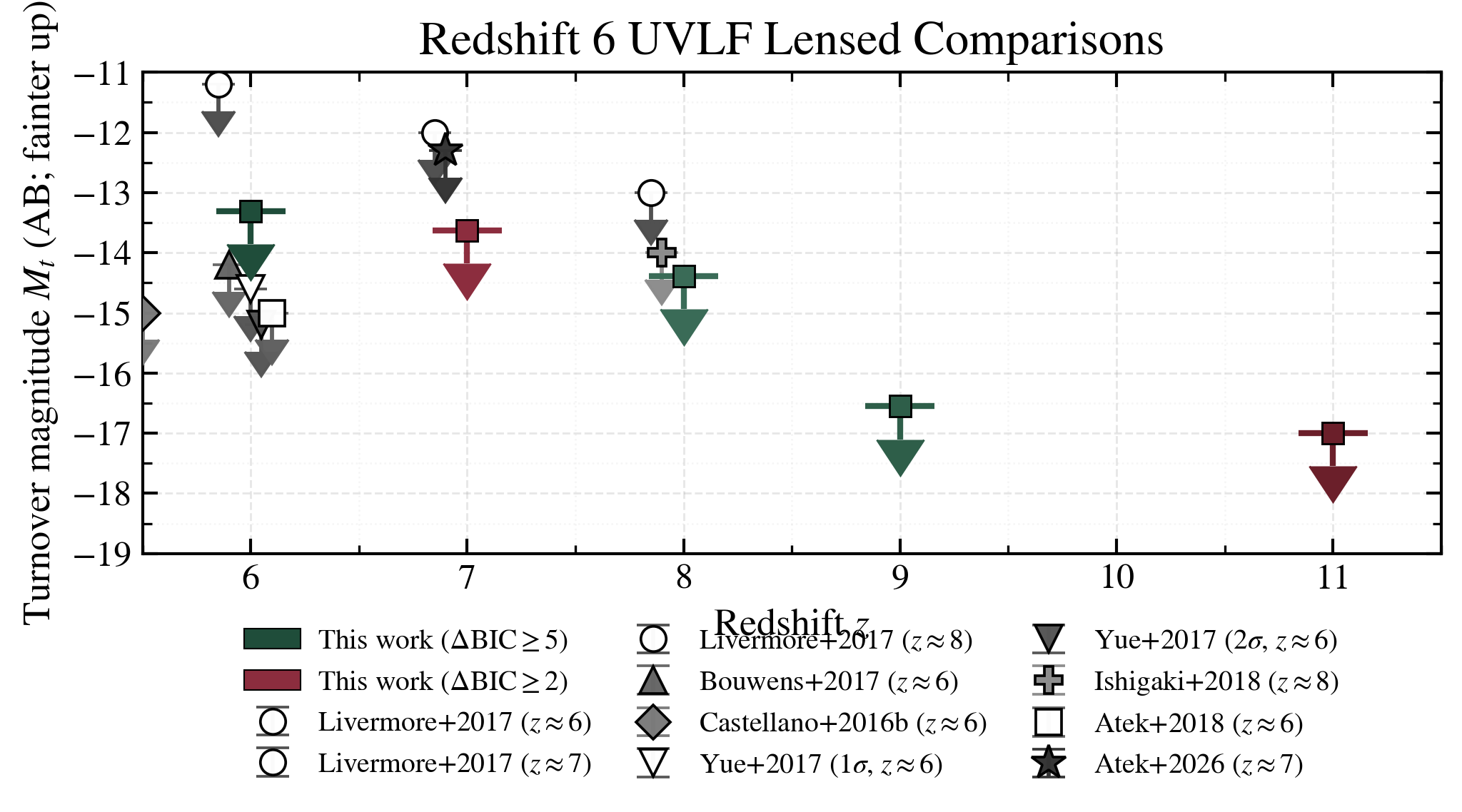}
        \label{fig:8lit}
    \end{subfigure}
    \caption{This figure illustrates our attempt to constrain the turnover by directly comparing a turnover to a no-turnover model. In the top figure, you can see how by assuming a strong turnover, and by having the turnover strength ($\delta$) and the turnover magnitude as free parameters, we observe the $\Delta$ BIC significantly favoring the no-turnover model, despite the very low $\Delta \chi^2$. Applying this generally-used methodology to every redshift bin, we can ostensibly put a lower limit on the turnover magnitude at various redshifts, as seen in the lower figure. However, for a more robust treatment of the turnover, look at Figures \ref{fig:turnover_grid} and \ref{fig:turnover_heat}.}
    \label{fig:Faint_easy}
\end{figure}

\begin{table}
\centering
\caption{Updated UV luminosity density and cosmic star-formation-rate density lower-limits, evaluated from the DPL UVLF integrated down to the turnover-limit magnitude $M_{\rm t}$ in each redshift bin. The star-formation-rate densities are computed from $\rho_{\rm UV}$ using $K_{\rm UV}=1.15\times10^{-28}\,M_\odot\,\mathrm{yr}^{-1}\,(\mathrm{erg\,s^{-1}\,Hz^{-1}})^{-1}$.}
\label{tab:rho_uv_results}
\begin{tabular}{cccc}
\hline
$z$ & $\log_{10}\rho_{\rm UV}$ & $\log_{10}\rho_{\rm SFR}$ & $M_{\rm t}$ \\
& (erg\,s$^{-1}$\,Hz$^{-1}$\,Mpc$^{-3}$) & ($M_\odot$\,yr$^{-1}$\,Mpc$^{-3}$) & \\
\hline
6.0  & $26.41^{+0.25}_{-0.26}$ & $-1.53^{+0.25}_{-0.26}$ & $-13.51$ \\
7.0  & $26.42^{+0.30}_{-0.30}$ & $-1.52^{+0.30}_{-0.30}$ & $-13.63$ \\
8.0  & $25.83^{+0.33}_{-0.33}$ & $-2.11^{+0.33}_{-0.33}$ & $-14.39$ \\
9.0  & $25.47^{+0.33}_{-0.33}$ & $-2.47^{+0.33}_{-0.33}$ & $-16.55$ \\
11.0 & $25.15^{+0.63}_{-0.63}$ & $-2.79^{+0.63}_{-0.63}$ & $-17.00$ \\
\hline
\end{tabular}
\end{table}

\subsubsection{UVLF Turnover Modeling}
\label{subsubsec:turnover-modeling}

As discussed above, the UVLF is not expected to continue rising indefinitely towards arbitrarily faint magnitudes. In low-mass haloes, feedback, photoheating, inefficient gas accretion, and other baryonic suppression mechanisms are expected to reduce the efficiency of star formation, producing a flattening or turnover in the faint-end galaxy abundance \citep{Gnedin2000,Finlator2011}. We therefore test whether our measured UVLFs show evidence for a faint-end suppression relative to an unbroken DPL parameterisation.

We model the possible suppression relative to the DPL form used throughout this work. Following the quadratic turnover prescription used in previous high-redshift UVLF studies \citep[e.g.][]{Bouwens2017,Atek2018}, we multiply the DPL by a magnitude-dependent suppression factor,
\begin{equation}
    T(M_{\rm UV}\mid M_{\rm t},\delta) =
    \begin{cases}
    1, & M_{\rm UV} \leq M_{\rm t}, \\
    10^{-0.4\delta(M_{\rm UV}-M_{\rm t})^2}, & M_{\rm UV} > M_{\rm t},
    \end{cases}
    \label{eq:quadratic_turnover_factor}
\end{equation}
such that
\begin{equation}
    \phi_{\rm turn}(M_{\rm UV}) =
    \phi_{\rm DPL}(M_{\rm UV})\,
    T(M_{\rm UV}\mid M_{\rm t},\delta).
    \label{eq:dpl_turnover_model}
\end{equation}
Here \(M_{\rm t}\) sets the magnitude at which the turnover begins, while \(\delta\), the turnover parameter, controls the strength or curvature of the suppression. Larger values of \(\delta\) correspond to a sharper decline faintward of \(M_{\rm t}\), whereas smaller values produce a smoother, more gradual turnover. Since magnitudes increase towards fainter galaxies, the suppression is only applied for \(M_{\rm UV}>M_{\rm t}\).

We first compare a standard no-turnover DPL model against a DPL plus turnover model using the Bayesian Information Criterion,
\begin{equation}
    {\rm BIC} = \chi^2 + k\ln N,
\end{equation}
where \(k\) is the number of free parameters and \(N\) is the number of UVLF data points \citep{Schwarz1978}. In this comparison, the baseline DPL has only the faint-end slope \(\alpha\) varied, while the turnover model includes additional freedom through \(\alpha\), \(M_{\rm t}\), and \(\delta\). We define $\Delta{\rm BIC} ={\rm BIC}_{\rm turn} - {\rm BIC}_{\rm DPL}$ so that positive values favour the no-turnover DPL. Under the standard \citet{kass1995bayes} interpretation, \(\Delta{\rm BIC}\geq2\) and \(\Delta{\rm BIC}\geq5\) correspond to positive and strong evidence, respectively, against the additional turnover model.

Based on the confidence intervals discussed earlier, requiring a \(\Delta\mathrm{BIC}\) of 2 for 'positive evidence' and a \(\Delta\mathrm{BIC}\) of 5 for statistical significance, our results indicate that at no redshift or valid turnover magnitude that the turnover model fits the currently available data statistically significantly.

Instead, our data suggests that there is no turnover visible in any of our redshift bins, however, to varying degrees of significance. The results of this turnover modeling are seen in Figure \ref{fig:turnover} and in Table \ref{tab:rho_uv_results}. However, the most significant of these is in our $z=6$ bin, which, at the \(\Delta\mathrm{BIC}\) of 5 significance level, shows no turnover down to $M_{UV} = -13.5$.

However, this direct BIC comparison provides an incomplete and potentially misleading measure of the turnover constraint. Because the turnover model contains additional freedom, it is penalised even when it produces a modest improvement in \(\chi^2\). The no-turnover DPL can therefore be selected unless the turnover produces a sufficiently large gain in goodness of fit to overcome the additional parameter penalty. While this is useful as a conservative model-selection test, it can overstate the apparent significance of a turnover non-detection if interpreted as ruling out all possible turnover shapes.

A more robust approach is therefore to ask which specific turnover shapes are incompatible with the data. This is necessary because the strength of the non-detection is strongly dependent on the assumed turnover curvature. A sharp turnover produces a clear departure from the DPL and is comparatively easy to identify or exclude. In contrast, a shallow turnover produces a much smoother suppression, which can be partially absorbed by a small change in the fitted faint-end slope. Such models may remain nearly indistinguishable from an unbroken DPL over the magnitude range probed by the data, even if the physical UVLF has already begun to turn over.

\begin{figure*}
    \centering
    \includegraphics[width=\textwidth]{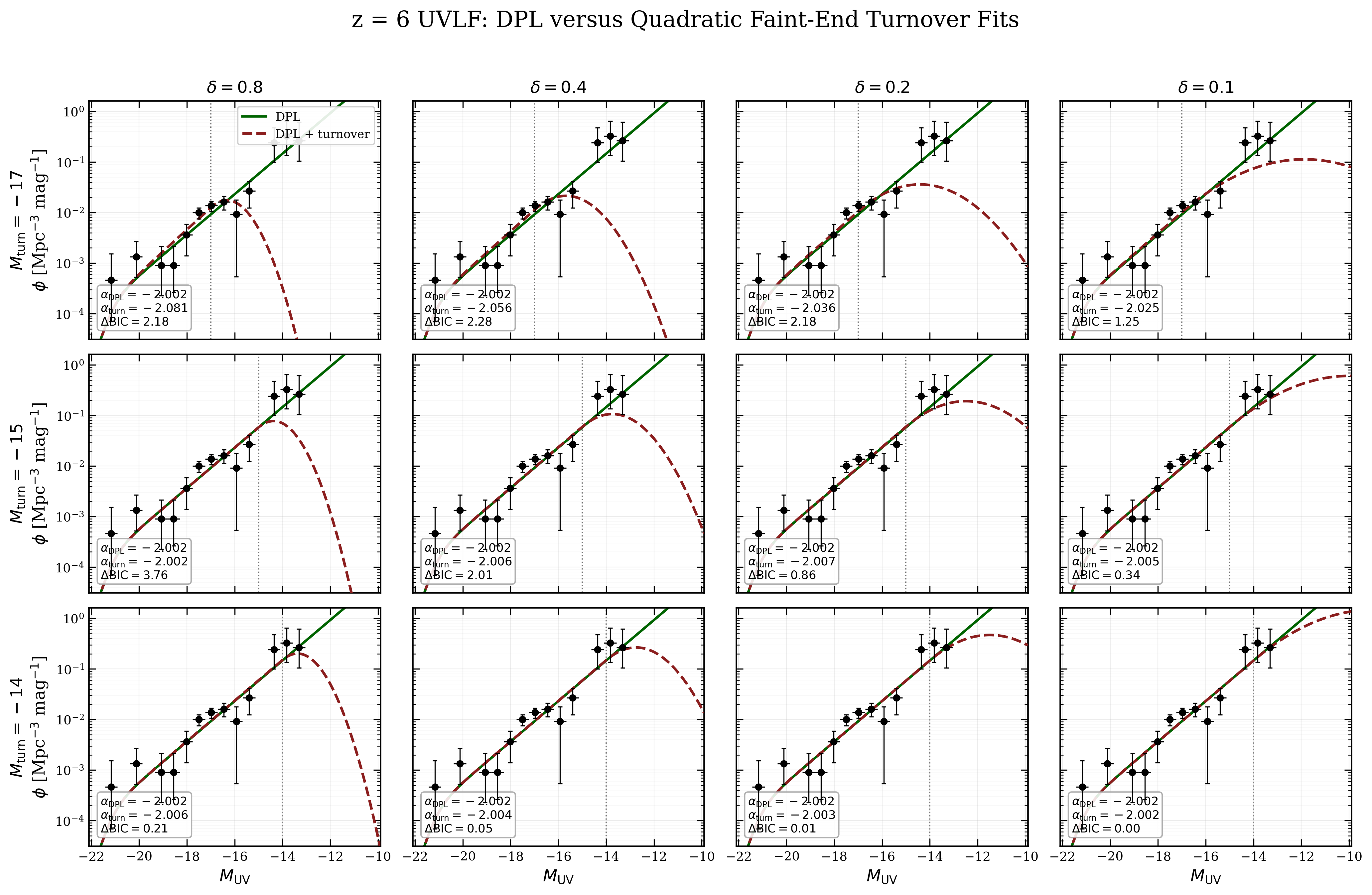}
    \caption{A comparison of the best-fit Double Power Law (DPL) fit of the Redshift 6 UVLF with the UVLF DPL with various turnover models with $\delta\ = 0.1, 0.2, 0.4,$  and 0.8 and $M_{\rm t} = -14, -15,$  and -17. This illustrates how little variation there is in the faint end between the turnover and no turnover models, especially when you assume a weak suppression mechanism.}
    \label{fig:turnover_grid}
\end{figure*}

We illustrate this behaviour in Figure \ref{fig:turnover_grid}, where we show a fixed grid of turnover models spanning different turnover magnitudes and turnover strengths. This figure demonstrates that models with the same turnover magnitude can imply substantially different faint-end behaviour depending on the value of the turnover parameter. In particular, the shallowest turnovers remain visually and statistically difficult to distinguish from the no-turnover DPL. These cases cannot simply be ignored, since a weak or gradual suppression is physically plausible and would not necessarily be captured by a model-selection criterion that favours the simpler DPL.

We therefore adopt a fixed-grid \(\Delta\chi^2\) analysis as our primary constraint on the allowed turnover parameter space. For each point in a grid of turnover magnitude and turnover strength, we keep the turnover parameters fixed and refit only the DPL faint-end slope. We then compare the resulting \(\chi^2\) to that of the no-turnover DPL. This removes the ambiguity associated with a fully free turnover model being penalised for its additional parameters, and instead directly identifies which specific turnover locations and curvatures provide an unacceptable fit to the observed UVLF.

\begin{figure}
    \centering
    \includegraphics[width=0.5\textwidth]{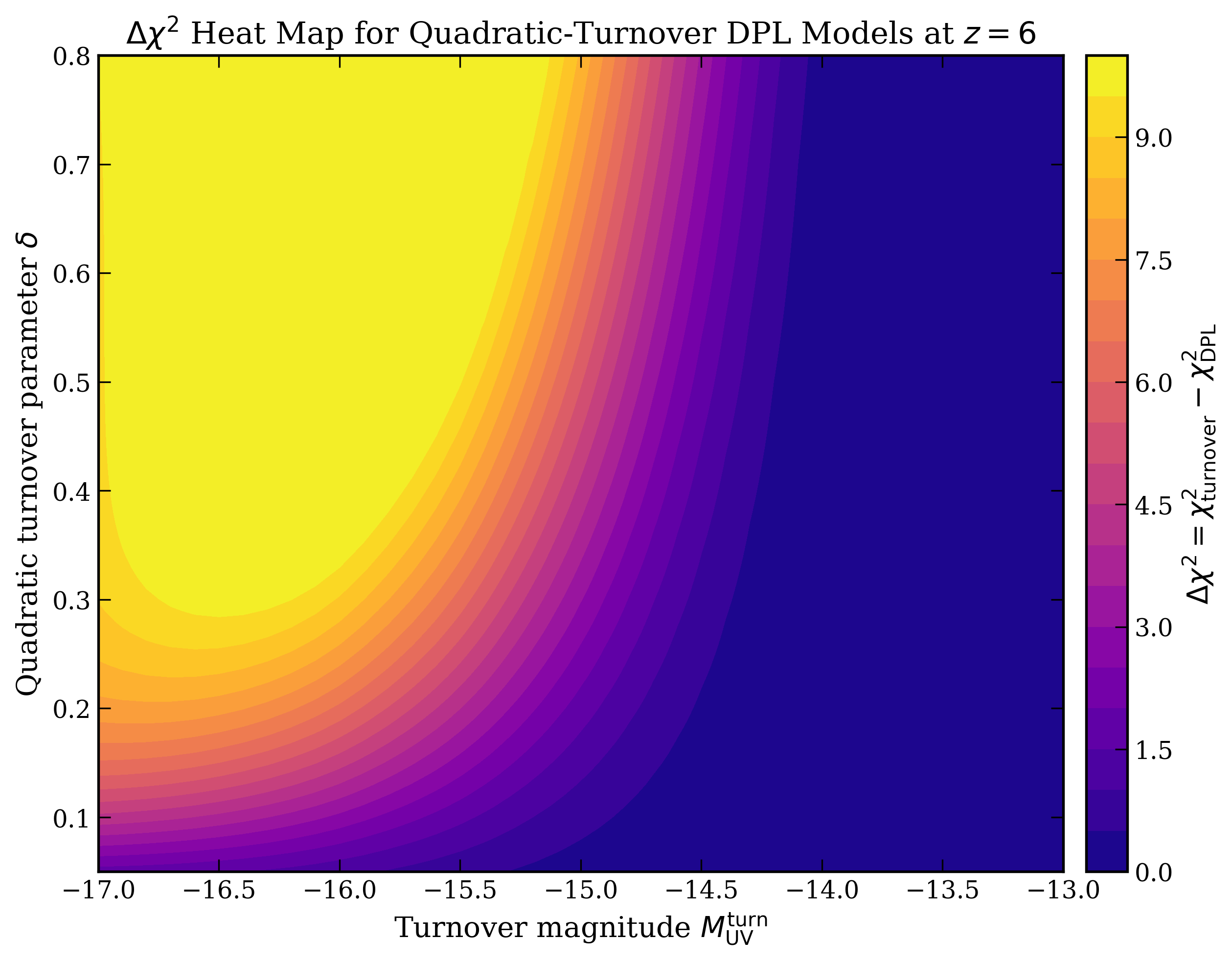}
    \caption{A heat map of the $\Delta \chi^2$, of how much better the no-turnover model fits the observed redshift $z=6$ UVLF when compared to a turnover model with a given $\delta$ and $M_{\rm t}$. Stronger  (higher $\delta$) and higher mass/ lower magnitude (lower $M_{\rm t}$) turnovers are strongly excluded by the comparison, while it is difficult to robustly exclude shallow turnovers or turnovers at higher magnitudes as the faint end of the UVLF is not constrained enough yet. }
    \label{fig:turnover_heat}
\end{figure}

The resulting \(\Delta\chi^2\) map is shown in Figure \ref{fig:turnover_heat}. Regions with large positive \(\Delta\chi^2\) correspond to turnover models that fit the data substantially worse than the no-turnover DPL and can therefore be robustly excluded. Conversely, regions with low \(\Delta\chi^2\) remain consistent with the data, either because the turnover occurs fainter than the current observational limit or because the suppression is sufficiently gradual that it is effectively degenerate with the fitted DPL slope. This fixed-grid analysis provides a more physically informative constraint than a single BIC comparison, or even by providing a 'best-fit' turnover model, 

\subsection{Cosmic Star-Formation-Rate Density and $\rho_{\rm UV}$ }
\label{subsec:CSFRD}

We compute conservative lower limits on the rest--UV luminosity density, $\rho_{\rm UV}$, by integrating the luminosity-weighted UVLF only down to the faintest turnover magnitude still allowed by the data for a given turnover strength. For the Schechter form in \autoref{eq:schechter_mag}, this gives
\begin{equation}
\rho_{\rm UV}(M_{\rm lim}) \;=\; \phi^\star\,L_\nu(M^\star)\,\Gamma\!\big(\alpha+2,\;10^{-0.4(M_{\rm lim}-M^\star)}\big),
\end{equation}
where $\Gamma$ is the upper incomplete gamma function. We evaluate this quantity in each UVLF redshift bin using the best-fitting parameters reported in Table~\ref{tab:MCMC}, setting the integration limit to $M_{\rm lim}=M_{\rm t}$ from the turnover analysis described in $\S$ \ref{subsubsec:turnover-modeling}. The resulting values are listed in Table~\ref{tab:rho_uv_results}.

\begin{table*}
\centering
\scriptsize
\caption{Minimum turnover magnitudes inferred for fixed turnover strengths. The 90\% and 95\% confidence limits correspond to $\Delta\chi^2 = 2.706$ and $\Delta\chi^2 = 3.841$, respectively. The quoted uncertainties on $\rho_{\rm UV}$ are the symmetric quadrature errors from the numerical integration, propagated directly to the SFRD using $K_{\rm UV}=1.15\times10^{-28}$. A dash indicates that the target $\Delta\chi^2$ threshold was not reached.}
\label{tab:turnover_limits_rhouv_sfrd_errors}
\resizebox{\textwidth}{!}{
\begin{tabular}{cc|ccc|ccc}
\hline
 &  & \multicolumn{3}{c|}{90\% confidence} & \multicolumn{3}{c}{95\% confidence} \\
\hline
$z$
& Turnover strength
& $M_{\rm t}$
& $\rho_{\rm UV}$
& SFRD
& $M_{\rm t}$
& $\rho_{\rm UV}$
& SFRD \\
&
$\delta$
& mag
& $10^{25}$ erg s$^{-1}$ Hz$^{-1}$ Mpc$^{-3}$
& $10^{-3}$ $M_\odot$ yr$^{-1}$ Mpc$^{-3}$
& mag
& $10^{25}$ erg s$^{-1}$ Hz$^{-1}$ Mpc$^{-3}$
& $10^{-3}$ $M_\odot$ yr$^{-1}$ Mpc$^{-3}$ \\
\hline
6 %\multirow{6}
& weak, $\delta=0.2$
& $-15.95$
& $22.410^{+2.88\times10^{-7}}_{-2.88\times10^{-7}}$
& $25.771^{+3.31\times10^{-7}}_{-3.31\times10^{-7}}$
& $-16.25$
& $21.884^{+2.62\times10^{-7}}_{-2.62\times10^{-7}}$
& $25.166^{+3.01\times10^{-7}}_{-3.01\times10^{-7}}$ \\
& medium, $\delta=0.5$
& $-15.11$
& $22.441^{+5.73\times10^{-7}}_{-5.73\times10^{-7}}$
& $25.807^{+6.59\times10^{-7}}_{-6.59\times10^{-7}}$
& $-15.24$
& $22.215^{+4.25\times10^{-7}}_{-4.25\times10^{-7}}$
& $25.548^{+4.89\times10^{-7}}_{-4.89\times10^{-7}}$ \\
& strong, $\delta=0.8$
& $-14.81$
& $22.485^{+1.24\times10^{-6}}_{-1.24\times10^{-6}}$
& $25.857^{+1.43\times10^{-6}}_{-1.43\times10^{-6}}$
& $-14.92$
& $22.353^{+8.98\times10^{-7}}_{-8.98\times10^{-7}}$
& $25.705^{+1.03\times10^{-6}}_{-1.03\times10^{-6}}$ \\
\hline
7 %\multirow{7}
& weak, $\delta=0.2$
& -- & -- & -- & -- & -- & -- \\
& medium, $\delta=0.5$
& $-17.61$
& $19.373^{+6.16\times10^{-7}}_{-6.16\times10^{-7}}$
& $22.279^{+7.08\times10^{-7}}_{-7.08\times10^{-7}}$
& $-17.74$
& $19.421^{+7.64\times10^{-7}}_{-7.64\times10^{-7}}$
& $22.334^{+8.79\times10^{-7}}_{-8.79\times10^{-7}}$ \\
& strong, $\delta=0.8$
& $-16.67$
& $20.005^{+5.26\times10^{-8}}_{-5.26\times10^{-8}}$
& $23.005^{+6.05\times10^{-8}}_{-6.05\times10^{-8}}$
& $-16.79$
& $19.469^{+5.84\times10^{-8}}_{-5.84\times10^{-8}}$
& $22.389^{+6.72\times10^{-8}}_{-6.72\times10^{-8}}$ \\
\hline
8 %\multirow{8}
& weak, $\delta=0.2$
& -- & -- & -- & -- & -- & -- \\
& medium, $\delta=0.5$
& $-17.20$
& $5.301^{+1.26\times10^{-7}}_{-1.26\times10^{-7}}$
& $6.096^{+1.44\times10^{-7}}_{-1.44\times10^{-7}}$
& $-17.62$
& $5.116^{+2.86\times10^{-7}}_{-2.86\times10^{-7}}$
& $5.883^{+3.29\times10^{-7}}_{-3.29\times10^{-7}}$ \\
& strong, $\delta=0.8$
& $-16.51$
& $5.651^{+7.42\times10^{-8}}_{-7.42\times10^{-8}}$
& $6.498^{+8.53\times10^{-8}}_{-8.53\times10^{-8}}$
& $-16.73$
& $5.485^{+9.77\times10^{-8}}_{-9.77\times10^{-8}}$
& $6.307^{+1.12\times10^{-7}}_{-1.12\times10^{-7}}$ \\
\hline
9 %\multirow{9}
& weak, $\delta=0.2$
& -- & -- & -- & -- & -- & -- \\
& medium, $\delta=0.5$
& $-18.62$
& $3.695^{+2.09\times10^{-7}}_{-2.09\times10^{-7}}$
& $4.249^{+2.41\times10^{-7}}_{-2.41\times10^{-7}}$
& $-18.86$
& $3.819^{+2.78\times10^{-7}}_{-2.78\times10^{-7}}$
& $4.392^{+3.20\times10^{-7}}_{-3.20\times10^{-7}}$ \\
& strong, $\delta=0.8$
& $-17.92$
& $3.404^{+9.11\times10^{-10}}_{-9.11\times10^{-10}}$
& $3.915^{+1.05\times10^{-9}}_{-1.05\times10^{-9}}$
& $-18.11$
& $3.488^{+1.35\times10^{-7}}_{-1.35\times10^{-7}}$
& $4.012^{+1.55\times10^{-7}}_{-1.55\times10^{-7}}$ \\
\hline
11 %\multirow{11}
& weak, $\delta=0.2$
& -- & -- & -- & -- & -- & -- \\
& medium, $\delta=0.5$
& $-20.76$
& $1.090^{+5.15\times10^{-7}}_{-5.15\times10^{-7}}$
& $1.253^{+5.92\times10^{-7}}_{-5.92\times10^{-7}}$
& $-20.83$
& $0.980^{+4.81\times10^{-7}}_{-4.81\times10^{-7}}$
& $1.127^{+5.54\times10^{-7}}_{-5.54\times10^{-7}}$ \\
& strong, $\delta=0.8$
& $-19.82$
& $1.872^{+1.31\times10^{-6}}_{-1.31\times10^{-6}}$
& $2.152^{+1.51\times10^{-6}}_{-1.51\times10^{-6}}$
& $-20.04$
& $1.669^{+2.35\times10^{-7}}_{-2.35\times10^{-7}}$
& $1.919^{+2.70\times10^{-7}}_{-2.70\times10^{-7}}$ \\
\hline
\end{tabular}
}
\end{table*}

We also compute $\rho_{\rm UV}$ for brightest potential turnover given various turnover models in our robust turnover search methodology described in $\S$ \ref{subsubsec:turnover-modeling}. These values are displayed in Table~\ref{tab:turnover_limits_rhouv_sfrd_errors} and Figure \ref{fig:rho_redshift}. These values should be interpreted as lower limits on $\rho_{\rm UV}$ for each assumed turnover strength, rather than as measurements of the total UV luminosity density. For a fixed turnover prescription, the adopted $M_{\rm t}$ corresponds to the brightest, and therefore most suppressive, faint-end cutoff still permitted at the 90 per cent confidence level. Any turnover occurring at fainter magnitudes, or any weaker suppression of the UVLF, would increase the integrated luminosity density. Thus, the weak, medium, and strong turnover cases provide conservative minimum $\rho_{\rm UV}$ estimates conditional on the assumed quadratic turnover strengths.

We also convert these UV luminosity-density limits into corresponding limits on the cosmic star-formation-rate density, $\rho_{\rm SFR}$, using
\begin{equation}
\rho_{\rm SFR} = K_{\rm UV}\,\rho_{\rm UV},
\end{equation}
with $K_{\rm UV}=1.15\times10^{-28}\,M_\odot\,{\rm yr}^{-1}\,({\rm erg}\,{\rm s}^{-1}\,{\rm Hz}^{-1})^{-1}$ \citep{Madau2014}. Because these $\rho_{\rm SFR}$ values are derived directly from the turnover-limited $\rho_{\rm UV}$ estimates, they should likewise be interpreted as lower limits for the assumed turnover model. 

We do not apply a dust correction to the reported luminosity densities. This choice is conservative: the dust attenuation of the faint galaxy population is not robustly constrained in our sample, and these systems are expected to have relatively low dust content. Moreover, any correction for dust attenuation would only increase the inferred intrinsic $\rho_{\rm UV}$ and hence $\rho_{\rm SFR}$. Neglecting dust therefore preserves the lower-limit interpretation of the values reported in Table~\ref{tab:turnover_limits_rhouv_sfrd_errors} and illustrated in Figure \ref{fig:rho_redshift}, where we compare our turnover-dependent limits to literature measurements.

\subsection{Ionizing Photon Production}
\label{subsec:ion}

We estimate the ionizing photon production rate density, $\dot{n}_{\rm ion}$, by combining our turnover-constrained UV luminosity densities with prescriptions for the ionizing photon production efficiency and the escape fraction. Specifically, we compute
\begin{equation}
    \dot{n}_{\rm ion}
    =
    \rho_{\rm UV}\,\xi_{\rm ion}\,f_{\rm esc},
\end{equation}
where $\rho_{\rm UV}$ is the integrated non-ionizing UV luminosity density, $\xi_{\rm ion}$ converts this UV luminosity into an intrinsic ionizing photon production rate, and $f_{\rm esc}$ is the fraction of ionizing photons that escape into the IGM.

For this calculation, we adopt the updated luminosity-dependent $\xi_{\rm ion}$ relation from \citet{Duncan2025_inprep}, parameterised as
\begin{equation}
    \log_{10}(\xi_{\rm ion}/{\rm Hz\,erg}^{-1})
    =
    m M_{\rm UV} + c,
\end{equation}
with $m=-0.006^{+0.019}_{-0.017}$ and $c=25.05^{+0.39}_{-0.34}$. We also use the updated UV-continuum slope relation from the same work,
\begin{equation}
    \beta_{\rm UV}(M_{\rm UV})
    =
    a + b(M_{\rm UV}+19.5),
\end{equation}
with $a=-2.88^{+0.43}_{-0.44}$ and $b=-0.040\pm0.022$ to estimate the escape fraction is then estimated using the \citet{Chisholm_2022} relation between $f_{\rm esc}$ and $\beta_{\rm UV}$,
\begin{equation}
    f_{\rm esc}
    =
    1.3\times10^{-4}\,10^{-1.22\beta_{\rm UV}}.
\end{equation}
Uncertainties in $\rho_{\rm UV}$, $\xi_{\rm ion}$, $\beta_{\rm UV}$, and $f_{\rm esc}$ are propagated through Monte Carlo sampling to obtain the final uncertainty on $\dot{n}_{\rm ion}$.

These estimates should be interpreted with caution. The empirical relations are currently calibrated only down to approximately $M_{\rm UV}\simeq-17$, whereas our turnover-constrained luminosity densities can include contributions from substantially fainter galaxies. We therefore extrapolate the $\xi_{\rm ion}(M_{\rm UV})$, $\beta_{\rm UV}(M_{\rm UV})$, and hence $f_{\rm esc}(\beta_{\rm UV})$ prescriptions beyond their directly studied magnitude range. The resulting $\dot{n}_{\rm ion}$ values should therefore be viewed as illustrative estimates of the possible ionizing emissivity implied by our faint-end UVLF constraints, rather than direct empirical measurements at these extremely faint luminosities.

\section{Discussion}
\label{sec:disc}

\begin{table}
\centering
\caption{Ionizing photon emissivity lower limits inferred using the \citet{Duncan2025_inprep} luminosity-dependent $\xi_{\rm ion}$ relation and the \citet{Chisholm_2022} $f_{\rm esc}(\beta_{\rm UV})$ prescription. Values are quoted as $\log_{10}(\dot{n}_{\rm ion}/{\rm s}^{-1}\,{\rm Mpc}^{-3})$.}
\label{tab:ndot_chisholm}
\footnotesize
\setlength{\tabcolsep}{3.5pt}
\begin{tabular}{cccc}
\hline
$z$ & $\delta$ & $M_{\rm turn}$ & $\log_{10}\dot{n}_{\rm ion}$ \\
\hline
6  & 0.2 & -15.95 & $51.03^{+0.66}_{-0.80}$ \\
6  & 0.5 & -15.11 & $51.02^{+0.68}_{-0.79}$ \\
6  & 0.8 & -14.81 & $51.03^{+0.67}_{-0.78}$ \\
7  & 0.5 & -17.61 & $50.96^{+0.68}_{-0.80}$ \\
7  & 0.8 & -16.67 & $50.99^{+0.67}_{-0.80}$ \\
8  & 0.5 & -17.20 & $50.39^{+0.68}_{-0.80}$ \\
8  & 0.8 & -16.51 & $50.43^{+0.67}_{-0.80}$ \\
9  & 0.5 & -18.62 & $50.25^{+0.67}_{-0.79}$ \\
9  & 0.8 & -17.92 & $50.21^{+0.67}_{-0.80}$ \\
11 & 0.5 & -20.76 & $49.71^{+0.68}_{-0.80}$ \\
11 & 0.8 & -19.82 & $49.95^{+0.67}_{-0.80}$ \\
\hline
\end{tabular}%
\end{table}

\subsection{The UVLF Turnover}

Our binned UVLF shows no direct evidence for a steep downturn down to the faintest bin reached by the AS1063 data, $M_{\rm UV}=-13.5$ at $z=6$. This is a useful empirical statement, especially when comparing this work to other UVLF turnover studies such as \citet{Livermore2017,Bouwens2017, atek2026}, but it cannot and should not be interpreted as a robust non-detection of a potential turnover. As mentioned in $\S$ \ref{subsubsec:turnover-modeling}, much of the significance of this result is reliant upon the $\Delta BIC$ criterion, which artificially favors the non-turnover model due to the added complexity of the turnover parameters without said model actually representing a better fit to the data. Moreover, a turnover is unlikely to correspond to an abrupt luminosity threshold below which galaxy formation ceases entirely; more physically, it should represent a gradual suppression in the faint regime \citep{Yue2016, Gnedin2016}. Galaxies fainter than the turnover can therefore persist, albeit with reduced number densities, and would therefore continue to contribute to $\rho_{\rm UV}$ and, consequently, to the ionizing photon budget. Further, weak turnovers can remain consistent with the observations down to fainter $M_{\rm t}$ values than stronger turnovers.

Our fixed-grid turnover analysis, in which the DPL UVLF is modified by a quadratic faint-end suppression, more robustly tests a range of turnover magnitudes and suppression strengths, without being forced to lean on the $\Delta BIC$ criterion (see Fig. \ref{fig:turnover_heat}). We report three representative cases, corresponding to weak, medium, and strong suppression with $\delta=0.2$, $0.5$, and $0.8$, respectively, for each redshift bin considered (see Fig. \ref{fig:turnover_grid} and Tab. \ref{tab:turnover_limits_rhouv_sfrd_errors}). 

With this methodology, we can be $90\%$ confident that there is no turnover, even a shallow turnover, brighter than  $M_{UV}  \approx -16$ at $z = 6$, rather than merely failing to positively observe a strong turnover with the given data. This increase of the robustness of turnover constraints is important not only to consider shallow turn-over models against observations, but also to make the significant distinction between a non-observation of a turnover and the positive observation of no turnover.  

Thus, while the non-detection of an observed downturn to $M_{\rm UV}=-13.5$ is consistent with previous work, the more physically informative and unique result is that gradual-turnover models can be excluded only to model-dependent limits, with moderate and strong suppression ruled out to approximately $M_{\rm UV}\simeq-15$ at $z=6$ as seen in Figure \ref{fig:turnover_heat}. 

\subsubsection{The Redshift Evolution of the Turnover}

While simulations generally predict a faint-end downturn, often at $M_{\rm UV}\sim-8$ to $-12$,  both the location and sharpness of this feature depend strongly on the assumed feedback physics and radiation field \citep{Jaacks2013,2016ocvirk,Yue2016,Gnedin2016}. One of the most critical physical motivations of the turnover is photoheating feedback from the ionized intergalactic medium (IGM), whose pressure can theoretically limit the accretion of baryonic matter in a low-mass dark matter halo \citep{Gnedin2000, Okamoto2008}. 

A natural consequence of this is that as Reionization progresses, and as redshifts approach  $z \sim 6$, this feedback mechanism would become more significant as the IGM becomes more and more ionized. Thus, we may expect the nature and the location of the UVLF turnover to alter during the Epoch of Reionisation and evolve across redshifts. Moreover, if this logic holds, then this turnover would be expected to be most prominent at the latter-end of reionisation, around $z=6$, making these younger redshift bins the most promising for astronomers to observe a turnover. For this reason, treating the turnover as a redshift-independent quantity is likely an oversimplification. 

Of course, with the data currently available and the lack of any observation of a turnover, we cannot constrain this evolution in any way. While our lower limits on the turnover change across redshift, this is due to limiting factors of the data and our inability to robustly identify faint galaxies at higher and higher redshifts. Nevertheless, because the turnover's location and size directly set the faint-end contribution to $\rho_{\rm UV}$ and hence to the ionizing photon budget, its possible evolution should be explored more systematically with deeper lensing samples and physically motivated simulations.

\begin{figure*}
    \centering
    \includegraphics[width=0.8\textwidth]{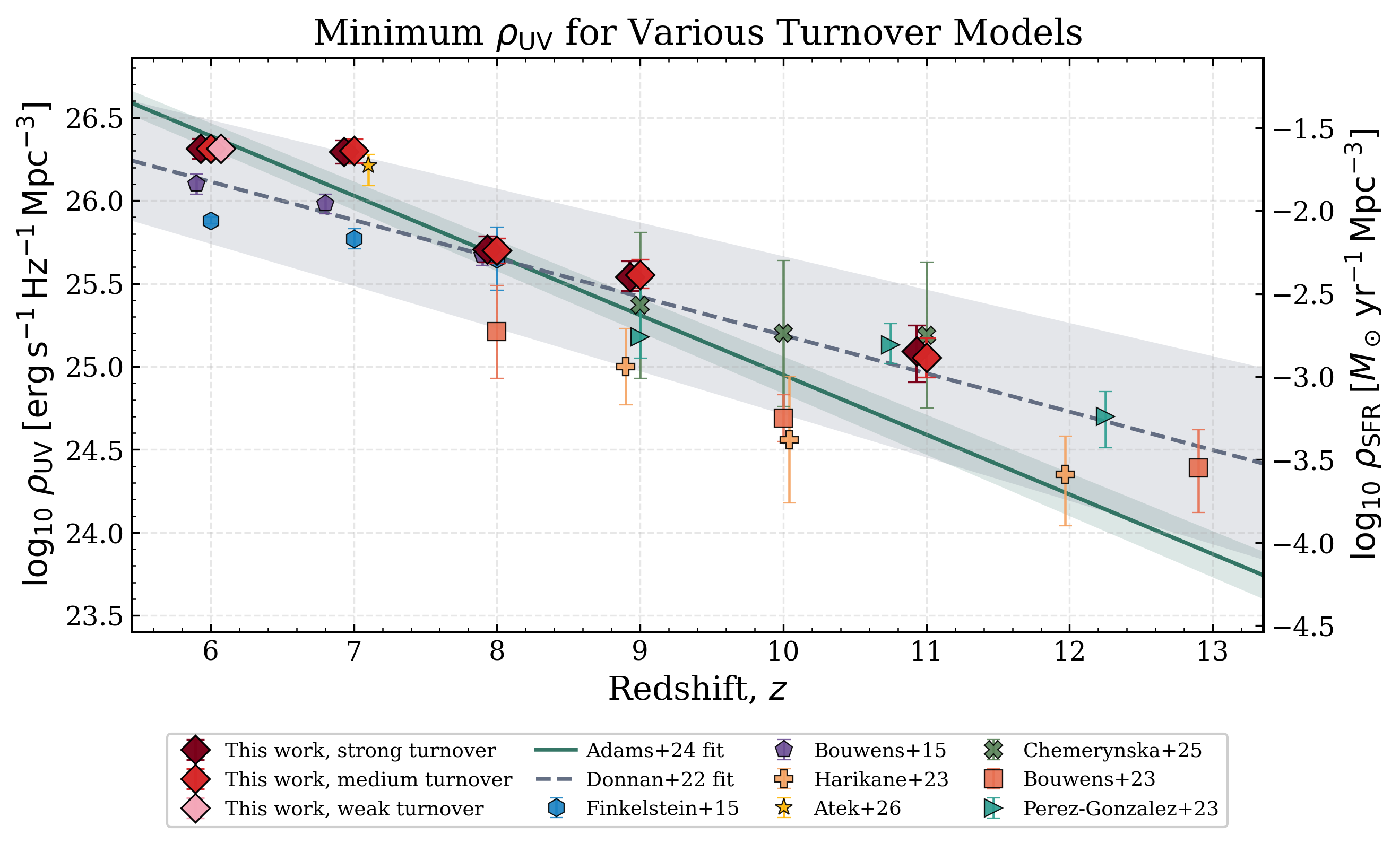}
    \caption{Redshift evolution of the minimum allowed non-ionizing UV luminosity density, $\rho_{\rm UV}$, inferred from fixed-strength faint-end turnover models. Coloured diamond markers show the lower-limit values from this work, obtained by placing the turnover at the limiting magnitude still allowed by the data at the 90 per cent confidence level. The plotted values thus represent conservative minimum estimates of $\rho_{\rm UV}$. The three turnover prescriptions correspond to a quadratic suppression applied to the DPL faint end, with weak, medium, and strong turnover strengths given by $\delta = 0.2$, $0.5$, and $0.8$, respectively. Error bars indicate the 16th--84th percentile range from the bootstrap realisations. Literature measurements of $\rho_{\rm UV}$ are shown for comparison. The right-hand axis gives the corresponding star-formation-rate density, $\rho_{\rm SFR}$, obtained using the UV-to-SFR conversion factor $K_{\rm UV}=1.15\times10^{-28}\,M_\odot\,{\rm yr}^{-1}\,({\rm erg}\,{\rm s}^{-1}\,{\rm Hz}^{-1})^{-1}$. }

    \label{fig:rho_redshift}
\end{figure*}

\subsection{Theoretical Comparisons}
\label{subsec:thoery}

\begin{figure} 
    \centering 
    \begin{subfigure}{0.5\textwidth} 
    \centering 
    \includegraphics[width=\linewidth]{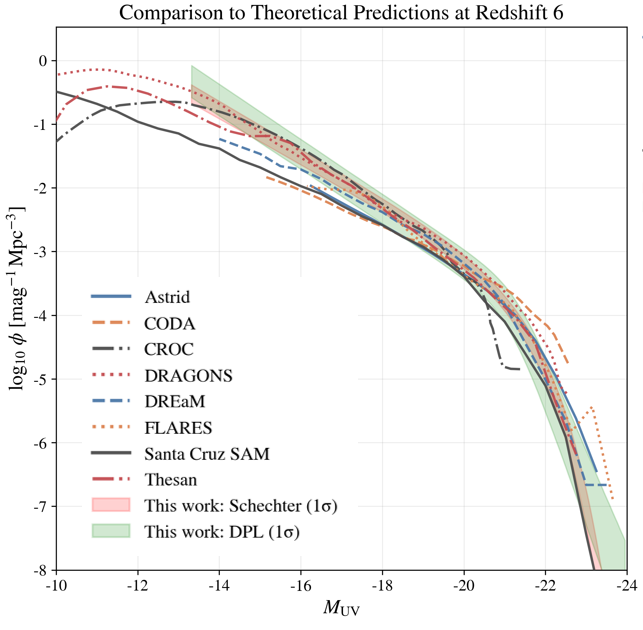} \label{fig:6lit} 
    \end{subfigure}
    \hfill 
    \begin{subfigure}{0.5\textwidth}
    \centering 
    \includegraphics[width=\linewidth]{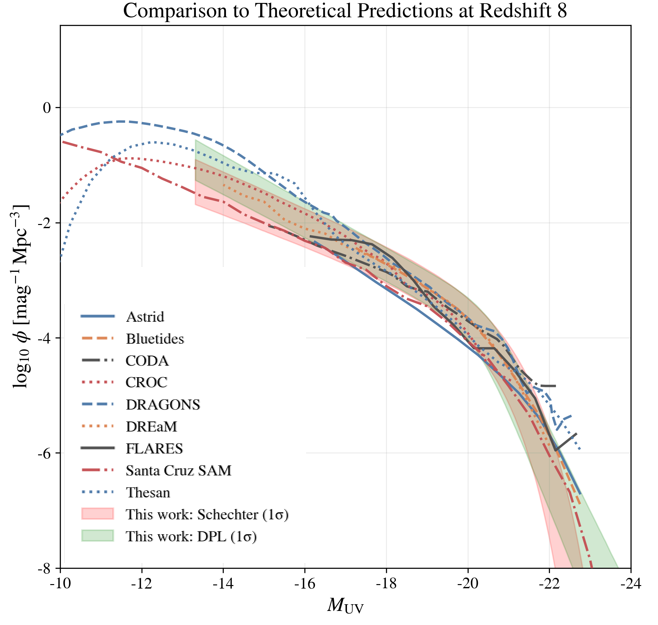} \label{fig:8lit} 
    \end{subfigure} 
    \caption{Constraints on the faint-end UV galaxy population from this work. Left: comparison of turnover magnitude constraints with previous studies, showing that, aside from \citet{Livermore2017}, this work places among the tightest constraints on the UVLF turnover to date. Right: comparison of the inferred $\rho_{\rm UV}$ lower limits from this work with literature values from \citet{Oesch2013, Bouwens2015, Finkelstein2015, Harikane2023, Donnan_2023, Atek2024Nature, Adams2024EPOCHS}.} 
    \label{fig:theory} 
\end{figure}

In order to better understand the significance of these results, we further compare our measured UVLFs to a range of theoretical predictions in Figure \ref{fig:theory}. As seen in the figure, at redshift $z = 6 $, our results are in broad agreement with theoretical predictions for the UVLF in the faint regime, particularly the FLARES and CROC simulations \citep{ Gnedin2016, FLARES}. It is worth mentioning, however, that we observe more faint sources that many of the other simulations predict, especially in the ultra-faint regime beyond $M_{UV}=-16$ \citep{2016ocvirk,Thesean}. 

It is widely accepted that suppression will occur at at some faint magnitude and mass, but it is still up for scientific discussion the nature, shape, and size of that suppression, and thus how much it will impact the faint end of the UVLF. Several mechanisms have been proposed: stellar feedback can drive outflows and lower the instantaneous star formation efficiency; photo-heating by the reionizing ultraviolet background can suppress gas accretion in shallow potential wells; limited atomic or molecular cooling can reduce the efficiency of star formation in the smallest haloes; and alternative dark matter models can further suppress the abundance of low-mass hosts \citep{FLARES, Thesean, darkmatter}. Both analytic and numerical studies show that sufficiently strong radiative feedback can flatten or turn over the UVLF at magnitudes as bright as $M_{\rm UV}\sim -15$, whereas weaker or more gradual suppression preserves a steep faint end to substantially fainter magnitudes \citep{yue2018, Thesean}. Within this broad context, our results suggest that any suppression of star formation in low-mass halos must either occur at magnitudes fainter than our current limit of $M_{UV} = -13.5$ or proceed more gradually. However, the present uncertainties remain too large to isolate a unique physical mechanism or to claim direct evidence for a turnover.

It is also worth briefly mentioning the work of \citet{darkmatter}, which used the HST lensed UVLFs to put constraints on dark matter models, specifically constraining the size of dark matter particles, comparing single versus double power laws, and comparing cold dark matter versus weak dark matter and fuzzy dark matter models, as each one would suppress the low-mass halos in slightly different ways. Our data is not a significant enough improvement on those previous HST studies, even though it probes higher redshifts, and will not be able to sufficiently constrain the parameters further. However, moving forward, future JWST lensing surveys should strongly consider its potential applications in cosmology in that regard. 

\begin{figure*}
    \centering
    \includegraphics[width=0.75\textwidth]{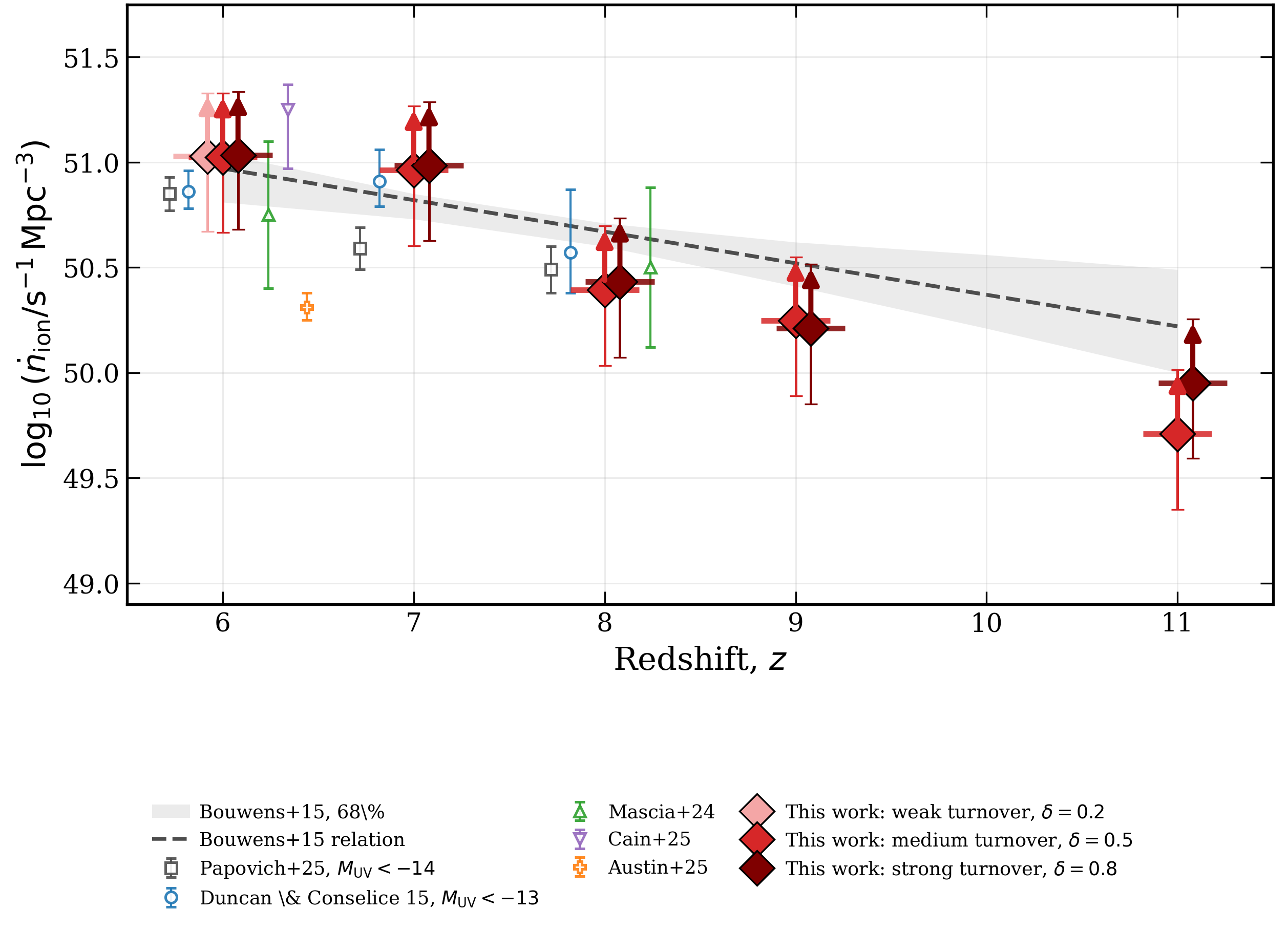}
    \caption{Lower limits on the ionizing photon emissivity, $\dot{n}_{\rm ion}$, inferred from the turnover-constrained UV luminosity densities using a luminosity-dependent Duncan-style $\xi_{\rm ion}(M_{\rm UV})$ relation and a Chisholm-style escape-fraction prescription, $f_{\rm esc}(\beta_{\rm UV})$. Literature measurements and model constraints are shown for comparison. These estimates should be interpreted cautiously, since the calculation extrapolates the $\xi_{\rm ion}$ and $f_{\rm esc}$ relations to extremely faint galaxies, beyond the magnitude range over which these empirical relations were originally calibrated.}
    \label{fig:ndot}
\end{figure*}

\subsection{Implications for Reionization}
\label{sec:reion_implications}

\subsubsection{The Ionizing Photon Budget}

A proper treatment of the UVLF turnover is essential when assessing the ionizing photon budget, because a turnover should not be interpreted as a hard luminosity cutoff. Even if the abundance of faint galaxies begins to decline, a suppressed population can still contribute substantially to the integrated UV luminosity density and hence to $\dot{n}_{\rm ion}$. To illustrate this, using our $z=6$ DPL parameters with a shallow quadratic turnover of $\delta=0.15$ at $M_{\rm t}=-15$, we find that galaxies fainter than the turnover would still contribute approximately $34\%$ of the total $\rho_{\rm UV}$ when integrated to $M_{\rm UV}=-8$. This illustrates that galaxies beyond the turnover can remain an important component of the reionization photon budget, particularly when the faint-end slope is steep. Consequently, models that impose a sharp cutoff at the turnover magnitude may significantly underestimate the contribution of low-luminosity galaxies to the ionizing emissivity.

To properly consider and model these effects, we calculated, from our turnover and faint-end slope measurements, conservative lower limits on $\rho_{\rm UV}$ and  Star Formation Rate Density $(SRFD)$ for various turnover models, which we then converted into $\dot{n}_{\rm ion}$ with the \citet{Duncan2025_inprep} and \citet{Chisholm_2022} relations for $\xi_{\rm ion}$ and $f_{\rm esc}$ respectively, as discussed in $\S$ \ref{subsec:ion} and $\S$ \ref{subsec:CSFRD}. Obviously, these relations were calculated from high redshift bright galaxy samples, and may not necessarily hold down to the faint magnitudes we observe and assume. 

Interestingly, while we are able to significantly better constrain stronger turnover models than weaker, the added contribution of photons from galaxies fainter than the turnover in weak-suppression models means that in a given redshift bin, the lower limits on the $\dot{n}_{\rm ion}$, $\rho_{\rm UV}$, and $SRFD$ are all broadly in line with one another for each suppression model (see Fig. \ref{fig:ndot}).

These lower limits on $\dot{n}_{\rm ion}$ and $\rho_{\rm UV}$ are elevated when compared to previous observational estimates as seen in previous works, notably \citet{Bouwens2015, Mascia_2023, Duncan2025_inprep, Atek2024Nature}, as seen in Figure \ref{fig:ndot} and Figure \ref{fig:rho_redshift}, particularly at $z=6$ where we are able to place the strongest constraints on the population of faint galaxies. At higher redshifts, our lower limits fall below previous calculations. The primary source of this discrepancy is the turnover: while previous works mostly assume a flat $M_{UV} = -15$ or $-17$ turnover, ours is physically motivated, with our turnover lower limit being brighter at higher redshifts and much fainter at lower redshifts. 

In the context of current discussions of the ionizing photon budget, our observations at $z=6$ and $z=7$ are potentially significant. Recent work has highlighted a possible ``photon-budget crisis'', in which galaxy-based estimates of the ionizing emissivity can become difficult to reconcile with constraints from the timing and duration of reionization if galaxies are assumed to have high $\xi_{\rm ion}$ and non-negligible escape fractions \citep{Munoz2024, Simmonds2024, Atek2024Nature}. Our results place stronger lower limits on the contribution from galaxies below the conventional $M_{\rm UV}=-15$ integration limit, suggesting there may be more UV photons than previously assumed by those who assume such a bright turnover.

Moreover, while we assume the \citet{Duncan2025_inprep} near-flat relationship between $M_{UV}$ and $\xi_{\rm ion}$ in calculating $\dot{n}_{\rm ion}$ (see Tab. \ref{tab:ndot_chisholm}), if the faint systems have ionizing photon production efficiencies higher than those inferred for brighter JWST-selected galaxies, as observed in \citet{Endsley2023_1,Simmonds2024,Atek2024Nature}, then $\dot{n}_{\rm ion}$ would be significantly higher than our lower limit.

\begin{figure*}
    \centering
    \includegraphics[width=0.85\textwidth]{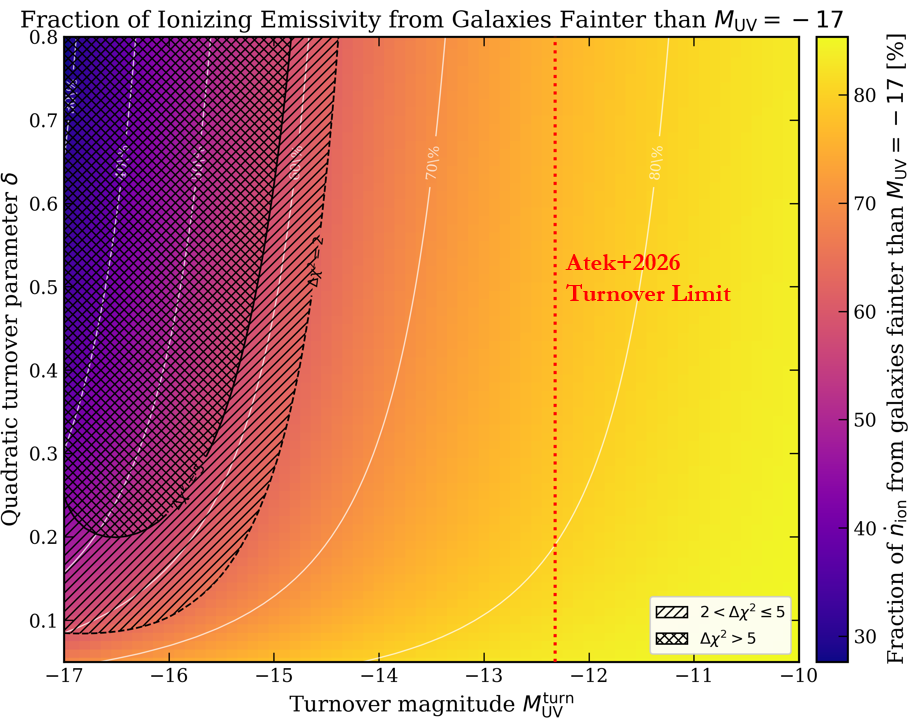}
    \caption{Fraction of the ionizing photon emissivity, $\dot{n}_{\rm ion}$, contributed by galaxies fainter than $M_{\rm UV}=-17$ at redshift 6 as a function of turnover magnitude and quadratic turnover strength. The calculation combines the DPL UVLF with the luminosity-dependent $\xi_{\rm ion}$ and $\beta_{\rm UV}$ relations from \citet{Duncan2025_inprep}, together with the $f_{\rm esc}(\beta_{\rm UV})$ prescription from \citet{Chisholm_2022}. Across the allowed parameter space, at least $\sim64\%$ of the ionizing photons are produced by galaxies fainter than $M_{\rm UV}=-17$, highlighting the dominant role of very faint galaxies in sustaining the ionizing photon budget required for reionization.}
    \label{fig:turnover}
\end{figure*}

However, we caution against interpreting this result as, by itself, establishing or resolving any ionizing photon ``overproduction'' tension. The mapping from $\rho_{\rm UV}$ to reionization history depends sensitively on the poorly constrained quantities $f_{\rm esc}$ and $\xi_{\rm ion}$, as well as on the recombination sink in the intergalactic medium. In the standard volume-averaged description,
\begin{equation}
\frac{{\rm d}Q_{\rm HII}}{{\rm d}t} =
\frac{\dot{n}_{\rm ion}}{\langle n_{\rm H}\rangle}
-
\frac{Q_{\rm HII}}{t_{\rm rec}},
\end{equation}
where $Q_{\rm HII}$ is the ionized hydrogen filling factor and $\langle n_{\rm H}\rangle$ is the mean comoving hydrogen number density. The recombination timescale may be written approximately as
\begin{equation}
t_{\rm rec}^{-1} \simeq
C\,\alpha_{\rm B}(T)\,\langle n_{\rm H}\rangle\,(1+z)^3
\left(1+\frac{Y}{4X}\right),
\end{equation}
where $C \equiv \langle n_{\rm HII}^2\rangle/\langle n_{\rm HII}\rangle^2$ is the effective clumping factor, $\alpha_{\rm B}(T)$ is the case-B recombination coefficient, and the final term accounts for the contribution of helium to the electron density \citep[e.g.][]{Madau1999,Davies2024}. Commonly adopted values in recent reionization models span approximately $f_{\rm esc}\sim0.05$--$0.2$ and $C\sim2$--$6$ at $z\gtrsim6$ \citep[e.g.][]{Mascia_2023,Asthana2025}, although inferences tied to the short mean free path near the end of reionization can favour substantially larger effective clumping factors, such as $C\approx12$ at $z\simeq5$--$6$ \citep{Davies2024}. Quasar-absorption based studies similarly suggest escape fractions at the $\sim10$--$20\%$ level by $z\sim5$--$6$, depending on the assumed faint-end contribution \citep{Cain2025}.

Our results therefore constrain the emissivity side of the reionization problem, but they do not uniquely determine the reionization history. By pushing any allowed turnover to fainter magnitudes, we increase the minimum plausible $\rho_{\rm UV}$ supplied by galaxies. This narrows the range of viable combinations of $(\rho_{\rm UV},\xi_{\rm ion},f_{\rm esc},C)$ that can simultaneously satisfy the galaxy UVLF constraints and independent probes of the ionization state of the IGM.

\subsubsection{Faint Galaxies Dominate Reionization}

A consequence of our turnover analysis is that faint galaxies must be treated as a major component of the reionization photon budget. Previous work has already argued that low-luminosity galaxies are likely to provide a substantial fraction of the ionizing photons required for reionization, particularly once galaxies below the conventional $M_{\rm UV}=-17$ integration limit are included \citep[e.g.][]{Robertson2022,Atek2024Nature}. Our results support this picture at the late stages of reionization. At $z=6$, where our faint-end constraints are strongest, galaxies fainter than $M_{\rm UV}=-17$ contribute more than half of the UV luminosity density in our turnover models, implying that these systems are not a negligible extrapolation but a central part of the star-forming galaxy emissivity.

Converting this UV luminosity density into ionizing photon emissivity introduces additional model dependence. The faint-galaxy contribution to $\dot{n}_{\rm ion}$ depends on the turnover magnitude, the turnover strength, the adopted $\xi_{\rm ion}(M_{\rm UV})$ relation, and the mapping between $M_{\rm UV}$, $\beta_{\rm UV}$, and $f_{\rm esc}$. Using the \citet{Duncan2025_inprep} luminosity-dependent $\xi_{\rm ion}$ and $\beta_{\rm UV}$ relations together with the \citet{Chisholm_2022} $f_{\rm esc}(\beta_{\rm UV})$ prescription, we find that galaxies fainter than $M_{\rm UV}=-17$ produce at least $\sim64\%$ of the ionizing photons from the star-forming galaxy population at $z=6$. This fraction refers only to the galaxy contribution, rather than any additional AGN component; however, current constraints suggest that AGN are unlikely to dominate the hydrogen-ionizing budget at these redshifts \citep{Trebitsch2021,Jiang2025}.

This result should still be interpreted as a lower limit under the adopted assumptions. The empirical $\xi_{\rm ion}$ and $f_{\rm esc}$ prescriptions are not directly calibrated for the extremely faint galaxies that dominate the extrapolated contribution, and the inferred fraction is therefore sensitive to how these relations behave below the luminosities currently accessible to spectroscopy. Nevertheless, the conclusion that faint galaxies are important is robust in a more general sense: even when considering only $\rho_{\rm UV}$, galaxies below $M_{\rm UV}=-17$ contribute more than half of the available UV light in our $z=6$ model. Moreover, if the true turnover occurs at fainter magnitudes than we can currently exclude, a possibility that is entirely probable, the contribution from these sources would increase further: our observations are simply lower limits. 

Such an interpretation is broadly consistent with recent JWST-era spectroscopic studies of low-luminosity or strongly lensed galaxies, which provide more direct constraints on nebular emission, $\xi_{\rm ion}$, and in some cases the conditions associated with significant LyC escape \citep[e.g.][]{Endsley2023_1,Simmonds2024,Atek2024Nature,asada2026}. These spectroscopic measurements support the plausibility of the stable $\xi_{\rm ion}$ and non-negligible $f_{\rm esc}$ values adopted in our emissivity calculation. While our photometric \textsc{Bagpipes} fits do not robustly constrain $\xi_{\rm ion}$ or $f_{\rm esc}$ for the full faint population, the galaxies that extend our UVLF to very faint magnitudes are consistent with the same broad picture of young, actively star-forming, weakly dust-obscured systems. It is difficult to know for certain if the $\xi_{\rm ion}$ or $f_{\rm esc}$ relationships adopted in this work are accurate, but even before applying luminosity-dependent prescriptions for $\xi_{\rm ion}$ and $f_{\rm esc}$, more than half of the UV luminosity density in our $z=6$ model arises from galaxies fainter than $M_{\rm UV}=-17$. Moreover, recent spectroscopic observations have even suggested that our relationships may be underestimating the $\xi_{\rm ion}$ and $f_{\rm esc}$ of this faint sample of galaxies \citep{asada2026, jecmen2026}. 

Taken together, these results support a picture in which faint galaxies provide a large, and potentially dominant, contribution to the ionizing photon budget. When these empirical ionizing-efficiency and escape-fraction scalings are included, the corresponding faint-galaxy contribution to $\dot{n}_{\rm ion}$ rises to at least $\sim64\%$ of all ionizing photons emitted by galaxies. It is crucial, then, that these sources are explicitly considered in modern modeling of reionization, rather than treated as an uncertain extrapolation with negligible impact. Even the treatment of the turnover as an integration limit after which there are no more UV luminous sources neglects a potential and unconstrained source of reionisation. At the same time, the conversion from faint-end UV light to escaping ionizing photons remains a major uncertainty. Thus, future spectroscopic, lensing, and multi-field studies are needed to extend empirical constraints on $\xi_{\rm ion}$, $\beta_{\rm UV}$, and $f_{\rm esc}$ into the extremely faint regime beyond $M_{UV} = -17$ and $-15$, where much of the reionization-relevant emissivity may reside. 

\begin{figure*}
    \centering
    \includegraphics[width=\textwidth]{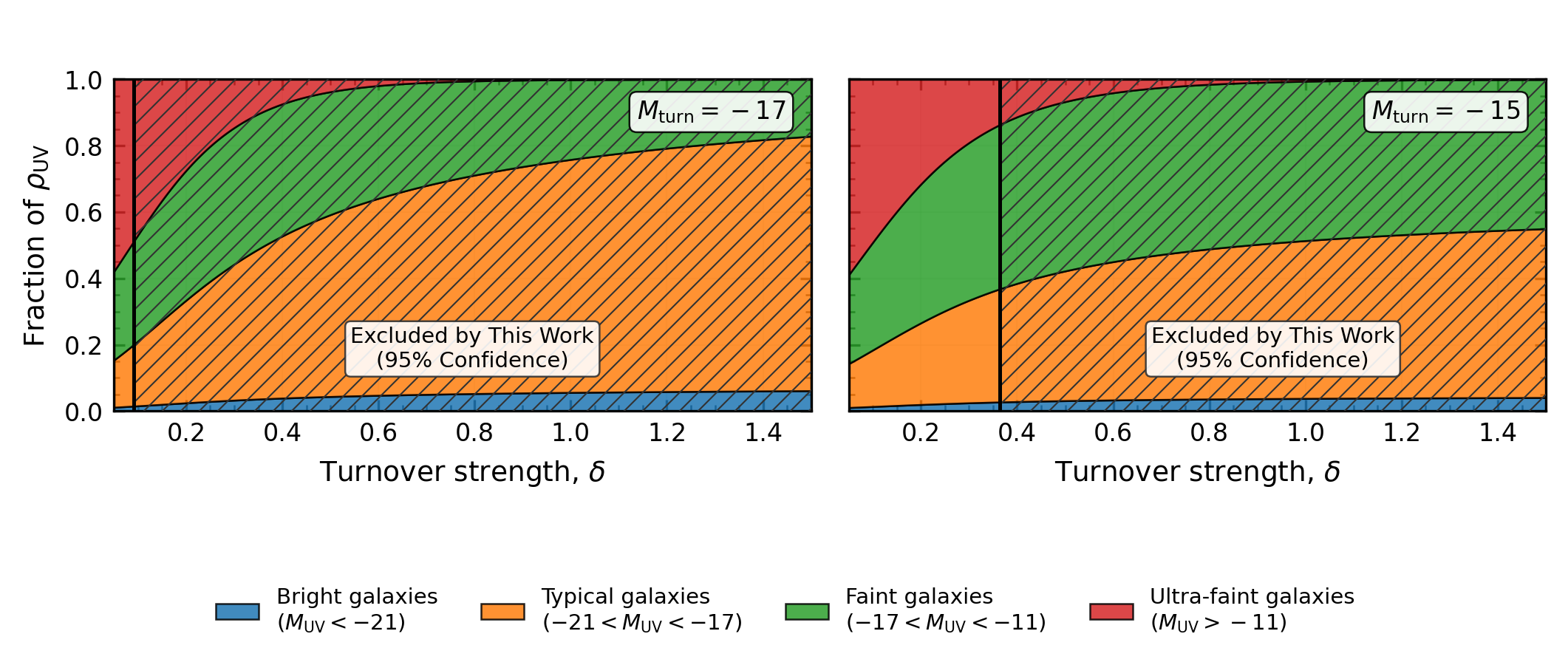}
    \caption{
    Estimated f-ractional contribution to the total UV luminosity density, $\rho_{\rm UV}$, as a function of the adopted turnover strength, $\delta$, for the $z=6$ double power-law luminosity function. The calculation uses the DPL parameters $M^{\ast}_{\rm UV}=-21.20$, $\log_{10}(\phi^{\ast}/{\rm Mpc^{-3}\,mag^{-1}})=-3.72$, faint-end slope $\alpha=-2.071$, and bright-end slope $\beta=-5.1$. The two panels show different assumed turnover magnitudes, $M_{\rm turn} = -17$ and $-15$. The stacked regions separate the relative contribution from bright galaxies, $M_{\rm UV}<-21$, typical galaxies, $-21<M_{\rm UV}<-17$, faint galaxies, $-17<M_{\rm UV}<-11$, and ultra-faint galaxies, $M_{\rm UV}>-11$. Hatched regions indicate turnover strengths excluded by this work at 95 per cent confidence, defined by $\Delta\chi^{2}=3.8$.  The vertical line shows the limit from our observations and the shaded is excluded. 
    }
    \label{fig:rhoUV_fraction_delta_turnover}
\end{figure*}

\subsection{Comparison to Previous GLIMPSE Works}
\label{subsec:compGLIMPSE}
The GLIMPSE survey, as one of the most unique and deepest surveys in JWSTs history, has already been the subject of much analysis and research, especially by those looking at the population of faint, highly magnified galaxies \citep{Chem2025, fujimoto2025, atek2026, Korber_2026, asada2026}. Much of their work overlaps with this one, but this provides an opportunity for independent confirmation of one of the most sensitive observations in reionization astrophysics. 

Our results, specifically properties (faint-end slope) of the UVLF, are in good agreement with the observations made in \citet{Chem2025,atek2026} (See Fig. \ref{fig:alpha}). At $z = 9$ and $11$, our DPL alphas are consistent within $1\sigma$ with \citet{Chem2025}. While it may seem like there is a tension between \citet{atek2026} and our observed alpha at $z=7$, their redshift bin spans our $z=6-8$ bins. Taking the weighted average of my Schechter alphas in those three bins, gives a weighted $\bar{\alpha}_{\mathrm{Sch}} = -1.976 \pm 0.025$, which is within $0.1\sigma$ of their reported $z=7$ Schechter faint end slope. The discrepancy in our $z=7$ bin may be due to cosmic variance (see $\S$ \ref{subsubsec:cosmicv}). Due to the independence of these results and studies, this serves as robust confirmation of the veracity of both of our results, despite the separate lensing models, image reduction, ICL handling, selection, photometry, and completeness pipelines used. 

As for our our calculations of $\rho_{UV}$ and $SFRD$, we continue to find strong agreement ($>1\sigma$ variation) with both despite our different handling of the turnover (see Fig. \ref{fig:rho_redshift}).

\subsection{Caveats and Limitations}
\label{subsec:caveats}

The results presented in this work are subject to several important systematic uncertainties that are inherent to studies of intrinsically faint galaxies behind massive lensing clusters. Although we have taken care to construct a robust high-redshift sample, propagate observational uncertainties, and account for the effects of lensing magnification where possible, the interpretation of the faint-end UVLF remains limited by both the complexity of the foreground mass distribution and the small effective survey volumes probed at the highest magnifications. These caveats are particularly important for any attempt to identify, or rule out, a turnover in the UVLF, since the relevant constraints are driven by the faintest luminosity bins, where the number statistics are poorest and the dependence on the lensing model is strongest.  As these are important issues we describe them in some detail in the following sub-sub-sections.

\subsubsection{Lensing-Induced Errors}

The dominant systematic uncertainty in this analysis arises from the lensing model of Abell~S1063. Although AS1063 is among the most extensively studied strong-lensing clusters, the magnification field remains uncertain, especially near critical curves and in regions of high magnification. These are precisely the regions that provide access to the intrinsically faintest galaxies. As a result, uncertainties in the lensing model propagate directly into the inferred intrinsic luminosities, effective survey volumes, and therefore the derived UVLF.

First, errors in the magnification factor, $\mu$, affect the inferred intrinsic absolute magnitudes of individual galaxies. This is particularly important at the faint end, where individual luminosity bins may contain only one or two sources. In such cases, even a modest uncertainty in $\mu$ can move a galaxy between adjacent magnitude bins, while larger uncertainties can significantly alter whether a source contributes to the faintest part of the UVLF at all. Since the turnover analysis is sensitive to the behaviour of precisely these faint bins, lensing-induced magnitude uncertainties represent a major limitation on the strength of any inferred constraint.

Second, the lensing model enters directly into the calculation of the effective survey volume. Highly magnified regions allow intrinsically fainter galaxies to be detected, but they correspond to very small source-plane areas. Consequently, the faintest galaxies in our sample are associated with effective $V_{\max}$ values of only tens of cubic megaparsecs. Any uncertainty in the magnification map therefore affects not only the inferred luminosity of a source, but also the volume over which such a source could have been detected. This coupling between luminosity and volume is an unavoidable feature of cluster-lensing UVLF measurements and becomes increasingly severe toward the faintest magnitudes.

Third, uncertainties in the cluster mass model propagate into the final number densities. The magnification, source-plane area, completeness correction, and $V_{\max}$ estimate are not independent quantities; all depend, either directly or indirectly, on the adopted lensing solution. This introduces degeneracies that are difficult to fully capture with a single error term. While our analysis incorporates lensing uncertainties as carefully as possible, the resulting UVLF should still be interpreted with caution, particularly in the faintest bins where the dependence on the lensing model is strongest.

These effects do not invalidate the use of AS1063 as a gravitational telescope, but they do limit the precision with which the intrinsic faint-end galaxy population can be reconstructed. In particular, they make it difficult to distinguish between a genuine physical turnover in the UVLF and fluctuations induced by magnification uncertainties, bin migration, incompleteness, or small-number statistics. For this reason, our turnover constraints should be interpreted as limits conditional on the adopted lensing model and its associated uncertainties, rather than as definitive measurements of the intrinsic turnover scale. 

A more robust discussion and treatment of the impact of these lens models is available in \citet{atek2026}. However, as mentioned in $\S$ \ref{subsec:compGLIMPSE}, the strong agreement between our observations and those of \citet{Chem2025, atek2026} are positive indicators as to the veracity of the model adopted in this paper. 

\subsubsection{Cosmic Variance}
\label{subsubsec:cosmicv}

The second major caveat is the impact of cosmic variance, which we do not include in our UVLF error budget. To estimate the cosmic variance errors in this work, we use the ``cosmic-variance'' python calculator from \citet{Jespersen2025}, which implements the cosmic variance prescription given in \citet{Moster2010} to determine the variance of dark matter, $\sigma_{\rm dm}$, from our survey area which is assumed to be square. This is related to the total cosmic variance error via the galaxy bias, $\sigma_{\rm CV}(z, M_{\rm UV})=b_{\rm g}(z, M_{\rm UV})\sigma_{\rm dm}$, which itself has a redshift and $M_{\rm UV}$ dependence. Although numerous clustering studies have shown that the galaxy bias is larger for the brightest galaxies at the highest redshifts \citep[e.g.][]{Harikane2016,Dalmasso2024a,Paquereau2025}, an improved cosmic variance prescription utilizing $b_{\rm g}(z, M_{\rm UV})$ has not yet been provided. We therefore choose to adopt the $b_{\rm g}(M_{\star} = 10^7\,\rm M_{\odot})$ prescription from \citet{Jespersen2025} for simplicity, noting that this will over/under-estimate $\sigma_{\rm CV}$ at the faint/bright end of the UVLF respectively.

Since cluster-lensing surveys probe very small source-plane volumes, especially at the faintest luminosities where high magnification is required, we expect the cosmic variance in this work to dominate the total error budget. Using the methodology explained above, the $6.15\,\rm arcmin^2$ total unmasked area from the \textit{entire} Abell S1063 lensing cluster suggests a cosmic variance $\sigma_{\rm CV}\sim 65\%$. While this by itself is substantial, the $\sim 14\,\rm arcsec^2$ total detectable area in the $M_{\rm UV}\simeq-13.5$ bin at $z\simeq7$ (with $\langle \mu \rangle=24.6$), leads to significantly larger $\sigma_{\rm CV}\sim 83\%$. Incorporating such large errors into our analysis would lead to a substantially greater uncertainty in the estimation of the UVLF faint-end slope and inferred turnover magnitude/slope.

While $b_{\rm g}$ estimates have historically been produced by halo occupation distribution (HOD) constraints from wide area galaxy clustering studies \citep[e.g.][]{Hatfield2018}, it is perhaps more relevant for us to use estimates derived on the scale of an individual NIRCam pointing. The variance in number counts from the PArallel wide-area Nircam Observations to Reveal And Measure the Invisible Cosmos \citep[PANORAMIC;][]{Williams2025} survey, has recently revealed $b_{\rm g}\gtrsim25-40$ at $z\simeq10$. This is a factor $\sim2-3$ times larger than suggested by galaxy clustering in the COSMOS-Web field \citep{Paquereau2025}, reflecting the galaxy bias scale dependence, and leading to an equivalent increase in $\sigma_{\rm CV}$. 

This caveat is especially pertinent for the $z\simeq7$ UVLF shown in Figure \ref{fig:uvlf_secondary}, where the faintest bins appear to lie slightly below the extrapolation expected from the lower-redshift behaviour and from comparison UVLF measurements. Using pure-parallel inferred $b_{\rm g}$ estimates yields a monumental $\sigma_{\rm CV}\sim 165-250\%$ in the faintest magnitude bin. Without consideration of these errors, or the assumed smooth evolution of $M_{\rm t}$, this could lead to an improper interpretation of the data.

It is important to note that these contributions are mere extrapolations to extremely faint magnitudes. As has been repeatedly established, we are currently unable to observe the turnover, much less understand the population of galaxies beyond. However, this analysis is done to properly model, understand, and put a robust lower limit on the ionizing photon population during the Epoch of Reionisation.

\section{Conclusion}
\label{sec:conc}

In this paper, we have presented an analysis of the galaxy population behind Abell S1063 using recent JWST imaging together with archival HST data. Leveraging the strong gravitational potential of Abell S1063, we construct a photometric catalogue of faint, low-mass, lensed galaxies reaching intrinsic magnitudes of approximately $M_{\rm UV}\simeq-13$ to $-20$. Our \textsc{Bagpipes} SED fits place many of these sources in the low-mass regime, with typical stellar masses of $\log_{10}(M_\star/M_\odot)\sim5$--$8$.

Using this lensed faint galaxy sample, and in particular the highly magnified low-mass sources, we constrain the faint-end slope, $\alpha$, of the rest-frame UV luminosity function in five redshift slices centered at $z \simeq 6,7,8,9$, and $11$. Since the GLIMPSE field does not strongly constrain the bright end of the UVLF, we anchor $(M^\star,\phi^\star)$ to literature measurements and refit only the faint-end slope. We perform this analysis using both Schechter and double power-law parameterisations, but primarily use the DPL model for comparison because it has been found to better describe the bright-end behaviour of the UVLF at high redshift \citep{Bowler2015}. These observations of the evolution of the faint-end slope are in strong agreement with literature values, including of other UVLFs taken from AS1063 \citep{Livermore2017, Adams2024EPOCHS, Chem2025, atek2026}. 

We then use these UVLF fits to investigate whether the data requires a turnover at the faintest magnitudes. While we find 'no evidence' of a turnover down to $M_{UV}=-13.5$ (Fig. \ref{fig:turnover}), we adopt a new more robust search of the turnover parameter space, allowing for far shallower turnover models that would include larger populations of galaxies beyond the turnover. While it is harder to prove the non-existence of the turnover, especially a shallow turnover, we are able confidently exclude even a weak turnover model at $z=6$ down to $M_{UV}=-15.9$, with stronger models being statistically excluded down to $M_{UV}=-14.8$ (see Fig. \ref{fig:turnover_heat}). These results are most robust at $z=6$ both due to the limitations of observational data and the theoretical evolution and strengthening of UVLF suppression mechanisms at the end of reionization.

Rather than treating the turnover as a hard luminosity cutoff, we model it as a gradual quadratic suppression, allowing galaxies fainter than the turnover magnitude to still contribute substantially to the integrated luminosity density. We use this to derive conservative lower limits on the UV luminosity density, $\rho_{\rm UV}$, and the corresponding star-formation-rate density (Table \ref{tab:rho_uv_results}). These values should be interpreted as lower bounds because any turnover occurring at fainter magnitudes would increase the integrated contribution from faint galaxies.

We also convert our UV luminosity-density limits into ionizing photon emissivity ($\dot{n}_{ion}$ ) estimates using a luminosity-dependent $\xi_{\rm ion}$ relation, an updated $\beta_{\rm UV}(M_{\rm UV})$ relation from \citet{Duncan2025_inprep}, and the \citet{Chisholm_2022} $f_{\rm esc}(\beta_{\rm UV})$ prescription. These again provide conservative lower limits on $\dot{n}_{ion}$ across our redshift bins and for different turnover strengths.

Under these assumptions, galaxies fainter than $M_{\rm UV}=-17$ produce at least $\sim64\%$ of the ionizing photons from the star-forming galaxy population at $z=6$ (Fig. \ref{fig:turnover}). While our adopted $\xi_{\rm ion}$ and $f_{\rm esc}$ relations must be extrapolated into a magnitude regime where they are not yet well calibrated, recent spectroscopic studies have found that our assumptions may even be conservative in the faint-regime, leading to an under-estimation of the importance of the faint galaxy population \citep{jecmen2026, asada2026}.

Our observations therefore demonstrate that faint galaxies made a substantial, and potentially dominant, contribution to the ionising photon budget at the end of reionisation, even under conservative and statistically robust constraints on the UVLF turnover. By treating the turnover as a gradual suppression in galaxy abundance rather than as an artificial hard integration limit, we recover a persistent population of low-mass, low-luminosity systems whose cumulative contribution to reionisation is potentially quite significant while still poorly understood.

\section*{Acknowledgements}

We acknowledge support from the ERC Advanced Investigator Grant EPOCHS (788113), as well as two studentships from the STFC.  This work is based on observations made with the NASA/ESA \textit{Hubble Space Telescope} (HST) and NASA/ESA/CSA \textit{James Webb Space Telescope} (JWST) obtained from the \texttt{Mikulski Archive for Space Telescopes} (\texttt{MAST}) at the \textit{Space Telescope Science Institute} (STScI), which is operated by the Association of Universities for Research in Astronomy, Inc., under NASA contract NAS 5-03127 for JWST, and NAS 5–26555 for HST. The authors thank all involved with the construction and operation of JWST, without whom this work would not be possible.

%%%%%%%%%%%%%%%%%%%%%%%%%%%%%%%%%%%%%%%%%%%%%%%%%%
\section*{Data Availability}



\bibliographystyle{mnras}
\bibliography{example,clustering,cosmic_variance,lensing} % if your bibtex file is called example.bib

%\begin{thebibliography}{99}
%\end{thebibliography}

%%%%%%%%%%%%%%%%%%%%%%%%%%%%%%%%%%%%%%%%%%%%%%%%%%

%%%%%%%%%%%%%%%%% APPENDICES %%%%%%%%%%%%%%%%%%%%%

\appendix
\section{Foreground Cluster Subtraction}
\label{app:icl_subtraction}

\begin{figure*}
    \centering
    \includegraphics[width=\textwidth]{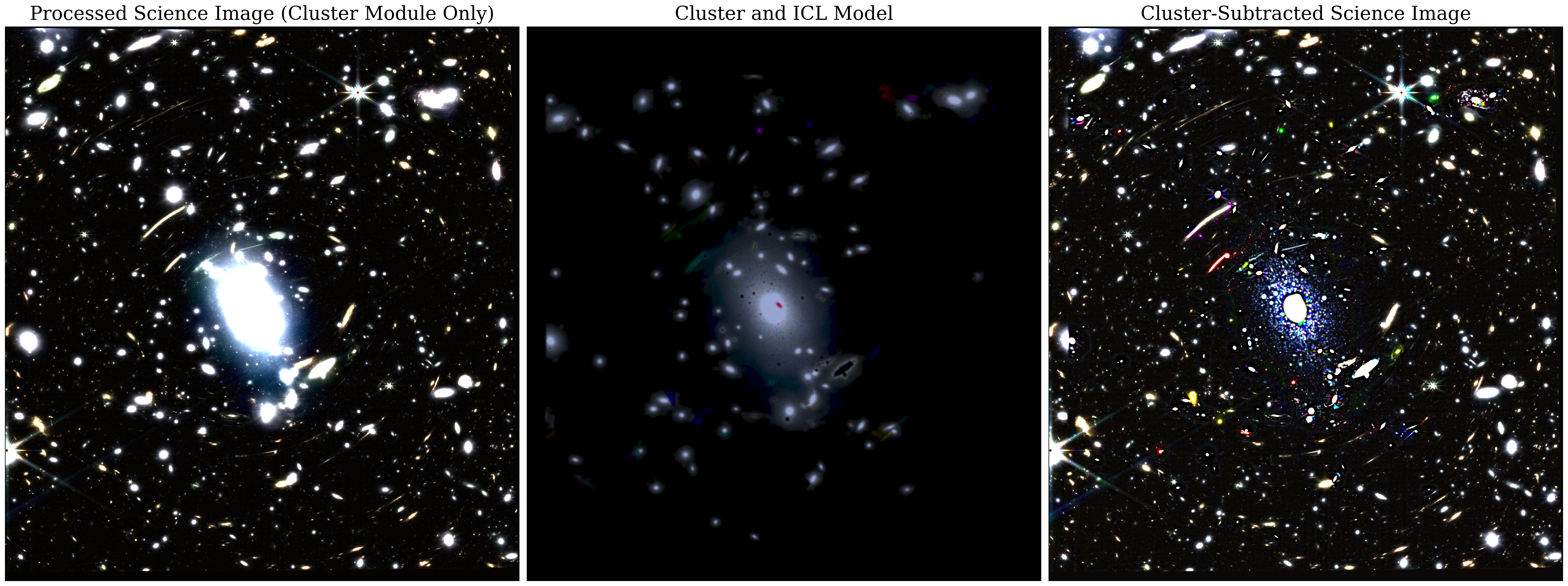}
    \caption[ICL Subtraction]{RGB Composite of JWST NIRCam F200W, F277W and F356W filter images of the cluster module prior to cluster subtraction, of the cluster model itself, and of the subtracted science image.}
    \label{fig:icl}
\end{figure*}

The detection and photometric measurement of faint, strongly lensed galaxies behind Abell S1063 is complicated by foreground emission from the cluster itself. The dominant sources of this foreground light are the brightest cluster galaxy (BCG), the surrounding intracluster light (ICL), and other bright cluster members. In AS1063, the BCG is particularly dominant and is embedded within an extended ICL component, producing strong low-spatial-frequency structure across the central region of the mosaic. This emission increases the local background, reduces the effective depth, and introduces spatially varying gradients that can bias the photometry of faint background sources. 

Before applying the custom subtraction described here, the JWST science mosaics were processed using the mesh-based background subtraction described in $\S$~\ref{subsec:glmpse}. This procedure estimated and subtracted the local background on a $32\times32$ pixel grid. While this removed a substantial fraction of the diffuse background and large-scale ICL contribution, significant residual foreground light remained in the vicinity of the BCG, bright cluster members, and the central ICL-dominated region. We therefore applied an additional cluster-light subtraction step designed specifically to isolate and remove only the largest-scale diffuse component.

The diffuse cluster light was modelled independently in each JWST and HST band using an undecimated `a trous wavelet decomposition, chosen because it preserves the native pixel grid at all wavelet levels, allowing the final diffuse model to be subtracted directly from the science mosaic without resampling. For each input image, the decomposition was constructed using the separable B3-spline scaling kernel
\begin{equation}
h[n] = \frac{1}{16}[1,4,6,4,1],
\qquad
H(x,y) = h[x]h[y],
\end{equation}
where the kernel at wavelet level $j$ is dilated by inserting $2^{j-1}-1$ zeros between adjacent coefficients. For an initial masked image $S_0$, the smoothed image at level $j$ is given by
\begin{equation}
S_j = S_{j-1} \ast H^{(j)},
\end{equation}
where $H^{(j)}$ is the dilated two-dimensional smoothing kernel and $\ast$ denotes convolution. The corresponding wavelet plane is
\begin{equation}
W_j = S_{j-1} - S_j .
\end{equation}
The original image can then be expressed as
\begin{equation}
S_0 = S_J + \sum_{j=1}^{J} W_j ,
\end{equation}
where $S_J$ is the residual smooth component after $J$ successive smoothing operations, and $W_j$ contains structure on the characteristic spatial scale associated with wavelet level $j$.

In this work we adopt $J=10$ and retain only the coarsest plane, $S_J$, as the basis of the diffuse-light model. This choice deliberately restricts the model to the largest-scale components, not the small, compact objects that may be higher redshift galaxies. This method is thus not intended to provide a physically complete model of the cluster, but rather just a smooth cluster model that we can use to improve the quality of our photometry. 

Prior to constructing the final diffuse model, a support mask, denoted $\mathcal{H}_J(\mathbf{x})$, was defined from the coarsest wavelet plane. This mask determines where the smooth component is allowed to contribute to the subtraction model. Pixels were included in $\mathcal{H}_J(\mathbf{x})$ only if the value of the coarsest plane was positive and exceeded $0.5\sigma_J$, where $\sigma_J$ is the robust dispersion of $S_J$ measured on unmasked pixels. This threshold ensures that only significant positive large-scale residuals are treated as foreground cluster light. In addition, pixels were required to belong to a connected component with an area of at least $5000$ pixels. This connected-area criterion removes small isolated structures, which are more likely to correspond to noise, compact sources, or local artefacts rather than the extended BCG+ICL morphology. A $50\sigma$ outlier mask was also applied during the modelling stage to prevent extremely bright or pathological pixels from influencing the smooth component.

The final diffuse-light model is then given by
\begin{equation}
M_{\rm diff}(\mathbf{x}) =
\mathcal{H}*J(\mathbf{x}) S_J(\mathbf{x}) .
\end{equation}
This model is subtracted from the input science mosaic,
\begin{equation}
I*{\rm final}(\mathbf{x}) =
I_{\rm in}(\mathbf{x}) - M_{\rm diff}(\mathbf{x}) ,
\end{equation}
where $I_{\rm in}$ denotes the science image entering the cluster-light subtraction step. Since the model is generated on the native pixel grid and is subtracted directly from the input mosaic, the final products retain the original WCS and pixel scale of the science images, apart from the removal of the smooth diffuse-light component.

The procedure was applied independently to each JWST and HST band. We chose not to impose a single common diffuse-light morphology across all filters because the BCG, ICL, and cluster members have wavelength-dependent surface-brightness distributions and colour gradients. 

The adopted configuration was fixed empirically after testing a range of wavelet depths, support thresholds, connected-component areas, and outlier masks. The final configuration uses 10 wavelet levels, retains only the coarsest plane, applies a $0.5\sigma_J$ support threshold, requires a minimum connected-component area of $5000$ pixels, and uses a $50\sigma$ outlier mask. These choices were selected to homogenize the background depth between the two NIRCam modules while minimizing visible over-subtraction around compact sources. Shallower wavelet decompositions or the inclusion of finer wavelet planes produced models that began to follow intermediate-scale structure around cluster members and faint sources. Conversely, more restrictive support masks left significant residual BCG+ICL light in the central region. The adopted parameters therefore represent a compromise between diffuse-light removal and photometric conservatism.

Although the subtraction substantially improves the uniformity of the background, it is not intended to model the detailed stellar light profiles of all cluster members. In regions immediately surrounding very bright galaxies, the foreground light can vary on smaller spatial scales than those captured by the coarsest wavelet plane. Such regions are more susceptible to residuals or over-subtraction. We therefore manually mask areas requiring extreme subtraction, particularly close to bright cluster members and the BCG core, during the subsequent source-selection analysis. This ensures that the final high-redshift sample is not driven by sources whose photometry is dominated by uncertain foreground-light modelling.

\begin{figure}
    \centering
    \includegraphics[width=\columnwidth]{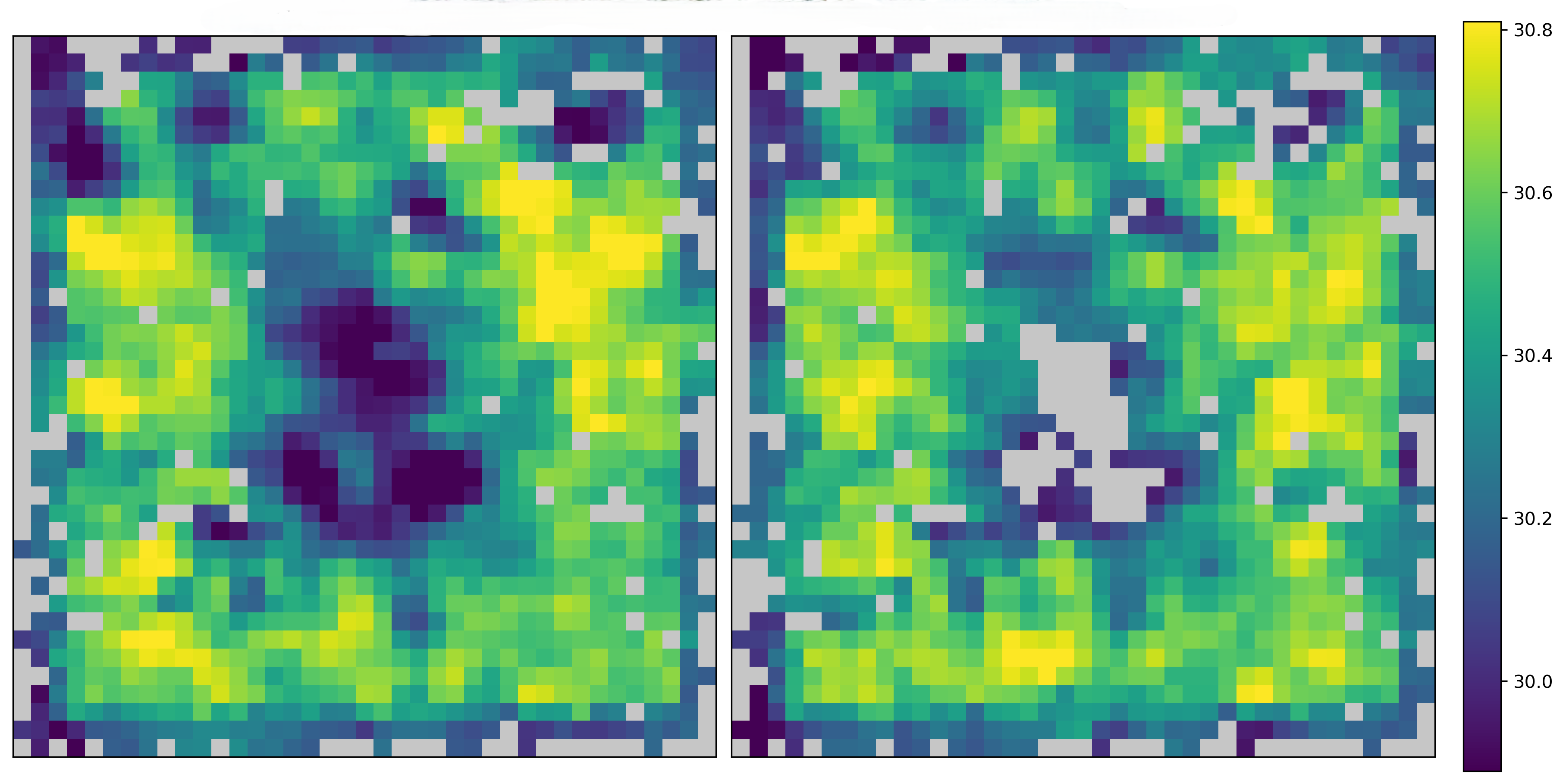}
    \caption[ICL Subtraction]{5-sigma rolling depth of the cluster module of the F277W JWST imaging of the AS1063 cluster prior to (Left Panel) and after the custom cluster light subtraction pipeline (right panel). Each bin is 120x120 pixels, and greyed out regions are regions with fewer than 2 objects identified by \texttt{SExtractor} after masking.  }
    \label{fig:icl2}
\end{figure}

Figure~\ref{fig:icl} illustrates the subtraction procedure. The figure shows the input image, the derived diffuse-light model, and the residual image after subtraction. The model follows the extended BCG+ICL+cluster member emission while avoiding subtracting compact sources, and the residual image showed a more spatially uniform background suitable for faint-source detection and photometry (see Figure \ref{fig:icl2}).

%%%%%%%%%%%%%%%%%%%%%%%%%%%%%%%%%%%%%%%%%%%%%%%%%%

% Don't change these lines
\bsp	% typesetting comment
\label{lastpage}
\end{document}